\documentclass[12pt]{article}

\newlength{\abstractwidth}
\usepackage{epsf}
\usepackage{color}
\usepackage{graphicx}
\usepackage{tikz}
\usetikzlibrary{arrows,shapes,positioning}
\usetikzlibrary{decorations.markings}
\usepackage[rightcaption]{sidecap}
\tikzstyle arrowstyle=[scale=1]
\tikzstyle directed=[postaction={decorate,decoration={markings,
    mark=at position .65 with {\arrow[arrowstyle]{stealth}}}}]
\tikzstyle reverse directed=[postaction={decorate,decoration={markings,
    mark=at position .65 with {\arrowreversed[arrowstyle]{stealth};}}}]
\usetikzlibrary{positioning}

\usepackage{amssymb}
\usepackage{latexsym}

\renewcommand{\thefootnote}{\fnsymbol{footnote}}
\renewcommand{\thanks}[1]{\footnote{#1}}
\newcommand{\starttext}{
\setcounter{footnote}{0}
\renewcommand{\thefootnote}{\arabic{footnote}}}

\newcommand{\bea}{\begin{eqnarray}}
\newcommand{\eea}{\end{eqnarray}}
\newcommand{\be}{\begin{eqnarray}}
\newcommand{\ee}{\end{eqnarray}}

\def\cB{{\cal B}}
\def\cC{{\cal C}}
\def\cD{{\cal D}}

\def\cF{{\cal F}}
\def\cG{{\cal G}}

\def\cL{{\cal L}}
\def\cM{{\cal M}}

\def\cO{{\cal O}}

\def\cR{{\cal R}}

\def\cV{{\cal V}}

\def\mC{\mathfrak{C}}

\def\mS{\mathfrak{S}}

\def\mm{\mathfrak{m}}

\def\mpp{\mathfrak{p}}

\def\ZZ{{\mathbb Z}}
\def\RR{{\mathbb R}}
\def\NN{{\mathbb N}}
\def\CC{{\mathbb C}}

\def\Im{{\rm Im \,}}
\def\tr{{\rm tr}}

\def\half{{1\over 2}}
\def\p{\partial}
\def\DD{\nabla  }

\def\no{\nonumber}
\def\sm{\smallskip}

\def\L{R}

\begin{document}
\starttext
\setcounter{footnote}{0}

\begin{flushright}
2016 August 15
\end{flushright}

\vskip 0.3in

\begin{center}

{\Large \bf Hierarchy of Modular Graph Identities}\footnote{The research reported in this paper was supported in part by the National Science Foundation under the grants PHY-13-13986 and PHY-16-19926.}

\vskip 0.5in

{\large \bf Eric D'Hoker  and Justin Kaidi } 

\vskip 0.2in

{\sl Mani L. Bhaumik Institute for Theoretical Physics }\\
{ \sl Department of Physics and Astronomy }\\
{\sl University of California, Los Angeles, CA 90095, USA}

{\tt \small dhoker@physics.ucla.edu; jkaidi@physics.ucla.edu}

\vskip 0.3in

\end{center}

\begin{abstract}

The low energy expansion of Type II superstring amplitudes at genus one  is organized in terms of modular graph functions associated with Feynman graphs of a conformal scalar field on the torus. In earlier work, surprising identities between two-loop graphs at all weights, and between higher-loop graphs of weights four and five were constructed. In the present paper, these results are generalized in two complementary directions. First, all identities at weight six and all dihedral identities at weight seven are  obtained and proven. Whenever the Laurent polynomial at the cusp is available, the form of these identities confirms the pattern by which the vanishing of the Laurent polynomial governs the full modular identity. Second, the family of modular graph functions is extended to include all graphs  with derivative couplings and worldsheet fermions. These extended families of modular graph functions are shown to obey a hierarchy of inhomogeneous Laplace eigenvalue equations. The eigenvalues for the extended family of dihedral modular graph functions are calculated analytically for the simplest infinite sub-families and obtained by Maple for successively more complicated  sub-families. The spectrum is shown to consist solely of eigenvalues $s(s-1)$ for positive integers $s$ bounded by the weight, with multiplicities which exhibit rich representation-theoretic patterns.

\end{abstract}

\newpage

\setcounter{tocdepth}{2}

\newpage

\section{Introduction}
\setcounter{equation}{0}
\label{intro}

The low energy expansion of Type IIB superstring amplitudes is constrained by $SL(2,\ZZ)$ duality which requires the dependence on the vacuum expectation value of the axion-dilaton field to be through modular functions \cite{Green:1997tv,Green:1997as}. For the lowest orders in the expansion in $\alpha '$ space-time supersymmetry  subjects these modular functions to homogeneous and inhomogeneous Laplace eigenvalue equations \cite{Green:1998by,Obers:1999um}. The resulting modular functions successfully pass various consistency checks against predictions from superstring perturbation theory at small string coupling, and their structure leads to powerful non-renormalization theorems for low energy effective interactions protected by supersymmetry
\cite{D'Hoker:2002gw,D'Hoker:2005dhp,D'Hoker:2005ht,GomezMafra,D'Hoker:2014gfa,Pioline:2015yea}.

\sm

Remarkably, a similar structure of modular functions obeying inhomogeneous Laplace eigenvalue equations has emerged in the superstring perturbation theory contributions to the low energy expansion of superstring theory. In this context, however, the dependence of the modular functions is not on the axion-dilaton field, but rather on the moduli of the worldsheet. The physical contribution to the low energy expansion is obtained by integrating these modular functions over the moduli space of the worldsheet. The pattern of Laplace eigenvalue equations as a function of moduli was first encountered in the study of supergravity amplitudes \cite{Green:2008bf}. More recently, a Laplace eigenvalue equation was shown to govern the genus-two contribution to the $D^6 \cR^4$ effective interaction \cite{D'Hoker:2014gfa}, and to be related to the Zhang-Kawazumi invariant \cite{D'Hoker:2013eea} introduced earlier in number theory  \cite{Zhang,Kawazumi}.

\sm

The most extensively studied situation is for superstring perturbation theory contributions at genus one where the $\alpha '$ expansion may be organized in terms of non-holomorphic modular functions of the complex modulus $\tau$ of the worldsheet torus. These contributions arise from correlators of a conformal scalar field on the torus for which each Feynman graph gives rise to a {\sl modular graph function} \cite{Green:2008uj,D'Hoker:2015qmf, D'Hoker:2016a}. The  {\sl number of loops} $L$ (also referred to as the {\sl depth}) and the {\sl weight} $w$ of the modular graph function respectively correspond to the number of loops $L$ and the number of edges $w$ of the associated Feynman graph for the conformal scalar  field theory on the torus. Modular graphs functions are thus given by multiple discrete sums over the momenta of the torus, namely one momentum for each edge constrained by  momentum conservation at each vertex. 

\sm

At one loop, namely for $L=1$, a modular graph function of weight $w$ is proportional to the non-holomorphic Eisenstein series $E_w(\tau)$. At two loops, namely for $L=2$, modular graph functions of weight $w$ were found to obey a hierarchy of inhomogeneous Laplace eigenvalue equations whose inhomogeneous parts are quadratic in Eisenstein series  \cite{D'Hoker:2015foa}. The eigenvalues were shown to be of the form $s(s-1)$ for positive integers $s$ bounded by the weight of the graph~$s \leq w-2$. A zero eigenvalue arises for each odd weight $w$ and the corresponding Laplace equation may be integrated to a linear algebraic identity between modular graph functions.  At higher loops, namely for $L \geq 3$, no systematic structure was found so far. Instead a number of algebraic identities were discovered at weights $w=4,5$ in \cite{D'Hoker:2015foa}, and subsequently proven in \cite{D'Hoker:2016a,D'Hoker:2015zfa}. 

\sm

The higher loop identities for weights four and five were conjectured by matching their Laurent polynomial in (both positive and negative) powers of $\Im (\tau)$ at the cusp $\tau \to i \infty$. In each case when such a matching was possible an identity between modular graph functions resulted that could  be proven by other methods. The observation of this pattern raises the question whether a polynomial combination of modular graph functions whose Laurent polynomial at the cusp vanishes must necessarily vanish as an identity  on the entire upper half $\tau$-plane. The general validity of the assertion remains an open problem. Recent calculations \cite{Zerbini} of the Laurent polynomials of certain modular graph functions at weights 6, 7, and 8 further challenge us to prove new conjectured identities \cite{DGV}, and to seek a general theorem.
The underlying nature of these algebraic identities remains to be fully uncovered, but there is a sense in which they generalize to modular functions the algebraic relations which exist between multiple zeta values
\cite{ZagierMZV,Hoffmann,W,BBBL,Zudilin,Blumlein:2009cf,Brown:2013gia}. For the role of multiple zeta values   in string amplitudes, see for example  \cite{Broedel:2013tta,Stieberger:2013wea,Broedel:2015hia,SCHLotterer:2012ny}.

\sm

In the present paper, we shall make progress on these problems on three different fronts. 

\sm

First, we shall use the sieve algorithm and holomorphic subgraph reduction, developed in \cite{D'Hoker:2016a}, to organize a systematic search for, and a complete proof of, all algebraic identities between modular graph functions at weight six, all dihedral modular graph functions at weight seven, and one trihedral weight seven identity between modular graph functions whose Laurent expansions involve irreducible multiple zeta values, as evaluated in \cite{Zerbini}. The methods are systematic and effective at any weight, though the combinatorial complication significantly increases with weight, even with the assistance of Mathematica or Maple. We confirm that, at these low weights and whenever the Laurent polynomials are available, their matching predicts identities which hold throughout the upper half $\tau$-plane.

\sm

Second, we shall  introduce a natural decomposition of modular graph functions into {\sl primitive modular graph functions}, obtained by subtracting those contributions to the momentum sums for which a subset of all the momenta entering any given vertex sums to zero. For those momentum configurations the graph effectively becomes disconnected. Remarkably, the modular graph identities proven so far at weights 4, 5, 6, and 7 become {\sl linear} when expressed in terms of primitive modular graph functions.  The decomposition is modular invariant, since the vanishing of a momentum is a modular invariant condition.  The decomposition is unfamiliar in customary quantum field theories with continuous momentum range since it relies entirely on the discreteness of the loop momentum spectrum on the torus. We conclude this part by showing that the combinatorics of the decomposition matches the combinatorics which appears in the holomorphic subgraph reduction procedure developed to search for and prove the identities, thereby promoting the effectiveness of the decomposition results to general weight and loop orders.

\sm

Third, the space of modular graph functions is {\sl extended} to include all graphs which arise in the low energy expansion  to genus-one order with derivative and non-derivative scalar couplings as well as worldsheet fermions. In the notations of \cite{D'Hoker:2016a}, the exponents of holomorphic and anti-holomorphic momenta may now vary independently, though their sums must equal one another to guarantee modular invariance. The motivation for this extension is that, for three loops and higher,  the action of the Laplacian closes on this extended space, but did not close on the space of modular graph functions whose exponents on holomorphic and anti-holomorphic momenta are pairwise equal to one another.  In this third part, we shall show that the modular graph functions in this extended space obey a  hierarchy of inhomogeneous Laplace eigenvalue equations, whose structure generalizes the one found for two-loop modular graph functions in \cite{D'Hoker:2015foa}. 

\sm

For the extended family of dihedral modular graph functions the eigenvalues of the system of inhomogeneous Laplace eigenvalue equations are calculated analytically for the simplest infinite sub-family which includes modular graph functions of arbitrarily high weight, and obtained by Maple for successively more complicated sub-families. Remarkably, on each one of these sub-families, the eigenvalues of the Laplacian are of the form $s(s-1)$ with $s $ a positive integer, bounded from above by an expression which involves the weight $w$ of the graph. The multiplicities of the eigenvalues, derived by Maple calculations up to sufficiently high weights, are found to exhibit rich representation-theoretic patterns, just as the multiplicities did in the case of two-loop modular graph functions in \cite{D'Hoker:2015foa}. In the two-loop case a complete proof of the multiplicity patterns was obtained by using methods based on generating functions. For higher loops complete proofs of the multiplicity patterns, using generating functions or other methods,  remains a challenging open problem.

\subsection{Organization}

The remainder of this paper is organized as follows. In section 2 we present a reasonably self-contained review of the structure of the genus-one low-energy expansion and its formulation in terms of modular graph functions. In section 3 we obtain and prove new identities for weights six and seven via the techniques developed in \cite{D'Hoker:2016a}. In addition to the dihedral and trihedral modular graph functions which generalize the simplest cases  in \cite{D'Hoker:2016a}, we will also examine the single tetrahedral case which first appears at weight six.  Implications for the Laurent polynomials near the cusp of these modular graph functions are obtained.  In section 4, we introduce {\sl primitive modular graph functions}, and  show that all the modular graph identities constructed in this paper  become linear when expressed in terms of primitive modular graph functions. In section 5 we show that the extended classes of modular graph functions obey a hierarchy of inhomogeneous Laplace eigenvalue equations. On a hierarchy of infinite families of dihedral modular graph functions the action of the Laplacian is diagonalized and shows that all eigenvalues are of the form $s (s-1)$ for $s$ integer. Discussions of our results, and the open problems they generate, are presented  in section 6. Appendix A provides a summary of holomorphic subgraph reduction formulas for the dihedral, trihedral, and tetrahedral graphs needed in this paper.

\newpage

\section{Review of modular graph functions and forms}
\setcounter{equation}{0}
\label{review}

In this section, we present a reasonably self-contained review of modular graph functions and forms which arise in the genus-one contribution to the low energy effective interactions of closed oriented superstring theories, such as Type IIB and Heterotic, and summarize some of the key results  derived in earlier publications \cite{Green:2008uj,D'Hoker:2015qmf, D'Hoker:2016a}.

\subsection{Genus-one superstring amplitudes}

Genus-one contributions to the low energy expansion of closed oriented superstrings are given by the integral over the moduli space for the torus of conformal correlators. Although worldsheet fermion and ghost correlators may appear at intermediate stages of the calculation, bosonization techniques allow us to reformulate these correlators in terms of conformal scalar fields, possibly after summation over spin structures. Throughout, we shall assume that this reformulation has been carried out. 

\sm

Concretely, a torus worldsheet $\Sigma$ with complex structure modulus $\tau$ may be represented in the complex plane by the quotient $\CC/\Lambda$ where the lattice $\Lambda $ is given by $\Lambda = \ZZ \oplus \tau \ZZ$. In terms of local complex coordinates $z, \bar z$ on $\Sigma$, the volume form is normalized as follows $d^2z = { i \over 2} dz \wedge d \bar z$. The moduli space $\cM_1$ of the torus may be represented by the quotient of the Poincar\'e upper half plane by $PSL(2,\ZZ)$. In terms of local complex coordinates $\tau, \bar \tau$ on $\cM_1$, the standard fundamental domain is given by $\cM_1= \{ \tau \in \CC, ~ 0< \tau_2 , ~ |\tau_1|\leq { 1 \over 2} , ~  1 \leq |\tau| \}$, with $\tau=\tau_1+i\tau_2$ and $\tau_1, \tau_2 \in \RR$, with the  Poincar\'e metric $ i d \tau \wedge d \bar \tau / 2 \tau_2^2$.

\sm

Given that bosonization formulas have been used to reformulate all worldsheet fermion and ghost correlators in terms of scalar correlators, the fundamental building block of the amplitudes is the scalar Green function $G$ on the torus $\Sigma$ with modulus $\tau$, defined by,
\bea
\p_{\bar z} \p_z G(z|\tau) = - \pi \delta ^{(2)} (z) +{ \pi \over \tau_2} 
\hskip 1in \int _\Sigma d^2 z \, G(z|\tau)=0
\eea
While explicit formulas for $G$ are available in terms of Jacobi $\vartheta$-functions, it will be convenient for our purpose to express $G$ as a Fourier sum over the lattice $\Lambda$,
\bea
\label{2a2}
G(z|\tau) = \sum _{p \in \Lambda}' { \tau _2 \over \pi |p|^2 } \, e^{2 \pi i (n \alpha - m \beta)}
\eea
where $z=\alpha + \beta \tau$ with $\alpha, \beta \in \RR/\ZZ$.  The integers $m,n$ parametrize the discrete momenta of the torus $p=m + n \tau \in \Lambda$, and the prime superscript on the sum  indicates that the contribution from the point $p=0$ in the lattice $\Lambda$ must be omitted from the summation.

\sm

The genus-one contribution to the low energy effective Lagrangian $\cL_{\rm eff}$ for fields in the supergravity multiplet  then takes the following schematic form,  
\bea
\label{2a4}
\cL_{\rm eff} = \sum _N \cV_N \int _{\cM_1} {i d \tau \wedge d \bar \tau \over 2 \tau_2^2} \, \cB_N (s_{ij}|\tau)
\eea
The sum  includes the basic Lorentz-invariant effective interactions $\cV_N$, such as $\cR^4$, $\cR^6$, and $\tr(\cF^4)$ which may be extracted from the scattering amplitude for $N$ massless strings. Successive higher space-time derivatives are accounted for by the dependence of the partial amplitude $\cB_N(s_{ij}|\tau)$ on the   Lorentz scalar combinations $s_{ij} = - \alpha ' (k_i+k_j)^2/4$ where $i,j = 1, \cdots , N$, and $k_i$ are massless space-time momenta. Momentum conservation $\sum_i k_i=0$ implies $\sum _i s_{ij}=0$ for all~$j$. The general form of $\cB_N$ is given as follows, 
\bea
\cB_N (s_{ij}|\tau) = \left ( \prod _{k=1}^N \int _\Sigma { d^2 z_k \over \tau_2} \right ) 
P(z_1, \cdots, z_N; s_{ij}|\tau)
\exp \left ( \sum _{1 \leq i < j \leq N} s_{ij} \, G(z_i - z_j |\tau) \right )
\eea
where $P$ is a polynomial in first derivatives $\p_{z_i} G(z_i-z_j|\tau)$, second derivatives $\p_{z_i} \p_{z_j}  G(z_i,z_j)$ and their complex conjugates. Since we have assumed that the effective interactions $\cV_N$ are built out of fields in the supergravity  multiplet only, each vertex $z_k$ will support no more than one derivative $\p_{z_k}$ and one derivative $\p_{\bar z_k}$. The partial amplitude $\cB_N (s_{ij}|\tau)$ is then a modular function of $\tau$. In the special case of the four-graviton scattering amplitude in Type II superstrings, the leading term corresponds to $N=4$, $\cV_4=\cR^4$ and $P=1$, but for amplitudes with a larger number of external gravitons, $P$ will be non-trivial.

\sm

For generic momenta, the integral over moduli space in (\ref{2a4}) will not be convergent, and needs to be defined by analytic continuation in $s_{ij}$ \cite{D'Hoker:1993ge,D'Hoker:1994yr}. The resulting non-analytical parts, such as poles and branch cuts in $s_{ij}$, correspond to on-shell intermediate states with respectively one or multiple strings, and need to be properly isolated before the analytic parts can be used to specify the strength of the effective interactions \cite{Green:1999pv}.

\subsection{Graphical representation}

The partial amplitude $\cB_N(s_{ij} |\tau)$ may be expanded in a power series in $s_{ij}$ whose coefficients are given by Feynman graphs for a conformal scalar field with Green function $G$. As usual, we represent the Green function graphically by a line or edge in a Feynman graph,  
\bea
\tikzpicture[scale=1.7]
\scope[xshift=-5cm,yshift=-0.4cm]
\draw (1,0) -- (2.5,0) ;
\draw (1,0) [fill=white] circle(0.05cm) ;
\draw (2.5,0) [fill=white] circle(0.05cm) ;
\draw(3.7,0) node{$= ~ ~ G(z_i-z_j|\tau) $};
\draw (1,-0.25) node{$z_i$};
\draw (2.5,-0.25) node{$z_j$};
\endscope
\endtikzpicture
\label{fig1}
\eea
The integration over the position of a vertex $z$ on which $r$  Green functions end will be denoted by an unmarked  filled black dot, in contrast with an unintegrated vertex $z_i$ which will be represented by a marked unfilled white dot. The basic ingredient in the graphical notation is depicted in the graph below,
\bea
\tikzpicture[scale=1.7]
\scope[xshift=-5cm,yshift=-0.4cm]
\draw (2,0.7) -- (1,0) ;
\draw (2,0.7) -- (1.5,0) ;
\draw (2,0.7) -- (2.5,0) ;
\draw (2,0.7) -- (3,0) ;
\draw (2,0.2) node{$\cdots$};
\draw [fill=black]  (2,0.68)  circle [radius=.05] ;
\draw (1,0)    [fill=white] circle(0.05cm) ;
\draw (1.5,0) [fill=white] circle(0.05cm) ;
\draw (2.5,0)    [fill=white] circle(0.05cm) ;
\draw (3,0) [fill=white] circle(0.05cm) ;
\draw(5.2,0) node{$\displaystyle= \ \int _\Sigma {d^2 z \over  \tau_2} \, \prod _{i=1}^r G(z-z_i|\tau)$};
\draw (1,-0.25) node{$z_1$};
\draw (1.5,-0.25) node{$z_2$};
\draw (2.57,-0.25) node{$z_{r-1}$};
\draw (3.1,-0.25) node{$z_r$};
\endscope
\endtikzpicture
\label{fig2}
\eea
Derivatives on the Green function $G$, which are required when $P$ is non-trivial, may be represented graphically as well, but we shall not introduce additional notation to do so here. For example, when $P=1$, the coefficient of the monomial $\prod _{i<j}s_{ij}^{\nu_{ij}}$ in the power series expansion of $\cB_N$  is given by a Feynman graph $\Gamma$ with associated  integral, 
\bea
\cC_{\Gamma}(\tau) = \left ( \prod_{k=1}^N \int_{\Sigma} {{d^2 z_k}\over \tau_2} \right ) 
\prod_{1\leq i < j \leq N} G(z_i - z_j | \tau)^{\nu_{ij}}
\eea
The graph $\Gamma$ has $N$ vertices, labelled by $k=1,\cdots, N$ and $\nu_{ij}$ edges between vertices $i$ and $j$, with the total number of edges  given by the {\sl weight} $w$ of the graph $\Gamma$, 
\bea
w=\sum_{1 \leq i < j \leq N} \nu_{ij}
\eea
In terms of the Fourier series (\ref{2a2}) for the Green function, this expression is given by, 
\bea
\label{2b3}
\cC_{\Gamma}(\tau) = \sum_{p_1,\ldots,p_w \in \Lambda}' 
\left ( \prod_{\alpha=1}^w {{\tau_2}\over {\pi |p_{\alpha}|^2}} \right ) 
\prod_{i=1}^N \delta\left(\sum_{\alpha=1}^w \Gamma_{i\alpha}p_{\alpha}\right)
\eea
All the information about the graph $\Gamma$ is contained in its connectivity matrix $\Gamma_{i \alpha}$ where the index $i=1,\cdots, N $ runs over all the vertices, and the index $\alpha=1,\cdots, w$ runs over all the edges of the graph $\Gamma$. When the edge $\alpha$ does not end on the vertex $i$, we have $\Gamma _{i \alpha}= 0$, while otherwise we have $\Gamma _{i \alpha} = \pm 1$, the sign depending on the orientation conventions for the momenta flowing into the vertices. The functions $\cC_\Gamma(\tau)$ are modular invariant and are referred to as {\sl modular graph functions}.

\subsection{Modular graph functions and forms}

The family of modular graph functions introduced above encompasses all cases in which none of the Green functions carry derivatives, i.e. when $P=1$. In those cases the exponents of the momenta $p_\alpha  = m_\alpha  + n_\alpha \tau$ which are holomorphic in $\tau$ and the exponents of their complex conjugates $\bar p_\alpha$ coincide for all edges $\alpha$. When derivatives on the Green functions occur, however, the  exponents of $p_\alpha $ and $\bar p_\alpha $ are allowed to differ from one another. The origin is easily seen in the Fourier series for the derivative of $G$,
\bea
\p_z G(z|\tau) = - \sum _{ p \in \Lambda} ' { 1 \over p} \, e^{2 \pi i ( n \alpha - m \beta)}
\eea
for $z=\alpha + \beta \tau$ and $\alpha, \beta \in \RR/\ZZ$. This leads us to generalize (\ref{2b3}) in terms of decorated graphs, which were introduced in \cite{D'Hoker:2016a}, and will now be briefly reviewed. 

\sm

In a decorated graph, every decorated edge must begin and end on distinct vertices. Each decorated edge $r$ with momentum $p_r$ carries a pair of exponents $(a_r, b_r)$ and is assigned the following momentum factor,
\bea
\tikzpicture[scale=1.7]
\scope[xshift=-5cm,yshift=-0.4cm]
\draw[thick] (0,0) -- (2,0) ;
\draw (0.65,-0.15) [fill=white] rectangle (1.35,0.15) ;
\draw (1,0) node{$a_r, b _r $};
\draw[fill=white] (0,0)  circle [radius=.05] ;
\draw[fill=white] (2,0)  circle [radius=.05] ;
\draw (3.5,0) node{$\approx \hskip 0.15in (p_r) ^{- a_r} \, (\bar p _r) ^{- b _r}$};
\endscope
\endtikzpicture
\label{fig3}
\eea
The effect of bivalent vertices may be represented by increasing the exponents $a_r$ and $ b_r$ on the decorated edge carrying the bivalent vertex. Therefore, decorated graphs are defined in terms of vertices with valence larger than or equal to three. We shall denote by $V$ the number of such vertices, and label the vertices by $i=1,\cdots, V$.  The number of decorated edges between two distinct vertices $i,j$ will be denoted by $\mu_{ij}$, and the pairs $i,j$ are ordered. Assembling the entire subgraph of edges between vertices $i,j$,
along with their decorations, gives rise to the following graphical representation, 
\bea
\tikzpicture[scale=1.5]
\scope[xshift=-5cm,yshift=-0.4cm]
\draw[thick] (-3,0.035) -- (-1.5,0.035);
\draw[thick] (-3,-0.035) -- (-1.5,-0.035);
\draw[fill=white] (-3,0)  circle [radius=.05] ;
\draw[fill=white] (-1.5,0)  circle [radius=.05] ;
\draw (-3, 0.3) node{$i$};
\draw (-1.5, 0.3) node{$j$};
\draw (-2.25,-0.3) node{$ A_{ij}, B_{ij} $};
\draw (-0.8,0) node{$=$};
\draw (0, 0.3) node{$i$};
\draw (4, 0.3) node{$j$};
\draw[thick] (0,0)  ..controls (2, 0.95) ..  (4,0) ;
\draw[thick] (0,0)  ..controls (2, -0.8)  ..  (4,0) ;
\draw[thick] (0,0)  ..controls (2, 0.2)   ..  (4,0);
\draw (1.3,0.46) [fill=white] rectangle (2.65,0.80) ;
\draw (2,0.64) node{$a_{ij\,1}, b _{ij\, 1}\, $};
\draw (1.3,-0.03) [fill=white] rectangle (2.65,0.31) ;
\draw (2,0.14) node{$a_{ij\, 2}, b _{ij\, 2} $};
\draw (1.3,-0.69) [fill=white] rectangle (2.65,-0.35) ;
\draw (2,-0.5) node{$a_{ij\, \mu_{ij}}, b _{ij\, \mu_{ij}} $};
\draw (2, -0.2) node{$ \cdots$};
\draw[fill=white] (0,0)  circle [radius=.05] ;
\draw[fill=white] (4,0)  circle [radius=.05] ;
\endscope
\endtikzpicture
\label{fig5}
\eea
The array of exponents $\{a_{ij\, \alpha}\}$ of holomorphic momenta $p_{ij \, \alpha}$   for $\alpha=1,\dots,\mu_{ij}$, and the array of exponents $\{b_{ij\, \alpha }\}$ of the complex conjugate momenta $\bar p_{ij \, \alpha}$ will be collected respectively in row matrices $A_{ij}$ and  $B_{ij}$, 
\bea
A_{ij} & = & 
\left [ \matrix{
a_{ij \, 1} & a_{ij \, 2 } & \cdots & a_{ij \, \mu_{ij}} \cr } \right ]
\no \\
B_{ij} & = & 
\left [ \matrix{
b_{ij \, 1} & \, b_{ij \, 2 } & \cdots & b_{ij \, \mu_{ij}} \cr} \, \right ]
\eea
For a general graph with $V$ vertices there are $V(V-1)/2$ sets of row matrices $A_{ij}$ and  $B_{ij}$ with $1\leq i < j \leq V$. The labels are assembled into a $ 2 \times \L$  matrix, in ascending order, 
\bea
\label{3d1}
\left [ \matrix{A \cr B \cr} \right ]  = \left [ 
\matrix{A_{12} \cr B_{12} \cr } \Bigg |  \cdots \Bigg | 
\matrix{A_{1V} \cr B_{1V} \cr } \Bigg | 
\matrix{A_{23} \cr B_{23} \cr } \Bigg |\cdots \Bigg | 
\matrix{A_{2V} \cr B_{2V} \cr } \Bigg | \cdots \Bigg | 
\matrix{A_{V-1 \, V} \cr B_{V-1 \, V} \cr }
\right ]
\eea
The vertical bars are used to separate the subsets of exponents for edges stretched between different pairs of vertices. We shall often replace the composite labels $(ij \, \mu_{ij})$ on the exponents $a$ and $b$ by a single enumeration label $r = 1, \cdots, \L$, where $\L$ is the total number of decorated edges in the graph.  

\sm

To a decorated graph $\Gamma$ with connectivity matrix $\Gamma _{i \, r}$ and with exponents given in (\ref{3d1}), we associate a modular graph form, given by the following expression,
\bea
\label{3b5}
\cC \left [ \matrix{A \cr B \cr} \right ]  (\tau) =  \sum_{p_1,\dots,p_\L \in \Lambda} ' ~ \prod_{r =1}^\L
  { (\tau _2 / \pi)^{\half a_r + \half b_r} \over  (p_r) ^{a_r} ~ (\bar p _r) ^{b_r} }\, \prod_{i =1}^V
  \delta \left ( \sum_{s =1}^\L \Gamma _{i \, s} \, p_s \right )
\eea 
Under a modular transformation in $PSL(2,\ZZ)$, the modular graph form transforms by, 
\bea
\cC \left [ \matrix{A \cr B \cr} \right ] \left  ({ \alpha \tau + \beta \over \gamma \tau + \delta} \right ) 
= \left ( { \gamma \tau+\delta \over \gamma  \bar \tau +\delta } \right ) ^{\half a - \half b}
\cC \left [ \matrix{A \cr B \cr} \right ] (\tau) 
\eea
where $\alpha, \beta , \gamma, \delta \in \ZZ$ and $\alpha \delta - \beta \gamma =1$, and   
the total exponents respectively of holomorphic and anti-holomorphic momenta are given by the following sums,
\bea
\label{ab}
a = \sum _{r =1}^\L  a_r \hskip 0.5in
b = \sum _{r =1} ^\L  b_r  
\eea
When $a=b$ the modular graph form is invariant under $PSL(2,\ZZ)$ and is referred to as a {\sl modular graph function}. When $a\not= b$, $\cC$ transforms as a modular form with weight $({a \over 2} -{b \over 2} , -{a \over 2} + {b\over 2} )$, and is referred to as a {\sl modular graph form}. For $a\not= b$ the normalization factor of $\tau_2$ is not canonical, and we  introduce the forms of modular weight $(0,b-a)$,
\bea
\label{Cplus}
\cC ^+\left [ \matrix{A \cr B \cr} \right ] (\tau) 
= (\tau_2)^{{a-b \over 2}} \cC \left [ \matrix{A \cr B \cr} \right ] (\tau) 
\eea
Further symmetry properties of modular graph functions are as follows.
\begin{itemize}
\itemsep=0in
\item Under complex conjugation,  the modular graph forms behave as follows, 
\bea
\label{3d2}
\cC \left [ \matrix{A \cr B \cr} \right ]  (\tau) ^* 
= \cC \left [ \matrix{B \cr A \cr} \right ]  (\tau)
\eea
\item The modular graph form is invariant under all permutations of pairs of exponents $(a_{ij \alpha}, b_{ij \alpha})$ associated with the edges  spanned between a given pair of vertices $i,j$. 
\item Under permutations of the vertices, the modular graph form is invariant.
\item Finally, modular graph forms obey momentum conservation identities, 
\bea
\label{3d3}
\sum _{r=1}^\L \Gamma _{k\, r} \, \cC \left [ \matrix{A  - S_r \cr B \cr} \right ] =
\sum _{r=1}^\L \Gamma _{k\, r} \, \cC \left [ \matrix{A   \cr B - S_r \cr} \right ] =0
\eea
where the $\L$-dimensional row-vector $S_r$ is defined to have zeroes in all slots except for the $r$-th, which instead has value 1, 
\bea
\label{Sr}
 S_r = [\underbrace{0,\ldots,0}_{r-1},1,\underbrace{0,\ldots,0}_{R-r}]
 \eea
\end{itemize}

\subsection{Differential operators on modular graph forms}

The action of the Cauchy-Riemann operator $\nabla = 2 i \tau_2^2 \p_{\tau}$ on modular graph forms $\cC^+$ defined and normalized  in (\ref{3b5}) and (\ref{Cplus})  for arbitrary weights $a,b$ is given by, 
\bea
\label{nab}
\nabla \cC^+ \left [  \matrix{ A \cr B \cr } \right] = \sum_{r=1}^R a_r \, \cC^+ \left [  \matrix{ A + S_r \cr B - S_r\cr } \right] 
\eea
where  $S_r$ was defined in (\ref{Sr}).  The action of the Laplace operator $\Delta = 4 \tau _2^2 \p_{\bar \tau} \p_\tau$ on modular graph functions (for which $a=b$) is given by,
\bea
\label{Lap1}
(\Delta + a ) \, \cC \left [ \matrix{ A \cr B \cr } \right ]
= 
\sum _{r,s=1}^R a_r b_s \, \cC \left [ \matrix{ A+S_r - S_s  \cr B -S_r + S_s \cr } \right ]
\eea
It is similarly possible to define the action of the Laplace operator on modular graph forms of arbitrary modular weight with $a \not= b$, but we shall not need them here.

\subsection{Modular graph forms with $V \leq 4$}

The combinatorial complexity of modular graphs forms increases rapidly with the number of vertices $V$ of valence three or higher. In the present paper, we shall deal only with {\sl dihedral}, {\sl trihedral}, and {\sl tetrahedral} graphs  respectively for $V=2,3,4$. Here, we present some of the basic simplifications which occur for graphs with $V \leq 4$.

\subsubsection{Graphs for $V=0$ and Eisenstein series}

Eisenstein series correspond to one-loop graphs with only bi-valent vertices, and thus  $V=0$. The non-holomorphic Eisenstein series $E_a(\tau)$, which is defined by,
\bea
\label{Es}
E_a(\tau) = \sum _{ p \in \Lambda}' \left ( { \tau_2 \over \pi \, p \, \bar p} \right )^a
= \sum _{(m,n) \not= (0,0)}  \left ( { \tau _2 \over \pi |m + n \tau|^2} \right )^a
\eea
satisfies the Laplace-eigenvalue equation $\Delta E_a = a(a-1) E_a$. Along with its successive Cauchy-Riemann  derivatives by $\DD = 2 i \tau _2 ^2 \p_\tau$, it may be represented in our notations  by, 
\bea
\DD ^k \, E_a(\tau)
= { \Gamma (a+k) \over \Gamma (a)} \, \cC^+ \left [ \matrix{ a+k & 0 \cr  a-k & 0 \cr } \right ] (\tau) 
\eea
For $a=k$ the multiple derivative is proportional to a holomorphic Eisenstein series $G_{2k}$,
\bea
\label{Gk}
\DD^k E_k (\tau)={ \Gamma (2k) \over \Gamma (k)}  \cG_{2k}(\tau) 
\hskip 1in
G_{2k} (\tau) = \sum _{p \in \Lambda } ' { 1 \over \pi ^k \, p^{2k}}
\eea
where we use the following abbreviation $\cG_{2k} (\tau) = (\tau_2)^{2k} G_{2k}(\tau)$.

\subsubsection{Dihedral graphs} 
 
Dihedral graphs have $V=2$ with only a single pair of vertices,  so that $\mu_{ij}=R$ and the momentum conservation equation simply becomes, 
\bea
\label{3d4}
\sum _{r=1}^\L  \cC \left [ \matrix{A  - S_r \cr B \cr} \right ] =
\sum _{r=1}^\L  \cC \left [ \matrix{A   \cr B - S_r \cr} \right ] =0
\hskip 1in 
\matrix{ A = [ \, a_1 & \cdots & a_R \, ] \cr B=[\, b_1 & \cdots & b_R\, ] \cr }
\eea
Dihedral graphs were discussed at great length in \cite{D'Hoker:2016a} to which we refer for more details.

\subsubsection{Trihedral graphs}

Trihedral graphs have  $V=3$, and the total number of  edges $\L$ is partitioned into three sets,  $\L=\L_1+\L_2+\L_3$,  where $\L_i $ for $i=1,2,3$  denote the numbers of edges  connecting pairs of vertices.  The labeling of the exponents is shown in the Figure below,
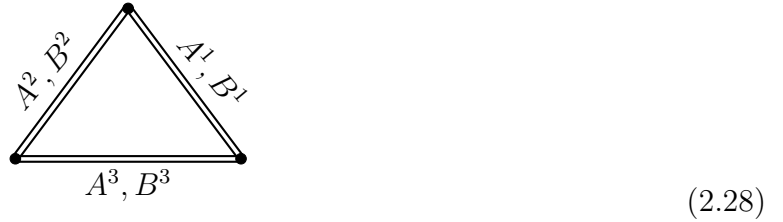
\bea
\tikzpicture[scale=1]
\scope[xshift=-5cm,yshift=-0.4cm]
\draw[thick] (0,0.035) -- (3,0.035);
\draw[thick] (0,-0.04) -- (3,-0.04);
\draw[thick] (-0.02,0.05) -- (1.47,2.05);
\draw[thick] (0,-0.06) -- (1.5,1.95);
\draw[thick] (3.02,0.05) -- (1.52,2.05);
\draw[thick] (3,-0.05) -- (1.5,1.95);
\draw[fill=black] (0,0)  circle [radius=.07] ;
\draw[fill=black] (3,0)  circle [radius=.07] ;
\draw[fill=black] (1.5,2)  circle [radius=.07] ;
\draw (1.5,-0.35) node{$ A^3, B^3 $};
\draw (0.4, 1.15) node [rotate=50] {$A^2,B^2$};
\draw (2.6, 1.15) node [rotate=-50] {$A^1,B^1$};
\endscope
\endtikzpicture
\eea
The graphical representation is based on the  double edge notation of Figure \ref{fig5} where
each line represents a collection of exponents,  given as follows,
\bea
\label{7a1}
\left [ \matrix{ A \cr B \cr } \right ] = 
 \left [ \matrix{ A^1 \cr B^1 \cr} \Bigg |  \matrix{ A^2 \cr B^2 \cr} \Bigg | \matrix{ A^3 \cr B^3 \cr} \right ] 
\hskip 1in 
\matrix{ A^i \, = \, [ \, a^i _1 ~  a^i _2 ~  \cdots ~  a^i _{\L_i} \, ]  \cr 
B^i \, = \, [ \,  b^i _2 ~ b^i _2 ~ \cdots ~ b^i _{\L_i} \, ] \cr } 
\eea
The associated modular graph form is given by,
\bea
\label{7a2}
\cC \left [ \matrix{ A \cr B \cr}  \right ] (\tau)
=
\sum _{  p^i_{k_i} \in \Lambda  } \delta _{\mpp^1,\mpp^2} \delta _{\mpp^2,\mpp^3} \delta _{\mpp^3,\mpp^1} 
\prod _{i=1}^3 \prod _{k_i =1}^{\L_i} 
{ (\tau_2/\pi) ^{\half a^i _{k_i} + \half b^i _{k_i}} \over (p^i _{k_i})^{a^i_{k_i}} \, (\bar p^i _{k_i})^{b^i_{k_i}} }
\hskip 0.5in 
\mpp^i = \sum _{k_i=1}^{\L _i} p^i _{k_i}
\eea
Trihedral graph forms are  invariant under the six permutations of the pairs $[A^i,B^i]$, and under the $R_i!$ permutations of the pairs of exponents within each pair $[A^i,B^i]$. Momentum conservation relations  follow from inserting $\mpp_1 - \mpp_2$ and $\mpp_2 - \mpp_3$ into the summands of (\ref{7a2}). The resulting relation for $\mpp_1-\mpp_2$  is as follows,
\bea
\sum _{k_1=1}^{\L_1} \cC  \left [ \matrix{ A^1 - S_{k_1}^{(1)} \cr B^1 \cr} \Bigg |  \matrix{ A^2 \cr B^2 \cr} \Bigg | \matrix{ A^3 \cr B^3 \cr} \right ] 
- \sum _{k_2=1}^{\L_2} \cC  \left [ \matrix{ A^1  \cr B^1 \cr} \Bigg |  \matrix{ A^2 - S_{k_2}^{(2)} \cr B^2 \cr} \Bigg | \matrix{ A^3 \cr B^3 \cr} \right ] = 0
\eea
The relation for $\mpp_2-\mpp_3$ is obtained by permuting entries 1 and 3, and two further relations are obtained on the lower exponents by complex conjugation.  Here, $S^{(i)}_{k_i}$ is a row vector of length $\L_i$, with $k_i =1, \cdots \L_i$, and is defined as in (\ref{Sr}) with $\L=\L_i$.

\subsubsection{Tetrahedral graphs}

Tetrahedral graphs have $V=4$, and the total number of edges $R$ is partitioned into six sets $R=\sum_{i=1}^6 R_i$ where $R_i$ counts the number of decorated edges between a given pair of vertices. In the present paper, we shall specialize to the case where $R_i=1$ for all $i$. In contrast to the cases of dihedral and trihedral graphs, there is no canonical  orientation for momentum flow, and we shall make the choice  illustrated in Figure (\ref{fig6}).
\bea
\label{fig6}
\tikzpicture[scale=1.1]
\scope[xshift=-5cm,yshift=-0.4cm]
%
\draw[directed, very thick]  (1.73,-1) ..controls (1.3,1) .. (0,2);
\draw[directed, very thick] (0,2) ..controls (-1.3,1) ..(-1.73,-1);
\draw[directed, very thick] (-1.73,-1)  ..controls (0,-1.6) .. (1.73,-1);
\draw[ directed, very thick] (0,2) node{$\bullet$} --  (0,0) node{$\bullet$};
\draw[ directed, very thick] (-1.73,-1) node{$\bullet$} --  (0,0) node{$\bullet$};
\draw[ directed, very thick] (1.73,-1) node{$\bullet$} --  (0,0) node{$\bullet$};
\draw (1.6,1) node{$p_2$};
\draw (-1.6,1) node{$p_3$};
\draw (0,-1.2) node{$p_1$};
\draw (0.3,1.1) node{$p_5$};
\draw (-0.85,-0.2) node{$p_6$};
\draw (0.85,-0.2) node{$p_4$};
\endscope
\endtikzpicture
\eea
The momenta $p_1, p_2, p_3$ are all oriented anti-clockwise. The  Kronecker $\delta$-function which enforces momentum conservation is given as follows,
\bea
\delta _p = \delta _{p_1,p_2+p_4} ~ \delta _{p_2,p_3+p_5} ~ \delta _{p_3,p_1+p_6}
\eea
The standard notation for the corresponding modular graph of (\ref{3d1}), with its vertical bars,  is excessively cumbersome, and we shall therefore introduce the following more convenient notation in terms of $\cD$-functions, defined by,\footnote{The functions $\cD^+$ are defined by replacing $\cC$ by $\cC^+$ of (\ref{Cplus}) in the  definition below.}
\bea
\label{DC}
\cD \left [ \matrix{ a_1 \, a_2 \, a_3 \, a_4 \, a_5 \, a_6 \cr b_1 \,  b_2 ~  b_3 ~ b_4 ~  b_5 \,  b_6 \cr } \right ]
=
\cC \left [ \matrix{ a_1 \cr b_1 \cr} \Bigg | \matrix{ a_2 \cr b_2 \cr} \Bigg | \matrix{ a_3 \cr b_3 \cr} \Bigg | \matrix{ a_4 \cr b_4 \cr} \Bigg | \matrix{ a_5 \cr b_5 \cr} \Bigg | \matrix{ a_6 \cr b_6 \cr}  \right ]
= 
\sum _{p_1, \cdots, p_6 \in \Lambda} ' \delta _p \prod _{i=1}^6 { (\tau _2/\pi) ^{ \half (a_i+b_i)} 
\over (p_i)^{a_i} \, (\bar p_i )^{b_i} }
\eea
The momentum conservation relations for holomorphic exponents are as follows,
\bea
\label{DCmom}
\cD \left [ \matrix{ a_1' ~ a_2 ~ a_3 ~ a_4 ~ a_5 ~ a_6 \cr b_1 \, ~ b_2 ~ b_3 ~ b_4 ~ b_5 ~ b_6 \cr } \right ]
- \cD \left [ \matrix{ a_1 ~ a_2' ~ a_3 ~ a_4 ~ a_5 ~ a_6 \cr b_1 ~ b_2 \, ~ b_3 ~ b_4 ~ b_5 ~ b_6 \cr } \right ]
- \cD \left [ \matrix{ a_1 ~ a_2 ~ a_3 ~ a_4' ~ a_5 ~ a_6 \cr b_1 ~ b_2 ~ b_3 ~\,  b_4 ~\,  b_5 ~ b_6 \cr } \right ]
& = & 0
\no \\
\cD \left [ \matrix{ a_1 ~ a_2' ~ a_3 ~ a_4 ~ a_5 ~ a_6 \cr b_1 ~ b_2 ~\, b_3 ~ b_4 ~ b_5 ~ b_6 \cr } \right ]
- \cD \left [ \matrix{ a_1 ~ a_2 ~ a_3' ~ a_4 ~ a_5 ~ a_6 \cr b_1 ~ b_2 ~ b_3 ~\,  b_4 ~ b_5 ~ b_6 \cr } \right ]
- \cD \left [ \matrix{ a_1 ~ a_2 ~ a_3 ~ a_4 ~ a_5' ~ a_6 \cr b_1 ~ b_2 ~ b_3 ~ b_4 ~\,  b_5 ~ b_6 \cr } \right ]
& = & 0
\no \\\cD \left [ \matrix{ a_1 ~ a_2 ~ a_3' ~ a_4 ~ a_5 ~ a_6 \cr b_1 ~ b_2 ~ \, b_3 ~ b_4 ~ b_5 ~ b_6 \cr } \right ]
- \cD \left [ \matrix{ a_1' ~ a_2 ~ a_3 ~ a_4 ~ a_5 ~ a_6 \cr b_1 ~\,  b_2 ~ b_3 ~ b_4 ~ b_5 ~ b_6 \cr } \right ]
- \cD \left [ \matrix{ a_1 ~ a_2 ~ a_3 ~ a_4 ~ a_5 ~ a_6' \cr b_1 ~ b_2 ~ b_3 ~ b_4 ~ b_5 ~\,  b_6 \cr } \right ]
& = & 0
\no \\\cD \left [ \matrix{ a_1 ~ a_2 ~ a_3 ~ a_4' ~ a_5 ~ a_6 \cr b_1 ~ b_2 ~ b_3 ~ b_4 \, ~ b_5 ~ b_6 \cr } \right ]
+ \cD \left [ \matrix{ a_1 ~ a_2 ~ a_3 ~ a_4 ~ a_5' ~ a_6 \cr b_1 ~ b_2 ~ b_3 ~ b_4 \, ~ b_5 ~ b_6 \cr } \right ]
+ \cD \left [ \matrix{ a_1 ~ a_2 ~ a_3 ~ a_4 ~ a_5 ~ a_6' \cr b_1 ~ b_2 ~ b_3 ~ b_4 ~ b_5 \, ~ b_6 \cr } \right ]
& = & 0 \qquad
\eea
where we have used the notation $a_i' = a_i-1$ in order to make each formula fit on one line.
The sum of the four relations vanishes by overall momentum conservation.

\newpage

\section{New algebraic identities up to weight seven}
\setcounter{equation}{0}
\label{six-seven}

In this section, we shall use the methods of \cite{D'Hoker:2016a} to systematically derive algebraic identities between certain classes of modular graph functions of the following type,
\bea
\label{3a1}
C_A= \cC \left [  \matrix{ A \cr A \cr } \right]  
\eea
The behavior of $C_A(\tau)$ near the cusp $\tau \to i \infty$ is given by a Laurent polynomial \cite{D'Hoker:2015qmf}, 
\bea
\label{3a11}
C_A (\tau) = \sum _{k=1-w} ^w f_A(k) \, \tau_2^k + \cO (e^{-2 \pi \tau_2})
\eea
where the coefficients $f_A(k)$ depend on the exponents $A$, but are independent of $\tau$. A first algebraic identity at weight three was conjectured in \cite{Green:2008uj},
\bea
\label{3a2}
C_{1,1,1} = E_3 + \zeta (3)
\eea
based on the matching of their Laurent polynomial at the cusp $\tau \to i \infty$, and proven by performing the lattice sums in \cite{Zagier:2014}. It was  shown in \cite{D'Hoker:2015foa}  that a linear algebraic identity exists between two-loop modular graph functions $C_{a_1, a_2, a_3}$ for every odd weight $w=a_1+a_2+a_3$. 

\subsection{Weight four and five identities}

For higher loop orders, the systematic structure of algebraic relations remains to be identified. For the low weights 4 and 5,  identities were conjectured in \cite{D'Hoker:2015foa} on the basis of their matching Laurent polynomial near the cusp, namely $F_4=F_5=F_{3,1,1}=F_{2,2,1}=0$ where,   
\bea
\label{4a1}
F_4 &=& D_4 - 24 C_{2,1,1} - 3 E_2^2 + 18 E_4
\no\\
F_5 &=& D_5 - 60 C_{3,1,1} - 10 E_2 C_{1,1,1} + 48 E_5 - 16 \zeta (5)
\no\\
40 F_{3,1,1} &=& 40 D_{3,1,1} -  300  C_{3,1,1} - 120 E_2 E_3 + 276 E_5 -  7  \zeta (5)
\no \\
10 F_{2,2,1} & = & 10 D_{2,2,1} - 20 C_{3,1,1} + 4 E_5 - 3 \zeta (5)
\eea
Relations (\ref{3a2}) and (\ref{4a1})  exhaust all possible dihedral and trihedral identities at weights 3, 4, and 5. They  were proven in \cite{D'Hoker:2016a} with the help of a sieve algorithm and holomorphic subgraph reduction, and subsequently proven by  Green function methods in \cite{Basu:2016kli}. Here, we have adopted the notations $D_\ell = C_{1_\ell}$ and $ D_{\ell, 1,1} = C_{2, 1_\ell} $ familiar from \cite{Green:2008uj}, and graphically represented as follows,
\bea
\label{fig8}
\tikzpicture[scale=1.2]
\scope[xshift=-5cm,yshift=-0.4cm]
\draw (-0.9,0) node{$D_\ell \, =$};
\draw[thick]   (0,0) node{$\bullet$} ..controls (1,0.7) .. (2,0) node{$\bullet$} ;
\draw[thick]   (0,0) node{$\bullet$} ..controls (1,-0.1) .. (2,0) node{$\bullet$} ;
\draw[thick]   (0,0) node{$\bullet$} ..controls (1,-0.7) .. (2,0) node{$\bullet$} ;
\draw[thick] (1,0.2) node{$\cdots$};
\draw (5,0) node{$D_{\ell ,1,1} \, =$};
\draw[thick]   (6,0) node{$\bullet$} ..controls (7,0.7) .. (8,0) node{$\bullet$} ;
\draw[thick]   (6,0) node{$\bullet$} ..controls (7,-0.1) .. (8,0) node{$\bullet$} ;
\draw[thick]   (6,0) node{$\bullet$} ..controls (7,-0.7) .. (8,0) node{$\bullet$} ;
\draw[thick] (7,0.2) node{$\cdots$};
\draw (7,-0.53) node{$\bullet$};
\endscope
\endtikzpicture
\eea
The ellipses in $D_\ell$ stand for  $\ell-3$ edges, while the ellipses in $D_{\ell,1,1}$ stand for $\ell-2$ edges. We use the notation $D_{\ell,m,n}= C_{1_\ell |1_m|1_n}$ for trihedral graphs, and graphically represent the example  $m=2$, $n=1$  as follows,
\bea
\label{fig9}
\tikzpicture[scale=1.2]
\scope[xshift=-5cm,yshift=-0.4cm]
\draw (-0.9,0) node{$D_{\ell, 2,1} \, = ~ $};
\draw[thick]   (0,0) node{$\bullet$} ..controls (1,0.7) .. (2,0) node{$\bullet$} ;
\draw[thick]   (0,0) node{$\bullet$} ..controls (1,-0.2) .. (2,0) node{$\bullet$} ;
\draw[thick]   (0,0) node{$\bullet$} ..controls (0.5,-0.5) .. (1,-0.7) node{$\bullet$} ;
\draw[thick]   (1,-0.7) node{$\bullet$} ..controls (1.5,-0.6) .. (2,0) node{$\bullet$} ;
\draw[thick]   (1,-0.7) node{$\bullet$} ..controls (1.5,-0.2) .. (2,0) node{$\bullet$} ;
\draw[thick] (1,0.2) node{$\cdots$};
\endscope
\endtikzpicture
\eea
The ellipses in $D_{\ell,2,1}$ stand for $\ell-2$ edges. 

\sm

In the present section, we shall extend the use of holomorphic subgraph reduction and the sieve algorithm of \cite{D'Hoker:2016a} to systematically derive all algebraic identities between modular graph functions of type (\ref{3a1}) at weight six, including dihedral, trihedral, and one tetrahedral modular graph function. We shall also derive all such identities at weight seven for dihedral graphs and one trihedral graph, and confirm that in each of these cases, when the Laurent polynomial is known, its vanishing uniquely leads to the identities proven here.

\subsection{Holomorphic subgraph reduction and sieve algorithm}
\label{sec:31}

We denote by $\cV_w$ the subspace of weight $w$ modular graph functions  contained in the polynomial ring generated by modular graph functions of the form (\ref{3a1}). We include in this ring the generators provided by the zeta-values $\zeta(s)$ for odd integer weight $s$.  The fundamental tools to prove the identities (\ref{4a1}) were developed in \cite{D'Hoker:2016a}
using a  sieve algorithm based on holomorphic subgraph reduction, and consist of the two key  ingredients below.
\begin{enumerate}
\itemsep=0in
\item \underline{Lemma:} ~ If $F$ is a non-holomorphic modular function in $\cV_w$ with polynomial growth near the cusp $\tau \to i \infty$, which satisfies the differential equation $\nabla ^n F=0$ with $\nabla = 2 i \tau_2^2 \p_\tau$ for some integer $n \geq 1$, then $F$ is independent of $\tau$.
\item We initiate the sieve algorithm by setting $\cV^{(0)} = \cV_w$ and define  the space $\cV ^{(n+1)}$ as the subspace of linear combinations in $ \cV^{(n)}$ for which all the holomorphic Eisenstein series contributions in $\nabla \cV ^{(n+1)}$, which result from holomorphic subgraph reduction,  cancel. The space $\cV ^{(w-1)}$ which emerges from this process contains all the elements $F \in \cV_w$ for which $\nabla ^{w-1} F=0$. By the Lemma, each $F$ must be independent of $\tau$ and the corresponding constant may be evaluated at the cusp $\tau \to i \infty$. 
\end{enumerate}

Holomorphic subgraph reduction formulas were given for dihedral graphs and for the specific trihedral graph $D_{2,2,1}$ in \cite{D'Hoker:2016a}; they will be given in Appendix A for all the graphs needed in the present paper, including trihedral and tetrahedral graphs. We illustrate the use of holomorphic subgraph reduction by recalling from \cite{D'Hoker:2016a} the proof of the $F_4$ relation in (\ref{4a1}). Formula (\ref{nab}) allows us to evaluate the derivatives of $D_4=C_{1,1,1,1}$, and the momentum conservation equations of (\ref{3d4}) allow us to simplify the outcome as follows,
\bea
\nabla ^2 D_4 = 
12 \, \cC^+ \! \left [  \matrix{  2\, 2 \, 1 \, 1  \cr 0 \, 0 \, 1 \, 1 \cr } \right]
- 24 \, \cC^+ \! \left [  \matrix{  3\, 1 \, 1 \, 1  \cr 0 \, 0 \, 1 \, 1 \cr } \right]
\eea
The dependence on the loop momentum through the closed loop formed by the first two edges in $\nabla ^2 D_4$ is purely holomorphic in $\tau$ and holomorphic subgraph reduction is based on using the methods of holomorphic modular forms to evaluate the corresponding sums. The result is a reduction in the number of loops in the graph by one. Indeed, the simplest holomorphic subgraph reduction formula applies in this case, and gives the identity,
\bea
\label{3d5}
\cC^+ \! \left [  \matrix{  2\, 2 \, 1 \, 1  \cr 0 \, 0 \, 1 \, 1 \cr } \right]
- 2 \, \cC^+ \! \left [  \matrix{  3\, 1 \, 1 \, 1  \cr 0 \, 0 \, 1 \, 1 \cr } \right]
= 2 \, \cC ^+ \! \left [  \matrix{  4 \, 1 \, 1  \cr  0 \, 1 \, 1 \cr } \right]
+ \half E_2 \nabla ^2 E_2
\eea
Note that the 3-loop graphs on the left  result in a two-loop graph in the first term on the right plus a product of one-loop terms. One further derivative leads to the relations, 
\bea
\nabla ^3 (D_4 - 3 E_2^2) & = & {18 \over 5} \nabla ^3 E_4 - 24 (\nabla ^2 E_2) (\nabla E_2)
\no \\
\nabla ^3 C_{2,1,1} & = & {9 \over 10} \nabla ^3 E_4 - (\nabla ^2 E_2) (\nabla E_2)
\eea
from which the relation $\nabla ^3 F_4=0$ immediately follows. Using  Lemma 1 we conclude that $F_4$ must be a constant, whose value may be determined at the cusp.

\subsection{All algebraic identities at  weight six}
\label{sec:32}

Using the tools of the preceding subsection, along with the holomorphic subgraph reduction formulas of Appendix A, one shows that the following dihedral weight six combinations are independent of $\tau$,
and that they are the only such algebraic identities of weight six,  
\bea
\label{6di}
F_6 & = & D_6 - 15 E_2 D_4 + 30 E_2^3 - 10 C_{1,1,1}^2 - 60 D_{4,1,1} + 720 C_{2,2,1,1} +240 E_3 C_{1,1,1} 
\no \\ && 
- 720 E_2 E_4 - 1440 E_3^2  - 5280 C_{3,2,1} 
+ 360 E_2 C_{2,1,1} - 1280 C_{2,2,2} + 3380 E_6 
\no \\ 
F_{3,1,1,1} & = & 2 C_{3,1,1,1}+ 3 C_{2,2,1,1} -9 E_{2} E_{4} - 6 E_{3}^2 -18 C_{4,1,1}
\no \\ &&
-24 C_{3,2,1}-2 C_{2,2,2}+32 E_{6}  
\no\\
F_{4,1,1} & = & 
- 3 D_{4,1,1} + 109 C_{2,2,2} + 408 C_{3,2,1} + 36 C_{4,1,1} +18 E_2 C_{2,1,1} 
\no \\ &&
 + 12 E_3 C_{1,1,1} - 211 E_6 
\eea
The graphical representation for $D_6$ and $D_{4,1,1}$ were already given in (\ref{fig8}), while those for $C_{2,2,1,1}$ and $C_{3,1,1,1}$ are as follows,
\bea
\label{fig10}
\tikzpicture[scale=1.2]
\scope[xshift=-5cm,yshift=-0.4cm]
\draw (-1.3,0) node{$C_{2,2,1,1} \, =$};
\draw[thick]   (0,0) node{$\bullet$} ..controls (1,0.8) .. (2,0) node{$\bullet$} ;
\draw[thick]   (0,0) node{$\bullet$} ..controls (1,-0.3) .. (2,0) node{$\bullet$} ;
\draw[thick]   (0,0) node{$\bullet$} ..controls (1,0.2) .. (2,0) node{$\bullet$} ;
\draw[thick]   (0,0) node{$\bullet$} ..controls (1,-0.8) .. (2,0) node{$\bullet$} ;
\draw (1,-0.21) node{$\bullet$};
\draw (1,-0.6) node{$\bullet$};
\draw (5.6,0) node{$C_{3,1,1,1} \, =$};
\draw[thick]   (7,0) node{$\bullet$} ..controls (8,0.8) .. (9,0) node{$\bullet$} ;
\draw[thick]   (7,0) node{$\bullet$} ..controls (8,-0.3) .. (9,0) node{$\bullet$} ;
\draw[thick]   (7,0) node{$\bullet$} ..controls (8,0.2) .. (9,0) node{$\bullet$} ;
\draw[thick]   (7,0) node{$\bullet$} ..controls (8,-0.8) .. (9,0) node{$\bullet$} ;
\draw (7.66,-0.5) node{$\bullet$};
\draw (8.33,-0.5) node{$\bullet$};
\endscope
\endtikzpicture
\eea
The following weight six combinations involving only trihedral and dihedral modular graph functions are also independent of $\tau$, and constitute all such algebraic identities, 
\bea
\label{6tri}
F_{2,2,2}
& = & 3 D_{2,2,2} - 18 C_{2,2,1,1}  - 58 C_{2,2,2} - 192 C_{3,2,1}   
\no \\ &&
- 3 E_2^3 + 24 E_3^2    + 18 E_2 E_4  + 46 E_6
\no\\
F_{3,2,1} & = &
2 D_{3,2,1} + 18 C_{2,2,1,1}  - 36 C_{4,1,1} - 69 C_{2,2,2} - 288 C_{3,2,1}
\no \\ &&
 - 6 E_2 C_{2,1,1} -18 E_2E_4 -36 E_3^2 +183 E_6
\no\\
F_{2,2,1,1}& = & 
3 D_{2,2,1,1} + 6 C_{2,2,1,1} - 10 C_{2,2,2} - 48 C_{3,2,1} -12 C_{4,1,1}
\no \\ &&
- 6 E_2 E_4 - 12 E_3^2   + 40 E_6 
\no \\
F_{2,1,1,1;1} & = & 
18 D_{2,1,1,1;1} - 9 C_{2,2,1,1}  - 20 C_{2,2,2}  - 60 C_{3,2,1} 
\no \\ &&
+ 9 E_2 E_4 + 18 E_3^2  - 10 E_6
\eea
The graphical representations of $C_{2,2,1,1}$ and $C_{3,1,1,1}$ were already given in (\ref{fig10}), and that of $D_{3,2,1}$ in (\ref{fig9}), while those of $D_{2,2,2}$ and $D_{2,1,1,1;1}$ are given by, 
\bea
\label{fig11}
\tikzpicture[scale=1.1]
\scope[xshift=-5cm,yshift=-0.4cm]
\draw (-1.3,0) node{$D_{2,2,2} \, =$};
\draw[thick]   (0,0.5) node{$\bullet$} ..controls (1,0.75) .. (2,0.5) node{$\bullet$} ;
\draw[thick]   (0,0.5) node{$\bullet$} ..controls (1,0.25) .. (2,0.5) node{$\bullet$} ;
\draw[thick]   (0,0.5) node{$\bullet$} ..controls (0.8,0) .. (1,-0.8) node{$\bullet$} ;
\draw[thick]   (0,0.5) node{$\bullet$} ..controls (0.5,-0.5) .. (1,-0.8) node{$\bullet$} ;
\draw[thick]   (2,0.5) node{$\bullet$} ..controls (1.2,0) .. (1,-0.8) node{$\bullet$} ;
\draw[thick]   (2,0.5) node{$\bullet$} ..controls (1.5,-0.5) .. (1,-0.8) node{$\bullet$} ;
\draw (5.6,0) node{$D_{2,1,1,1;1} \, =$};
\draw[thick]   (7,-0.7) node{$\bullet$} -- (8.5,-0.7) node{$\bullet$} ;
\draw[thick]   (7,-0.7) node{$\bullet$} -- (7,0.7) node{$\bullet$} ;
\draw[thick]   (7,0.7) node{$\bullet$} -- (8.5,0.7) node{$\bullet$} ;
\draw[thick]   (8.5,0.7) node{$\bullet$} ..controls (8.2,0) .. (8.5,-0.7) node{$\bullet$} ;
\draw[thick]   (8.5,0.7) node{$\bullet$} ..controls (8.8,0) .. (8.5,-0.7) node{$\bullet$} ;
\draw[thick]   (7,-0.7) node{$\bullet$} -- (8.5,0.7) node{$\bullet$} ;
\endscope
\endtikzpicture
\eea
The graph $D_{2,2,1,1}$ may be represented in two different graphical ways,
\bea
\label{fig12}
\tikzpicture[scale=1.1]
\scope[xshift=-5cm,yshift=-0.4cm]
\draw (5.6,0) node{$D_{2,2,1,1} \, =$};
\draw[thick]   (7,-0.7) node{$\bullet$} -- (8.5,-0.7) node{$\bullet$} ;
\draw[thick]   (7,-0.7) node{$\bullet$} ..controls (6.7,0) ..  (7,0.7) node{$\bullet$} ;
\draw[thick]   (7,-0.7) node{$\bullet$} ..controls (7.3,0) ..  (7,0.7) node{$\bullet$} ;
\draw[thick]   (7,0.7) node{$\bullet$} -- (8.5,0.7) node{$\bullet$} ;
\draw[thick]   (8.5,0.7) node{$\bullet$} ..controls (8.2,0) .. (8.5,-0.7) node{$\bullet$} ;
\draw[thick]   (8.5,0.7) node{$\bullet$} ..controls (8.8,0) .. (8.5,-0.7) node{$\bullet$} ;
\draw  (9.4,0) node{$=$} ;
\draw[thick]   (10,0.7) node{$\bullet$} -- (12,0.7) node{$\bullet$} ;
\draw[thick]   (11,-0.7) node{$\bullet$} ..controls (10.1,0) ..  (10,0.7) node{$\bullet$} ;
\draw[thick]   (11,-0.7) node{$\bullet$} ..controls (10.9,0) ..  (10,0.7) node{$\bullet$} ;
\draw[thick]   (11,-0.7) node{$\bullet$} ..controls (11.1,0) ..  (12,0.7) node{$\bullet$} ;
\draw[thick]   (11,-0.7) node{$\bullet$} ..controls (11.9,0) ..  (12,0.7) node{$\bullet$} ;
\draw  (11,0.7) node{$\bullet$} ;
\endscope
\endtikzpicture
\eea
In its first form, this graph was investigated in detail in \cite{Basu:2016xrt}.
Finally, there is a unique  weight six algebraic identity which involves also the tetrahedral modular graph function defined by $\cD_T = \cD_{1,1,1,1,1,1}= C_{1|1|1|1|1|1}$ in the notations of (\ref{DC}) and (\ref{3a1}),
\bea
\label{6tet}
F_T = 3 \cD_T - C_{2,2,2} - 12 C_{3,2,1} +4 E_6
\eea
which was represented graphically  in (\ref{fig6}). The relation was derived independently in \cite{Basu:2016kli}. The proofs of these identities will be discussed in subsections \ref{sec:33}, \ref{sec:34}, and \ref{sec:35}. The precise values of the constants $F$ will be obtained in subsection \ref{sec:36} through the use of asymptotics near the cusp and are given by,\footnote{We thank Johannes Br\"odel, Nils Matthes, Oliver Schlotterer, and Federico Zerbini  for questioning the validity of the non-zero values which had been obtained for these constants in an earlier version of this paper, and for suggesting that they might all vanish. 
The previous non-zero values arose due to troubles in the asymptotics of [15]. The corrections to the asymptotics required in the corresponding results of \cite{Green:2008uj} will be spelled out in subsection \ref{sec:36}, and in particular in footnote 8.}
\bea
\label{Fconst}
F_6=F_{3,1,1,1}=F_{4,1,1}=F_{2,2,2}=F_{3,2,1}=F_{2,2,1,1}=F_{2,1,1,1;1}=F_T=0
\eea

\subsection{Selected algebraic  identities at weight seven}
\label{sec:33}

At weight 7, we find that the following dihedral combinations are independent of $\tau$,
and these appear to yield the only algebraic purely dihedral identities at weight seven,\footnote{The weight seven identity $F_{3,2,2}$ being constant is a two-loop order identity, which was already obtained in   equation (3.34) of \cite{D'Hoker:2015foa}, in which a typo has been corrected in the present version of the equation.}
 \bea
 \label{7di}
 F_{3,2,2} &=& 7C_{3,2,2} + 7C_{3,3,1} -3E_{7} 
\no\\
F_{2,2,2,1} & = & 7C_{2,2,2,1} - 21 E_{4} E_{3} - 14C_{3,2,2} - 28C_{4,2,1} +31E_{7} 
\no\\
F_7 & = & 
D_7 - 21 E_2 D_5 +35 D_3 D_4 - 1680 D_3 C_{2,1,1} + 336 C_{2,2,1,1,1} 
\no \\ &&
- 1008 E_2 C_{2,2,1} - 2016 E_3 C_{2,1,1} - 4032 C_{4,1,1,1}+ 12096 E_2 E_5 
\no \\ &&
+ 30744 C_{5,1,1} + 924 D_3 E_4 + 14868 C_{3,3,1} - 22680 C_{4,2,1} - 22248 E_7
\no\\
F_{4,1,1,1}&=&28C_{4,1,1,1}+28C_{2,2,2,1}+84C_{3,2,1,1}-168E_2 E_5 
\no\\&&
-252 E_3 E_4 - 294 C_{5,1,1} + 105 C_{3,2,2} - 378 C_{4,2,1} + 654 E_7 
 \eea
 The graphical representations of $C_{2,2,2,1}$, $C_{2,2,1,1,1}, C_{3,2,1,1}$ and $C_{4,1,1,1}$ are immediate generalizations of those given in (\ref{fig10}), while $D_7$ is given by (\ref{fig8}). Amongst the many identities involving both dihedral and trihedral graphs we obtain only the following trihedral identity,
 \bea
 \label{7tri}
F_{3,3,1}& = & 
20 D_{3,3,1} - 9 D_{5,1,1} - 30 E_2 D_{3,1,1} + 45 E_3 E_2^2 - 840 C_{2,2,1,1,1} 
\no \\ &&
+ 2520 E_2 C_{2,2,1} + 6210 E_3 C_{2,1,1} - 360 C_{3,1,1,1,1} + 2160 E_2 C_{3,1,1} 
\no \\ &&
+ 720 C_{4,1,1,1} - 2160 C_{3,2,1,1} - 990 C_{5,1,1} + 5790 E_4 E_3 
\no \\ &&
+ 8265 C_{3,2,2} + 50910 C_{4,2,1} - 32100 E_7 
\eea
where the graphical representation of $D_{3,3,1}$ is given by, 
\bea
\label{fig13}
\tikzpicture[scale=1.1]
\scope[xshift=-5cm,yshift=-0.4cm]
\draw (-1.3,0) node{$D_{3,3,1} ~ =$};
\draw[thick]   (0,0.5) node{$\bullet$} ..controls (1,0.75) .. (2,0.5) node{$\bullet$} ;
\draw[thick]   (0,0.5) node{$\bullet$} ..controls (0.9,0) .. (1,-0.8) node{$\bullet$} ;
\draw[thick]   (0,0.5) node{$\bullet$} ..controls (0.3,-0.5) .. (1,-0.8) node{$\bullet$} ;
\draw[thick]   (0,0.5) node{$\bullet$} -- (1,-0.8) node{$\bullet$} ;
\draw[thick]   (2,0.5) node{$\bullet$} ..controls (1.2,0.1) .. (1,-0.8) node{$\bullet$} ;
\draw[thick]   (2,0.5) node{$\bullet$} ..controls (1.7,-0.4) .. (1,-0.8) node{$\bullet$} ;
\draw[thick]   (2,0.5) node{$\bullet$} -- (1,-0.8) node{$\bullet$} ;
\endscope
\endtikzpicture
\eea
This final case involving the trihedral modular graph function $D_{3,3,1}$ is an important piece of evidence in favor of the claim that equality of the Laurent polynomial of two modular graph functions implies their exact equality. The importance of this case stems from the fact that it involves the lowest weight modular graph functions whose Laurent polynomial contains the irreducible multi-zeta values $\zeta(3,5,3)$ and $\zeta(3, 5)$, as evaluated in  \cite{Zerbini}. These multi-zeta values occur in both the $D_{3,3,1}$ and $D_{5,1,1}$ asymptotics and can only be cancelled in the combination $20 D_{3,3,1} - 9 D_{5,1,1}$, which is exactly the combination observed in (\ref{7tri}).

\subsection{Proof for a dihedral identity}
\label{sec:34}

The search for all modular identities at a given weight, as well as the proof of each modular identity, may be carried out all at once. One begins by organizing the calculation of multiple derivatives of each modular function by subtracting, at each order of derivatives, all terms involving purely holomorphic  Eisenstein series $G_{2k}$, namely terms proportional to $\cG_{2k}= (\tau_2)^{2k} G_{2k}$, as they emerge from the holomorphic subgraph reduction procedure. 

\sm

For example, in taking successive $\DD$-derivatives of the two-loop modular graph functions $C_{3,2,1}$ and $C_{2,2,2}$, no holomorphic Eisenstein factors are encountered until their fourth derivatives. Subtracting the contributions involving $\cG_6$ then allows one to take one further derivative. Other two-loop graphs, such as $C_{4,1,1}$, require holomorphic subtractions at a lower order in derivatives. For weight six, the derivatives of all two-loop modular functions may be arranged as follows, 
 \bea
 \label{3e2}
\nabla   \left ( \nabla ^4 C_{3,2,1} + 80 \, \cG_6 \, \nabla E_3  \right)
& = & {11 \over 14} \, \nabla ^5 E_6 
- 340 \, \cG_6 \,\nabla ^2 E_3
\no \\
\nabla \left ( \nabla ^4 C_{2,2,2} - 240 \, \cG_6 \, \nabla E_3 \right )
& = & - { 9 \over 7} \, \nabla ^5 E_6 + 1200 \, \cG_6 \, \nabla ^2 E_3
\no\\
\nabla \left(\nabla \left(\nabla^3 C_{4,1,1}  + 3 \, \cG_4 \, \nabla E_4 \right) 
+ 9\, \cG_4 \, \nabla ^2 E_4  \right) 
& = &
{167 \over 126}\nabla ^5 E_6
-18\, \cG_4 \,\nabla ^3 E_4 
\no \\ &&
-20\, \cG_6 \,\nabla ^2 E_3  
-2520\, \cG_8 \,\nabla E_2 
\eea
Successive derivatives of products of modular graph functions of lower weights will also be needed, and may be calculated with the same procedure. There are many such products at weights six and seven, so we shall list just two example here at weight six, 
\bea
\label{3e3}
\nabla \left(\nabla \left( \nabla^3 E_3^2 - 120 \, \cG_6 \, E_3 \right) 
- 360\, \cG_6 \, \nabla E_3 \right) = 720\, \cG_6 \,\nabla ^2 E_3 
\eea
as well as,
\bea
\label{3e4}
&&
\nabla \left ( \nabla \Big  ( \nabla \left ( \nabla ^2 (E_2 E_4) - 6 \cG_4 E_4 \Big ) - 12 \cG_4 \nabla E_4 \right ) -18 \cG_4 \nabla ^2 E_4 - 840 \cG_8 E_2 \right ) 
\no \\ && \hskip 2.5in 
= 24 \cG_4 \nabla ^3 E_4 + 3360 \cG_8 \nabla E_2
\eea
Successive derivatives of higher loop graphs, such as the 3-loop graphs $C_{2,2,1,1}$ and $C_{3,1,1,1}$, may be taken following the same procedure, but the result is now quite a bit more involved, and given as follows, 
\bea
\label{3e5}
&& 
\nabla \left( \nabla \left(\nabla \left( \nabla^2 C_{2,2,1,1} - 6 \, \cG_4 E_4  \right)
-12 \, \cG_4 \, \nabla C_{2,1,1} 
-240 \, \cG_6 \, E_3 \right)\right.
\no\\&&
\qquad\qquad\left. 
-{72\over 5}\, \cG_4 \, \DD^2 E_4
+18\, \cG_4 \, \left(\DD E_2\right)^2
-400\, \cG_6 \, \nabla E_3 
-840\, \cG_8 \, E_2 \right ) 
\no\\&&
\qquad\qquad\qquad= 
-{73 \over 63} \nabla ^5 E_6 
+{84 \over 5} \, \cG_4 \, \nabla ^3 E_4 
+1000\, \cG_6 \, \nabla ^2 E_3 
+2352\, \cG_8 \, \nabla E_2 
\qquad
\eea
and
\bea
&&
\nabla \left(\nabla \left(\nabla \left( \nabla^2 C_{3,1,1,1} - 18 \, \cG_4 E_4 \right) 
-27 \, \cG_4 \nabla E_4 
+18 \, \cG_4 \, \nabla C_{2,1,1}  \right)\right.
\no\\&&
\qquad\qquad\left.
+{108\over 5}\, \cG_4 \, \DD^2 E_4
-27\, \cG_4 \, \left(\DD E_2\right)^2
+240\, \cG_6 \nabla E_3 
-2520\, \cG_8 E_2 \right) 
\no\\&&
\qquad\qquad
={122 \over 21} \nabla ^5 E_6
-{396 \over 5} \, \cG_4 \nabla ^3 E_4
-2400\, \cG_6 \nabla^2 E_3 
-11088 \, \cG_8 \nabla E_2 
\eea
Successive derivatives of higher loop functions, such as $D_6$ and $D_7$ are much more involved yet, and their analogous expressions have been evaluated with the help of Maple and Mathematica. We shall not reproduce them here. 

\sm

To obtain algebraic modular identities at weight $w$, we need to achieve a relation of the type $\nabla ^{w-1} F=0$, where $F$ is a polynomial in modular graph functions, which is of homogeneous weight $w$. For example, for dihedral graphs of weight six, $F$ will be a linear combination of the following monomials: $D_6, D_{4,1,1}=C_{2,1,1,1,1}, C_{3,1,1,1}, C_{2,2,1,1}, C_{4,1,1}, C_{3,2,1}, C_{2,2,2}, E_6$; as well as of the products $E_2C_{2,1,1}, E_2 E_4, E_2^3 , E_3^2, E_3 \zeta (3) $. Note that we may use the relation (\ref{3a2}) for $C_{1,1,1}$ and (\ref{4a1}) for $D_4$ to omit the terms $E_3C_{1,1,1}, C_{1,1,1}^2$ and $E_2 D_4$ from the list. As an example, we may inquire as to the existence of identities on the subspace,
\bea
F = a \, C_{2,2,1,1}+b \, C_{3,1,1,1}+c \, E_2E_4+ d \, E_3^2 + e \, C_{4,1,1}+f \, C_{3,2,1}+g \, C_{2,2,2}+ h \, E_6
\eea
We now proceed by requiring the cancellation of holomorphic Eisenstein contributions at each derivative order, up to order $w-1=5$. The resulting conditions are,  
\bea
\cG_4 E_4  & \hskip 0.5in & a+3b+c=0
\no \\
\cG_4 \nabla E_4  & \hskip 0.5in & 9b+4c-e=0
\no \\
\cG_6  E_3 & \hskip 0.5in & 2a+d=0
\no \\
\cG_4 \, \left(\DD E_2\right)^2 & \hskip 0.5in & 2a - 3b = 0
\no \\
\cG_4 \nabla^2 E_4& \hskip 0.5in & 8a - 12b +10c -5e =0
\no \\
\cG_6  \nabla E_3 & \hskip 0.5in & 10a -6b + 9d -2f + 6g=0=0
\no \\
\cG_8  \nabla E_2 & \hskip 0.5in & 98a -462b +140c-105e=0
\no \\
\cG_6  \nabla ^2 E_3 & \hskip 0.5in & 50a -120b +36 d - e -17 f+60 g =0
\no \\
\nabla ^5 E_6 && 146 a - 732b - 167 e - 99 f + 162 g - 126 h=0
\eea
The coefficient of $\cG_8 E_2$ coincides with the one for $\cG_4 E_4$ and does not yield an independent equation, and likewise for the coefficients of $\cG_4 \DD C_{2,1,1}$ and $\cG_4 \, \left(\DD E_2\right)^2$. Up to overall scaling, these equations have a single solution, given by the identity $F_{3,1,1,1}$ of (\ref{6di}). The process may be repeated including all weight six monomials for dihedral graphs as well, and produces precisely  all the identities of (\ref{6di}) and (\ref{6tri}). Similarly, the analysis may be extended to weight seven.

\subsection{Proof for a trihedral identity}

For trihedral graphs, we shall illustrate the key points of the calculation for the case of the identity $F_{2,2,2}$, for which all the ingredients are available, except the trihedral graph $D_{2,2,2}$ itself. Its representation in terms of $\cC$-functions is given by,
\bea
D_{2,2,2} = \cC \left [ \matrix{1 \, 1 \cr 1 \, 1 \cr} \Bigg | \matrix{1 \, 1 \cr 1 \, 1 \cr} \Bigg | \matrix{1 \, 1 \cr 1 \, 1 \cr} \right ]
\eea
Its first and second derivatives are easily computed, and upon use of the $a_0=4$ two-point holomorphic subgraph reduction formula of appendix \ref{sec:A21}, we obtain, 
\bea
\nabla ^2 D_{2,2,2} & = & 
12 \, \cC^+ \left [ \matrix{4 \cr 0  \cr} \Bigg | \matrix{1 \, 1 \cr 1 \, 1 \cr} \Bigg | \matrix{1 \, 1 \cr 1 \, 1 \cr} \right ]
+ 24 \, \cC^+ \left [ \matrix{3 \, 1 \cr 0 \, 1 \cr} \Bigg | \matrix{1 \, 1 \cr 0 \, 1 \cr} \Bigg | \matrix{1 \, 1 \cr 1 \, 1 \cr} \right ]
\no \\ && 
+ 24 \, \cC^+ \left [ \matrix{2 \, 1 \cr 0 \, 1 \cr} \Bigg | \matrix{2 \, 1 \cr 0 \, 1 \cr} \Bigg | \matrix{1 \, 1 \cr 1 \, 1 \cr} \right ]
+ 18 \, \cG_4 \, E_2^2 
\eea
The holomorphic Eisenstein series in the last term may be absorbed by subtracting $E_2^3$ from $D_{2,2,2}$.  A further derivative of this combination yields,
\bea
\nabla^3 (D_{2,2,2} - E_2^3) 
& = & 216 \, \cC^+ \left [ \matrix{5 \cr 0  \cr} \Bigg | \matrix{1 \, 1 \cr 0 \, 1 \cr} \Bigg | \matrix{1 \, 1 \cr 1 \, 1 \cr} \right ]
+ 144 \, \cC^+ \left [ \matrix{4 \cr 0  \cr} \Bigg | \matrix{2 \, 1 \cr 0 \, 1 \cr} \Bigg | \matrix{1 \, 1 \cr 1 \, 1 \cr} \right ]
- 6 (\nabla E_2)^3
\no \\ &&
+ 144 \, \cC^+ \left [ \matrix{4 \, 1 \cr 0 \, 1 \cr} \Bigg | \matrix{1 \, 1 \cr 0 \, 1 \cr} \Bigg | \matrix{1 \, 1 \cr 0 \, 1 \cr} \right ]
+ 288 \, \cC^+ \left [ \matrix{3 \, 1 \cr 0 \, 1 \cr} \Bigg | \matrix{2 \, 1 \cr 0 \, 1 \cr} \Bigg | \matrix{1 \, 1 \cr 0 \, 1 \cr} \right ]
\no \\ &&
+ 48 \, \cC^+ \left [ \matrix{2 \, 1 \cr 0 \, 1 \cr} \Bigg | \matrix{2 \, 1 \cr 0 \, 1 \cr} \Bigg | \matrix{2 \, 1 \cr 0 \, 1 \cr} \right ]
- 72 \, \cG_4 \, \cC^+ \left [ \matrix{1 \cr 0  \cr} \Bigg | \matrix{1 \, 1 \cr 0 \, 1 \cr} \Bigg | \matrix{1 \, 1 \cr 1 \, 1 \cr} \right ]
\eea
and may be evaluated using the three-point holomorphic subgraph reduction formulas given in appendix \ref{sec:A22}.
The remaining required formula for derivatives of $C_{2,2,2}$ and $ C_{3,2,1}$ were given in (\ref{3e2}), while those for $E_3^2$, and $E_2E_4$ are in (\ref{3e3}) and (\ref{3e4}), and for $C_{2,2,1,1}$ in (\ref{3e5}). Establishing the remaining trihedral identities proceeds along similar lines.

\subsection{Proof of a tetrahedral identity}
\label{sec:35}

In this subsection, we provide two different proofs of the weight six identity $F_T$ of (\ref{6tet}), which involves the unique weight six tetrahedral modular graph function,
\bea
\cD_T = \cD \left [ \matrix{ 1 \,  1\,  1\,  1\,  1\, 1 \cr 1 \,  1\, 1 \, 1 \,  1 \, 1  \cr } \right ]
\eea
using the notations of (\ref{DC}). The first proof is based on a simple manipulation of an earlier result of \cite{Basu:2015ayg}, while the second proof uses the general methods of holomorphic subgraph reduction of this paper. A third proof was provided in \cite{Basu:2016kli}.

\subsubsection{Proof by  Basu's Poisson equation}
\label{sec:371}

The following differential equation was derived  for the unique weight six tetrahedral graph (or ``Mercedes graph")  $\cD_T$  in \cite{Basu:2015ayg},
\bea
\label{7b1}
(\Delta + 6) \cD_T = 48 C_{3,2,1} + 12 E_6 - 12 E_3^2
\eea
We also have  Laplace equations for the following two-loop graphs $C_{2,2,2}, \, C_{3,2,1}$, and $C_{4,1,1}$,\footnote{We thank Br\"odel, Matthes, Schlotterer and Zerbini for pointing out a typo on the last line of (3.32).}
\bea
\label{7b2}
(\Delta -2) ( C_{2,2,2} + 4 C_{3,2,1} ) & = & 52 E_6 - 4 E_3^2
\no \\
(\Delta - 12) (C_{2,2,2} - 6 C_{3,2,1}) & = & - 108 E_6 + 36 E_3^2
\no \\
(\Delta - 12) (C_{2,2,2} + 6 C_{4,1,1}) & = &  120 E_6 +  12 E_3^2 - 36 E_2 E_4
\eea
which were proven  in \cite{D'Hoker:2015foa}. To obtain an algebraic equation for $\cD_T$, we need to  factor  $(\Delta +6)$ from the entire equation (\ref{7b1}).   The only way we can do so is if the contributions $E_3^2$ and $E_2 E_4$ can be eliminated from the inhomogeneous part  on the right side of (\ref{7b1}). Thus, $C_{4,1,1}$ cannot enter the identity, as it alone produces the product $E_2 E_4$. Next, we proceed to eliminating $E_3^2$ between (\ref{7b1}) and the first two equations of (\ref{7b2}). Taking a linear combination with free parameter $\alpha$ of the resulting equations gives, 
\bea
(\Delta + 6) \cD_T & = & 
((3+\alpha)  \Delta - 6 - 3 \alpha) C_{2,2,2}  
\no \\ &&
+ ( (12 + 3 \alpha) \Delta + 24 ) C_{3,2,1} -(144 + 36 \alpha) E_6 
\eea
The operators acting on $C_{3,2,1}$ and $C_{2,2,2}$ both become proportional to $(\Delta+6)$ upon setting $\alpha = - { 8 \over 3} $. Using the equation $(\Delta +6) E_6 = 36 E_6$, and factoring out $(\Delta +6)$, we find, 
\bea
(\Delta +6) \left ( 3 \cD_T - C_{2,2,2} - 12 C_{3,2,1} + 4 E_6 \right ) =0
\eea
Over the space of modular graph forms considered here, the kernel of $\Delta +6$ vanishes, so that we have proven the relation (\ref{6tet}) including the vanishing of $F_T$, as stated above (\ref{Fconst}).

\subsubsection{Proof by holomorphic subgraph reduction}

Since all edges are equivalent to one another, the first derivative of $\cD_T$  is  as follows,
\bea
\DD \, \cD_T = 6 \, \cD^+ \left [ \matrix{ 2 \,  1\,  1\,  1\,  1\, 1 \cr 0 \,  1\, 1 \, 1 \,  1 \, 1  \cr } \right ]
\eea
Three different contributions arise to its second derivative, depending on whether we differentiate the first edge (labelled by index 1), one of the four equivalent edges that share one vertex with the first edge, or the one remaining edge which has no vertices in common with the first edge. Using the momentum conservation identities of (\ref{DCmom}) we obtain, 
\bea
\nabla ^2 \, \cD_T = 
24 \, \cD^+ \left [ \matrix{ 3 \,  1\,  1\,  1\,  1\, 1 \cr 0 \,  1\, 1 \, 0 \,  1 \, 1  \cr } \right ]
+ 24 \, \cD^+  \left [ \matrix{ 2 \,  2\,  1\,  1\,  1\, 1 \cr 0 \,  1\, 1 \, 0 \,  1 \, 1  \cr } \right ]
+ 6 \, \cD^+ \left [ \matrix{ 2 \,  1\,  1\,  1\,  2 \, 1 \cr 0 \,  1\, 1 \, 1 \,  0 \, 1  \cr } \right ]
\eea
To collect like contributions appearing in different permutations of the exponents, we make use of the symmetry properties of the graphs upon interchange of their edges. These symmetries are non-canonical, because the orientation of the loop momenta for tetrahedral graphs is non-canonical. They are generated by the following three operations, 
\bea
\label{symtet}
R_0 \left \{ \matrix{ 
p_1  \leftrightarrow  -p_2 \cr 
p_3  \leftrightarrow  -p_3 \cr 
p_4  \leftrightarrow  +p_4 \cr 
p_5  \leftrightarrow  +p_6 \cr } \right .
\hskip 0.5in 
R_1 \left \{ \matrix{ 
p_1  \leftrightarrow  +p_1 \cr 
p_2  \leftrightarrow  +p_4 \cr 
p_3  \leftrightarrow  -p_6 \cr 
p_5  \leftrightarrow  -p_5 \cr } \right .
\hskip 0.5in 
R_2 \left \{ \matrix{ 
p_1  \leftrightarrow  -p_1 \cr 
p_2  \leftrightarrow  +p_6 \cr 
p_3  \leftrightarrow  -p_4 \cr 
p_5  \leftrightarrow  -p_5 \cr } \right .
\eea
which satisfy  $R_0^2=R_1^2=R_2^2=1$ and $R_1R_2=R_2R_1$. A further derivative is taken in a similar manner, and we obtain, 
\bea
\label{D3DT}
{ 1 \over 24 } \nabla ^3 \cD_T & = &
6 \, \cC^+ \left [ \matrix{ 6 \,  2\,  1 \cr 1 \,  1\, 1  \cr } \right ]
-3 \, \cD^+ \left [ \matrix{ 4 \,  1\,  1\,  1\,  1\, 1 \cr 0 \,  0 \, 1 \, 1 \,  1 \, 0  \cr } \right ]
+ \cD^+ \left [ \matrix{ 3 \,  2 \,  1\,  1\,  1\, 1 \cr 0 \,  0 \, 1 \, 1 \,  0 \, 1  \cr } \right ]
\no \\ &&
-4 \, \cD^+ \left [ \matrix{ 3 \,  2 \,  1\,  1\,  1\, 1 \cr 0 \,  0 \, 1 \, 1 \,  1 \, 0  \cr } \right ]
+ \cD^+ \left [ \matrix{ 3 \,  1\,  1\,  2\,  1\, 1 \cr 0 \,  0\, 1 \, 0 \,  1 \, 1  \cr } \right ]
+3 \, \cD^+ \left [ \matrix{ 3 \,  1\,  1\,  1\,  2 \, 1 \cr 0 \,  0 \, 1 \, 1 \,  0 \, 1  \cr } \right ]
\no \\ &&
+3 \, \cD^+ \left [ \matrix{ 2 \,  2\,  1\,  1\,  2\, 1 \cr 0 \,  0\, 1 \, 1 \,  0 \, 1  \cr } \right ]
+ \cD^+ \left [ \matrix{ 2 \,  2\,  1\,  2\,  1\, 1 \cr 0 \,  0\, 1 \, 0 \,  1 \, 1  \cr } \right ]
+ \cD^+ \left [ \matrix{ 3 \,  1\,  1\,  1\,  1\, 2 \cr 0 \,  0\, 1 \, 1 \,  1 \, 0  \cr } \right ]
\eea
The 3-point and 4-point holomorphic subgraph reduction formulas necessary to evaluate these combinations are presented in Appendix \ref{sec:A3}. The final result is as follows,
\bea
\nabla \left ( { 1 \over 24 } \nabla ^3 \cD_T + 10 \, \cG_6 \nabla E_3 \right ) 
= - 40 \cG_6 \, \nabla ^2 E_3 + {29 \over 504 } \, \nabla ^5 E_6
\eea
Using the first two equations of (\ref{3e2}) we eliminate $\cG_6 \nabla E_3$ and $\cG_6 \nabla^2 E_3$, and find $\nabla ^5 F_T=0$, where  $F_T$ was given in (\ref{6tet}), so that we conclude from the Lemma that $F_T$ is constant.

\subsection{Laurent polynomials at the cusp for weight 6}
\label{sec:36}

In this final subsection, we return to one of the issues raised in the Introduction, namely the relation between the algebraic identities between modular graph functions derived earlier in this section, and their Laurent expansion in powers of $\tau_2$ near the cusp $\tau \to i \infty$. The Laurent expansion for any modular graph function $C_A$ with equal exponents on holomorphic and anti-holomorphic momenta, of weight $w$, takes the  form given in (\ref{3a11}),  where the coefficients $f_A(k)$ depend on the exponents $A$ but are independent of $\tau$ \cite{D'Hoker:2015qmf}. 

\sm

The expression of the Laurent polynomial  is  familiar for the  non-holomorphic Eisenstein series, and is given as follows,
\bea
E_n = (-1)^{n-1}\frac{ 4^n B_{2 n} }{(2 n)!}y^n+\frac{4^{2-n} (2 n-3)! }{(n-2)! (n-1)!}y^{1-n} \zeta (2
   n-1) + \cO (e^{-2 \pi \tau_2})
\eea
where $B_{2n}$ are Bernoulli numbers.\footnote{The values needed here are given as follows, $B_4=B_8=-{1 \over 30}$, $B_6 ={1 \over 42}$, $B_{10} = { 5 \over 66}$, and $B_{12} = -{ 691 \over 2730}$.}

\sm

The values of the coefficients $f_A(k)$ for general exponents $A$ is not known, though an explicit formula for the Laurent polynomials of $D_\ell$ was derived in equation (B.6) of \cite{Green:2008uj}, with the help of results of Zagier on multiple zeta sums. To low weight order, the Laurent polynomial may be computed diagram by diagram \cite{D'Hoker:2015foa} using summations over lattice sums, Laplace eigenvalue equations, and low weight identities, such as the ones that have already been established in (\ref{3a2}) and (\ref{4a1}). 

\sm

Using the identities of (\ref{4a1}), these methods give all weight 4 and weight 5 modular graph functions, such as $D_4, D_5$ and $D_{2,2,1}$ in terms of $C_{2,1,1}$ and $C_{3,1,1}$, whose Laurent polynomials are given by,\footnote{In the remainder of this subsection, the addition of the symbol $\cO(e^{-2\pi \tau_2})$ to indicate that exponential correction are being omitted, will always be understood, but will not be exhibited explicitly.} 
\bea
C_{2,1,1}&=&
\frac{2 y^4}{14175}+\frac{ \zeta (3) y}{45}+\frac{5 \zeta (5)}{12 y}-\frac{\zeta (3)^2}{4 y^2}+\frac{9 \zeta (7)}{16 y^3}
\no \\ 
C_{3,1,1}&=&
\frac{2 y^5}{155925}+\frac{ 2\zeta (3) y^2}{945}-\frac{ \zeta (5)}{180}
+\frac{7 \zeta (7)}{16 y^2} - { \zeta (3) \zeta (5) \over 2 y^3} +\frac{43 \zeta (9)}{64 y^4}
\eea
At weight six, the Laurent polynomial of two-loop modular graph functions are given by,
\bea
C_{4,1,1}&=&\frac{808 y^6}{638512875}+\frac{ \zeta (3)  y^3}{4725}-\frac{ \zeta (5)  y}{1890} +\frac{\zeta (7)}{720 y}
+\frac{23 \zeta (9)}{64y^3}
  \no\\&&
-\frac{\zeta (5)^2+30 \zeta(3) \zeta (7)}{64 y^4}+\frac{167 \zeta (11)}{256 y^5}
  \no \\
C_{3,2,1}&=&\frac{43 y^6}{58046625}+\frac{y \zeta (5)}{630}+\frac{\zeta (7)}{144 y}+\frac{7 \zeta (9)}{64 y^3}-\frac{17 \zeta (5)^2}{64y^4}+\frac{99 \zeta (11)}{256 y^5}
  \no \\
C_{2,2,2}&=&\frac{38 y^6}{91216125}+\frac{\zeta (7)}{24 y}-\frac{7 \zeta (9)}{16 y^3}+\frac{15 \zeta (5)^2}{16y^4}-\frac{81 \zeta (11)}{128 y^5}
 \eea
They can be easily obtained from the inhomogeneous  Laplace-eigenvalue equations satisfied by these functions, and given in \cite{D'Hoker:2015foa}. Parts of their expressions were given in \cite{Green:2008uj} up to order $\cO(y^{-1})$ contributions. 

\sm

Next, we have the three-loop dihedral modular graph functions,\footnote{
In the expressions for $C_{2,2,1,1}$, $D_{2,2,1,1}$ and $D_{2,1,1,1;1}$ below, the coefficients of the constant terms in $\zeta (3)^2 $ have been corrected from their earlier versions of this paper, which in turn had been obtained from eqs (B.54-55), (D.31), and (D.38) in \cite{Green:2008uj}.}
 \bea
C_{3,1,1,1}&=&\frac{5 y^6}{567567}+\frac{2  \zeta (3)y^3}{945}-\frac{ \zeta(5)y}{252}+\frac{\zeta (3)^2}{60}+\frac{49 \zeta (7)}{240 y} -\frac{5 \zeta (3) \zeta (5)}{8 y^2}
 \no\\&&
+\frac{49 \zeta (9)+3\zeta (3)^3}{12y^3}-\frac{33 \zeta (3) \zeta (7)+30\zeta (5)^2}{16y^4}+\frac{183 \zeta (11)}{64 y^5}
  \no \\
  C_{2,2,1,1}&=&
  \frac{103 y^6}{13030875}
  +\frac{ \zeta (3)y^3}{2025}
  +\frac{\zeta(5)y }{54}
  -\frac{ \zeta (3)^2}{90} 
  -\frac{\zeta (7)}{360 y}  
  +\frac{5 \zeta (3) \zeta (5)}{12 y^2}
\no \\ &&
  +\frac{5 \zeta (9)-48\zeta (3)^3}{288y^3}+\frac{14 \zeta (3) \zeta (7)+25 \zeta (5)^2}{32y^4}-\frac{73 \zeta (11)}{128 y^5}
\eea 
These graphs were denoted by $C_{3,1,1,1}=D_{1,1,1,3}$ and $C_{2,2,1,1}=D_{1,1,1,1;2}$ in \cite{Green:2008uj}, and the Laurent polynomial of $C_{3,1,1,1}$  was evaluated up to order $\cO(y^{-1})$ in equation (B.56) of \cite{Green:2008uj}. The remaining terms were evaluated here using the $F_{3,1,1,1}$ and $F_{2,2,1,1}$ identities, along with their value at the cusp given by \cite{Green:2008uj}. 

\sm

The four and five-loop dihedral modular graph functions, 
\bea
D_{4,1,1}&=&\frac{284 y^6}{18243225}+\frac{2  \zeta (3) y^3}{135}+\frac{5  \zeta (5) y}{18}+\frac{\zeta (3)^2}{10} +\frac{51 \zeta (7)}{20y}
\no\\&&
  +\frac{11\zeta (3) \zeta (5)}{2 y^2}+\frac{79 \zeta (9)-36 \zeta (3)^3}{24 y^3}-\frac{9 \zeta (3) \zeta (7)}{4y^4}+\frac{45 \zeta (11)}{16 y^5}
  \no \\
D_{6}&=&\frac{53 y^6}{729729}+\frac{5  \zeta (3) y^3}{27}+\frac{140 \zeta (5) y}{9}+25 \zeta (3)^2+\frac{1005 \zeta (7)}{4 y}-\frac{135 \zeta (3) \zeta (5)}{y^2}
 \no\\&&
+\frac{405 \zeta (9)+90\zeta (3)^3}{2y^3}-\frac{1350 \zeta (3) \zeta (7)+675 \zeta (5)^2}{8y^4}+\frac{4725 \zeta (11)}{32 y^5}
\eea
were also evaluated in \cite{Green:2008uj} to order $\cO(y^{-1})$, and computed in full in \cite{Zerbini}.

\sm

The Laurent polynomials of the trihedral three-loop modular graph functions are, 
\bea
D_{2,2,2}&=&\frac{193 y^6}{11609325}+\frac{ \zeta (3)y^3}{315}+\frac{59  \zeta (5)y}{315}+\frac{23 \zeta (7)}{20 y}
  \no\\&&
+\frac{5\zeta (3) \zeta (5)}{2 y^2}-\frac{65 \zeta (9)}{48 y^3}+\frac{21 \zeta (5)^2-18 \zeta (3)\zeta (7)}{16 y^4}+\frac{99 \zeta (11)}{64 y^5}
  \no \\
D_{3,2,1}&=&\frac{298 y^6}{42567525}+\frac{ \zeta (3) y^3}{315}+\frac{173  \zeta(5) y}{1260}+\frac{3 \zeta (3)^2}{20}+\frac{53 \zeta (7)}{20 y}
  \no\\&&
-\frac{5 \zeta (3) \zeta (5)}{2 y^2}+\frac{96 \zeta (3)^3+223 \zeta(9)}{32 y^3}-\frac{99 \zeta (5)^2+162 \zeta(3) \zeta (7)}{32 y^4}+\frac{729 \zeta (11)}{128 y^5}
   \no\\
   D_{2,2,1,1}&=&
   \frac{68 y^6}{70945875}
   +\frac{4  \zeta (3)y^3}{14175}
   -\frac{ \zeta(5) y}{945}
 +\frac{\zeta (3)^2}{45} 
   +\frac{13 \zeta (7)}{45 y}
 \no\\&&
 -\frac{5 \zeta (3) \zeta (5)}{6 y^2}+\frac{61 \zeta (9)+12\zeta (3)^3}{36y^3}-\frac{3 \zeta (3) \zeta (7)+\zeta (5)^2}{2y^4}+\frac{81 \zeta (11)}{64 y^5}
  \no \\
D_{2,1,1,1;1}&=&
\frac{802 y^6}{638512875}
+\frac{2  \zeta (3) y^3}{14175}
+\frac{43  \zeta (5) y}{3780}
-\frac{\zeta(3)^2}{180}
+\frac{11 \zeta (7)}{180 y}
 \no\\&&
 +\frac{5 \zeta (3) \zeta (5)}{24y^2}-\frac{65 \zeta(9)+48\zeta (3)^3}{576 y^3}-\frac{6 \zeta (3) \zeta(7)+\zeta (5)^2}{64 y^4}+\frac{147 \zeta (11)}{256 y^5}
 \eea
 The Laurent polynomial  of the graphs $D_{4,1,1}$, $D_{3,2,1}$, and $D_{2,2,2}$ were evaluated in equations (B.36-38) of \cite{Green:2008uj} up to order $\cO(y^{-1})$ contributions, and were computed in full in \cite{Zerbini} whose expressions we have borrowed. The Laurent polynomial of $C_{3,1,1,1}=D_{3,1,1,1}$ was also evaluated in equation (B.56) of \cite{Green:2008uj} up to order $\cO(y^{-1})$ contributions; its full expression was derived here with the help of the  identity for $F_{3,1,1,1}$ and the value at the cusp provided by (B.56) of \cite{Green:2008uj}.

\sm

Finally, there is the tetrahedral modular graph function, which is three-loops,
 \bea
\cD_T&=&\frac{46 y^6}{212837625}+\frac{2  \zeta (5) y}{315}+\frac{\zeta (7)}{24 y}+\frac{7 \zeta (9)}{24 y^3}-\frac{3 \zeta (5)^2}{4y^4}+\frac{87 \zeta (11)}{128 y^5}
 \eea
Its value was derived here from the identity (\ref{6tet}) which $\cD_T$ satisfies, where the value $F_T=0$ may be derived either from our proof in (\ref{sec:371}) based on the result of \cite{Basu:2015ayg}, or directly on the results of \cite{Basu:2016kli}.

\subsection{Sieve algorithm versus Laurent polynomial matching}

The identities between modular graph functions at weights three, four and five, respectively in (\ref{3a2}) and (\ref{4a1}), were all at first conjectured on the basis of their matching Laurent polynomial near the cusp in \cite{D'Hoker:2015foa} , and subsequently proven using the sieve algorithm of \cite{D'Hoker:2016a}. Matching the Laurent polynomials obtained by \cite{Zerbini}  was further used in \cite{DGV} to conjecture the identities $F_6$ and $F_{4,1,1}$ in (\ref{6di}). This was possible because these two relations involve only four-vertex (including bivalent vertices in the counting) modular graph functions. 

\sm

The calculation of Laurent polynomials of modular graph functions directly by carrying out the Kronecker summations for graphs with more vertices, and for higher weight, appears to become prohibitively complicated, and has not been carried out so far.  Therefore, the remaining identities which we have presented at weight six and seven were obtained here by the sieve algorithm of \cite{D'Hoker:2016a}. In all cases where Laurent polynomials were known in full, the procedure of matching the Laurent polynomials at the cusp does produce the corresponding modular identities uniquely. 

\sm

Many of the Laurent polynomials newly obtained here are for modular graph functions with more than four vertices (including bivalent vertices). Higher point function amplitudes in string theory at genus-one, and their interrelations, are of considerable interest for higher effective actions, and were analyzed recently  in \cite{Green:2013bza} and \cite{Basu:2016mmk}.

\newpage

\section{Identities for primitive modular graph functions}
\setcounter{equation}{0}
\label{subgraphs}

In this section, we develop a natural decomposition of modular graph functions into {\sl primitive modular graph functions} in terms of which modular identities will greatly simplify. 

\sm

Primitive modular graph functions are defined by subjecting the momentum summation in the definition of  modular graph functions to the additional restriction that all non-trivial subsets of the momenta entering any vertex of the graph must sum to a non-zero value.  Since zero momentum is a modular invariant characterization, the additional restriction  is modular invariant, so that primitive modular graph functions are indeed modular functions. The primitive modular graph function associated to a  graph with exponents $A,B$ will be denoted by $\hat \cC$, and is given by a formula analogous to (\ref{3b5}),
\bea
\label{prim}
\hat \cC \left [ \matrix{A \cr B \cr} \right ]  (\tau) =  \sum_{p_1,\dots,p_\L \in \Lambda} ^{(\pi)} ~ \prod_{r =1}^\L
  { (\tau _2 / \pi)^{\half a_r + \half b_r} \over  (p_r) ^{a_r} ~ (\bar p _r) ^{b_r} }\, \prod_{i =1}^V
  \delta \left ( \sum_{s =1}^\L \Gamma _{i \, s} \, p_s \right )
\eea 
where the superscript $(\pi)$ on the momentum summation stands for the momentum restriction discussed above. The primitive modular graph functions $\hat C$, $\hat D$, and $\hat E$ are defined with the help of the same prescription on momentum summation. The formulas for momentum conservation in (\ref{3d3}), the action of the derivative $\DD$ in (\ref{nab}), and of the Laplacian $\Delta$ in (\ref{Lap1}), all have the same form in terms of primitive modular graph functions $\hat \cC$. The formulas for  algebraic and holomorphic subgraph reduction will, however, be different for $\hat \cC$ functions, as will be spelled out in Appendix \ref{holo}.

\sm

The decomposition of a modular graph function into primitive modular graph functions is obtained by partitioning the momentum sums into all cases in which a non-trivial subset of the momenta entering any given vertex sums to zero. These contributions effectively correspond to disconnected subgraphs, and evaluate to a product of two or more component subgraphs. Each of the factors in this sum gives rise to a separate primitive modular graph function. Therefore, the decomposition expresses a modular graph function of weight $w$ as a polynomial in primitive modular graph functions which is homogeneous of weight $w$. 

\sm

 The importance of primitive graphs is that they, rather than the modular graph functions from which they stem, provide the simplest and most natural building blocks of modular identities. Specifically, all the algebraic identities established in section~\ref{six-seven} will become {\sl linear in terms of primitive modular graph functions}. 
 
 \sm
 
 In the remainder of this section, we shall make the decomposition formulas explicit and express the  identities derived earlier in terms of primitive modular graph functions. Finally, we shall relate the combinatorics of the decomposition into primitive modular graph functions to holomorphic subgraph reduction and the sieve algorithm of \cite{D'Hoker:2016a} and investigate the origin of the  linearity of the identities observed at low weights.

\subsection{Decomposition into primitive modular graph functions}

We begin with some elementary examples. The simplest family of modular graph functions are the one-loop  non-holomorphic Eisenstein series $E_w$, which are automatically primitive, 
\bea
E_w = \hat E_w
\eea 
Indeed, the graph only contains bivalent vertices and the only non-trivial subset of the momenta entering a bivalent vertex is a single momentum, which is not allowed to vanish. In fact, this argument for bivalent vertices applies in any graph.

\sm

All two-loop modular graph functions, including $C_{a,b,c}$, are also automatically primitive,
\bea
C_{a,b,c} = \hat C_{a,b,c} 
\eea
The condition for primitives on bivalent vertices is trivially satisfied, while on the remaining trivalent vertices the non-trivial subsets of momenta entering a vertex are either on a single momentum (which is trivial), or on a pair of distinct momenta. The sum of a pair of distinct momenta entering a trivalent vertex gives the third momentum which cannot vanish, and hence the condition on any pair of distinct momenta is also trivial.
In fact, this argument for trivalent vertices applies in any graph. Thus all graphs whose vertices are either bivalent or trivalent are automatically primitive. The same holds for any graph in which at most a single vertex has valence greater than three. Examples of interest to us are, 
\bea
\label{Dhats}
D_{2,2,1} & = & \hat D_{2,2,1} \hskip 1.5in \cD_T = \hat \cD_T
\no \\
D_{2,2,1,1} & = & \hat D_{2,2,1,1}
\no \\
D_{2,1,1,1;1} & = & \hat D_{2,1,1,1;1}
\eea
where $\cD_T$ stands for any tetrahedral graph built of bivalent and trivalent vertices only.

\sm

Starting at three loops, the restriction to primitives becomes non-trivial, as now several higher valence vertices will enter a graph. It is straightforward to write a general decomposition formula for three-loop dihedral graphs, \footnote{Throughout, we omit hats on $E_w$, $D_3$, and all two-loop graphs since they are automatically primitive.} 
\bea
\label{3loop}
 C_{a,b,c,d} & = &  \hat C_{a,b,c,d} + E_{a+b} E_{c+d} + E_{a+c} E_{b+d} + E_{a+d} E_{b+c}  
\eea
The formula may be readily generalized to higher loops,  trihedral  graphs, and so on. The low weight formulas we shall need are given by,
\bea
\label{speDin}
D_4 &=& \hat  D_4 + 3 E_2^2
\no\\
D_5 &=& \hat  D_5 + 10E_2 D_3
\no\\
 D_6 &=&  \hat D_6 + 15 E_2 \hat D_4 + 15 E_2^3 + 10 D_3^2 
\no\\
 D_7 &=& \hat D_7 + 21 E_2 \hat D_5 + 105 E_2^2 D_3  + 35 D_3 \hat D_4
\eea
In each formula above the coefficients of  the monomials  on the right side correspond to the number of partitions. For example in $D_7$  the number of partitions of 7 into 2 and 5 is 21; into 3 and 4 is 35; and  into 2, 2, and 3 is 105. All dihedral graphs with four loops and higher needed here are given by, 
\bea
 C_{a,1,1,1,1} & =&   \hat C_{a,1,1,1,1} + 6 E_2 C_{a,1,1} + 4 E_{a+1} D_3 \hskip 1in a \geq 2
 \no\\
 C_{2,2,1,1,1} &=& \hat C_{2,2,1,1,1} + 6 D_3 C_{2,1,1} + D_3 E_4 + 3 E_2 C_{2,2,1}
\no\\
C_{2,1,1,1,1,1} & = & \hat  C_{2,1,1,1,1,1} + 10 E_2 C_{2,1,1,1} + 15 E_2^2 E_3 + 10 D_3 C_{2,1,1} + 5 E_3 \hat D_4
\eea
The decompositions of the remaining trihedral graphs are as follows,
\bea
\label{speTriin}
 D_{2,2,2} &=& \hat D_{2,2,2} + 4E_3^2 + E_2^3
\no\\
 D_{3,2,1} &=& \hat D_{3,2,1} + 3 E_2 C_{2,1,1}
\no\\
 D_{3,3,1} &=& \hat D_{3,3,1} + 6 E_2 D_{3,1,1} + 9 E_3 E_2^2
\eea
Again, the coefficients correspond to the number of partitions for each monomial on the right side. For example in $D_{2,2,2}$ there is only one partition of 6 into 2, 2, and 2, but four of 6 into 3 and 3.

\subsection{Identities in terms of primitive modular graph functions}
\label{sec:42}

In this subsection, we express all the modular identities of section \ref{six-seven} in terms of primitive modular graph functions.  Remarkably, the resulting identities are all linear in primitive modular graph functions, thereby producing a significant simplification.  Whether this phenomenon extends to higher weights will be discussed below and further in section \ref{sec:6}. 

\sm

The weight four and weight five identities of (\ref{4a1}) become, 
\bea
F_4 &=& \hat D_4 - 24 C_{2,1,1} + 18 E_4
\no\\ 
F_5 &=& \hat D_5 - 60 C_{3,1,1} + 48 E_5 - 16 \zeta(5)
\no\\
F_{3,1,1} &=& 40 \hat D_{3,1,1} - 300 C_{3,1,1} + 276 E_5 - 7 \zeta(5)
\no \\
F_{2,2,1} & = & 10 \hat D_{2,2,1} - 20 C_{3,1,1} + 4 E_5 - 3 \zeta (5)
\eea
The constant values  $F$ in the identities above and the identities to follow are the same as in the original identities  and therefore written without hats. The weight six identities for dihedral  graphs of (\ref{6di}) become, 
\bea
F_{6} &=& \hat D_6 + 720 \hat C_{2,2,1,1} - 3460 C_{2,2,2} - 13440 C_{3,2,1} - 720 C_{4,1,1} + 7600 E_{6}
\no\\
F_{3,1,1,1}  &=&  2 \hat C_{3,1,1,1}+ 3 \hat C_{2,2,1,1} -18 C_{4,1,1}-24 C_{3,2,1}-2 C_{2,2,2}+32 E_{6} 
\no\\
F_{4,1,1}  &=&  - 3 \hat D_{4,1,1} + 109 C_{2,2,2} + 408 C_{3,2,1} + 36 C_{4,1,1} - 211 E_6
\eea
while those for trihedral graphs of (\ref{6tri}) take the form,
\bea
F_{2,2,2} &=& 3 \hat D_{2,2,2} - 18 \hat C_{2,2,1,1}  - 58 C_{2,2,2} - 192 C_{3,2,1}  + 46 E_6 
\no\\
F_{3,2,1}  &=& 2 \hat D_{3,2,1} + 18 \hat C_{2,2,1,1}  - 36 C_{4,1,1} - 69 C_{2,2,2} - 288 C_{3,2,1} +183 E_6
\no\\
F_{2,2,1,1}&=& 3 \hat D_{2,2,1,1} + 6 \hat C_{2,2,1,1} - 10 C_{2,2,2} - 48 C_{3,2,1} -12 C_{4,1,1} + 40 E_6 
\no\\
F_{2,1,1,1;1}  &=& 18 \hat D_{2,1,1,1;1} - 9 \hat C_{2,2,1,1}  - 20 C_{2,2,2}  - 60 C_{3,2,1} - 10 E_6
\eea
The weight six tetrahedral modular graph function $\cD_T$ is already primitive by (\ref{Dhats}), so that the corresponding identity is unmodified, and we have,
\bea
F_T = 3 \hat \cD_T - C_{2,2,2} - 12 C_{3,2,1} +4 E_6
\eea
Finally, the weight seven identities of (\ref{7di}) and (\ref{7tri}) become, 
\bea
F_{3,2,2} &=& 7C_{3,2,2} + 7C_{3,3,1} -3E_{7} 
\no\\
F_{2,2,2,1} &=& 7 \hat C_{2,2,2,1} - 14C_{3,2,2} - 28C_{4,2,1} +31E_{7} 
\no\\
F_{4,1,1,1} &=& 28 \hat C_{4,1,1,1} + 28 \hat C_{2,2,2,1} + 84 \hat C_{3,2,1,1} - 294 C_{5,1,1} + 105 C_{3,2,2}
\no\\&&
 - 378 C_{4,2,1} + 654 E_7 
\no\\
F_{3,3,1}& = & 80 \hat D_{3,3,1} - 36 \hat D_{5,1,1} - 3360 \hat C_{2,2,1,1,1} - 1440 \hat C_{3,1,1,1,1} + 2880 \hat C_{4,1,1,1}
\no\\&&
 - 8640 \hat C_{3,2,1,1} - 3960 C_{5,1,1} + 33060 C_{3,2,2} + 203640 C_{4,2,1} - 124800 E_7 
 \no\\
F_7 &=& \hat D_7 + 336 \hat C_{2,2,1,1,1} - 4032 \hat C_{4,1,1,1}  + 30744 C_{5,1,1}  + 14868 C_{3,3,1}
\no\\&&
 - 22680 C_{4,2,1} - 22248 E_7 
\eea
Manifestly, all identities above are linear in the primitive modular graph functions.

\subsection{Holomorphic subgraph reduction and primitive graphs}

The goal of this subsection is to elucidate the striking difference between the modular identities expressed in terms of primitive versus non-primitive modular graph functions. In the primitive case all the algebraic identities derived in this paper are linear, while in the non-primitive case non-linear contributions are required. The difference may be traced back directly to the structure of the holomorphic subgraph reduction procedure, and formulas for both primitive and non-primitive cases are provided in Appendix \ref{holo}.

\sm

The source of all non-linear terms in the identities for the non-primitive case resides in the contributions from the holomorphic subgraph reduction which are of the form,
\bea
\label{sub1}
\cG_{2k} \, C _{A_k}
\eea
where $C_{A_k}$ is a modular graph function of vanishing modular weight with exponents $A_k$. All other terms resulting from holomorphic subgraph reduction  are of the form,
\bea
\label{sub2}
\cC^+ \left [ \matrix{ A_0 \cr B_0 \cr} \right ] \hskip 0.5in \hbox{or} \hskip 0.5in
\cG_{2 \ell} \, \cC^+ \left [ \matrix{ A_\ell \cr B_\ell \cr} \right ]
\eea
with $\ell \geq 2$.
The exponents are unequal, $B_0 \not = A_0$ and $B_\ell \not = A_\ell$, and the  $\cC$-functions are forms of  modular weight $(0, s)$ with $s <0$ since they result from applying successive $\DD$-derivative, whose modular weight is $(0,-2)$. The key deciding factor as to whether terms arising from holomorphic subgraph reduction require non-linear terms or not in the corresponding modular identity is provided by the simple observation that the purely holomorphic Eisenstein series in terms of the type (\ref{sub1}) may be eliminated by the addition of a non-linear term $E_k C_{A_k}$ to the modular identity, while the holomorphic Eisenstein series in terms of the type (\ref{sub2}) cannot be  eliminated in this manner.

\subsubsection{The example of $D_L$}

We illustrate the effect of successive derivatives and the use of holomorphic subgraph reduction on the family of dihedral graphs of the type $D_L$. The first holomorphic Eisenstein series $G_4$ is encountered at order $\nabla^2$, 
\bea
\label{d2DL}
\nabla^2 D_L = 2L(L-1) \, \cC^+  \left [ \matrix{ 4 & 1_{L-2} \cr  0 & 1_{L-2} \cr }\right ] + 3 L(L-1) \, \cG_4 D_{L-2}
\eea 
where $\cG_4=(\tau_2)^2 G_4$. The second term on the right side is of the form (\ref{sub1}). The sieve algorithm developed in \cite{D'Hoker:2016a} is based on the fact that, in its present form,  a further application of $\DD$ on (\ref{d2DL}) will produce a factor involving $\DD \cG_4$, which needs to be neutralized, as its presence presents an obstruction to terminating the sequence of successive derivatives required by the Lemma. There is only one way to eliminate this obstruction, namely by subtracting a non-linear term from $D_L$ under the $\DD^2$ symbol. Doing so with the help of   $\cG_4 = \nabla ^2 E_2$ of (\ref{Gk}) we obtain, 
\bea
\nabla^2 \left ( D_L -{{1}\over{2}}L(L-1) E_2 D_{L-2} \right ) 
=
 - {{1}\over{2} }L(L-1)E_2 \nabla^2 D_{L-2} + \cdots
\eea 
with the ellipses representing terms which contain no holomorphic Eisenstein series. The factor $\nabla ^2 D_{L-2}$ on the right side again gives rise to a holomorphic Eisenstein series and the procedure may be iterated to produce the following combination,
\bea
\label{Dhat1}
 D_L + \sum _{\ell =1} ^{[L/2]} { (-)^\ell \over \ell!} \, E_2^\ell \, D_{L-2\ell} \prod_{m=0}^{\ell-1} {{L-2m}\choose{2}} 
\eea
whose successive derivatives never produce the holomorphic Eisenstein series $G_4$. One could now extend this analysis to eliminating the holomorphic Eisenstein series $G_{2k}$ that arise in the holomorphic subgraph reduction procedure for successively higher values of $k$, and produce a formula which would generalize (\ref{Dhat1}) to all orders in $k$. 

\sm

Actually, it is much more convenient, and much more instructive to take successive derivatives directly of primitive modular graph functions, such as $\hat D_L$. The formula corresponding to (\ref{d2DL}) is given by,
\bea
\label{d2DLp}
\nabla^2 \hat D_L = 2L(L-1) \, \hat \cC^+ \left [ \matrix{ 4 & 1_{L-2} \cr  0 & 1_{L-2} \cr }\right ] 
\eea 
and no holomorphic Eisenstein series, and therefore no obstruction to taking higher $\DD$-derivatives, appears.
From the holomorphic subgraph reduction formulas of (\ref{apA3}), it is clear that successive $\DD$-derivatives 
will never produce contributions of the type (\ref{sub1}), and therefore non-linear terms in the relation will not be needed to any order.

\subsubsection{The general case}

By the very construction of the holomorphic subgraph reduction formulas for primitive modular graph functions in appendices \ref{sec:A11} and \ref{sec:A23}, it is clear terms of the form (\ref{sub1}) will never be produced. Therefore, non-linear terms which were required by terms of the type (\ref{sub1}) are not needed. Indeed, in our explicit examples of section \ref{sec:42} up to weight seven included, they never arise.

\subsection{Non-linear identities on primitive modular graph functions ?}

Given that all the identities we have constructed between modular graph functions up to weight seven included are linear, this raises the question as to whether all identities between primitive modular graph functions are necessarily linear. 

\sm

Posed as such, the answer to this question is trivially no. The reason is that from the linear identities between primitive modular graph functions, we may form quadratic and higher power identities by taking polynomial in the linear identities. For example, we clearly have the following weight six and seven quadratic relations, 
\bea
\label{D3D4}
&& \left ( D_3 - E_3-\zeta(3) \right )^2 = 0
\no \\
&& \left ( D_3 - E_3-\zeta(3) \right ) \left ( \hat D_4 - 24 C_{2,1,1} + 18 E_4 \right )=0
\eea
Thus, the non-trivial question is whether all polynomial identities between primitive modular graph functions are generated by the linear identities between them. Since the simplest identity is of weight 3, and the next identity is of weight 4, such as in (\ref{D3D4}), it is easy to imagine that one might have to analyze identities to relatively high weight and possibly high loop order to find the first non-trivial non-linear relation between primitive modular graph functions, if it exists at all. Thus, the question of their existence remains open, but could be investigated concretely with the tools of holomorphic subgraph reduction and the sieve algorithm, as applied to primitive modular graph functions.

\newpage

\section{Inhomogeneous Laplace eigenvalue equations}
\setcounter{equation}{0}
\label{Laplace}

In this section, we shall show that modular graph functions, for arbitrary loop order and arbitrary weight,  obey a hierarchy of inhomogeneous Laplace eigenvalue equations, and we shall provide a precise definition of this hierarchy in the subsections below. We shall obtain the spectrum of the Laplace operator, namely its eigenvalues and their multiplicities, of a hierarchy of infinite subspaces of dihedral modular graph functions and find that the eigenvalues are always of the form $s(s-1)$ for $s$ a positive integer bounded by the weight $w$ of the modular graph function. These results generalize to higher loop order the results that were obtained in \cite{D'Hoker:2015foa}  for the family of two-loop modular graph functions of the form $C_{a,b,c}$ and arbitrary weight $w=a+b+c$, a case that we will briefly review first.

\subsection{Laplacian on two-loop functions $C_{a,b,c}$}

The action of the Laplace-Beltrami operator $\Delta = 4 \tau_2^2 \p_\tau \p_{\bar \tau}$ in the Poincar\'e upper half $\tau$-plane on the modular graph functions $C_{a,b,c}$ was studied in \cite{D'Hoker:2015foa},
and is given by, 
\bea
\label{LapC}
&& \Big (\Delta - a(a-1)-b(b-1)-c(c-1) \Big ) C _{a,b,c} 
\\ && \hskip .2in 
=
 ab \Big ( C _{a-1,b+1,c}  + \half C _{a+1,b+1,c-2}  - 2 C _{a,b+1,c-1}  \Big )
+ \hbox{5 perm. of $a,b,c$}
\no 
\eea
 The Laplacian maps the space of functions $C_{a,b,c}$ with weight $w=a+b+c$ and $a,b,c \geq 1$ to itself (the homogeneous part), plus a linear combination of  the Eisenstein series $E_w$ and products $E_s E_{s'}$ with $w=s+s'$ and integers $s,s' \geq 2$  (the inhomogeneous part). The latter arise when one of the exponents equals 0 or $-1$, in which case we have the identities, 
 \bea
 C_{a,b,0}& = & E_a \, E_b - E_{a+b}
 \no \\
 C_{a,b,-1} & = & E_{a-1} \, E_b + E_a \, E_{b-1}
 \eea
The eigenvalues of the (homogeneous part of the) Laplacian are  $s(s-1)$, with multiplicity  $n_s$, as $s=w-2\mm$ and $\mm$ runs over all integers in the range,
\bea
1 \leq \mm \leq \left [ { w-1\over 2} \right ] \hskip 1in n_s = \left [ { s+2 \over 3} \right ]
\eea
and $[x]$ denotes the integral part of $x$. The structure of the spectrum was understood and proven with the help of generating function techniques and the representation theory of the permutation group $\mS_3$ acting on $C_{a,b,c}$ by permuting the exponents $a,b,c$. The goal of the present section is to show the existence of inhomogeneous Laplace eigenvalue equations for  modular graph functions with more than two loops and arbitrary weight, and obtain the spectrum of the Laplace operator which we shall do for dihedral graphs.

\subsection{Extending the space of modular graph functions}

The simplicity of the action of the Laplacian on the two-loop functions $C_{a,b,c}$, and the fact that the homogeneous part is a linear combination of functions of the same type, is due to the presence of vertices which are at most trivalent. Once we start tackling dihedral graphs with more than two loops, such as $C_{a,b,c,d}$ with $a,b,c,d \geq 1$, for example, vertices of valence larger than three occur and the action of the Laplacian no longer closes on such functions. The phenomenon is readily seen on the graph $D_4=C_{1,1,1,1}$ whose Laplacian takes the form,
\bea
\Delta C_{1,1,1,1} = 12 \, \cC \left [  \matrix{  2\, 0 \, 1 \, 1  \cr 0 \, 2 \, 1 \, 1 \cr } \right]
\eea
The right side is a modular graph function of weight 4, but it is not of the form $C_{a,b,c,d}$ and it cannot be brought to such functions by using the momentum conservation relations of (\ref{3d3}) or (\ref{3d4}). Its presence reveals  two key differences with $C_{a,b,c,d}$ functions,  the first being that upper and lower exponents are not pairwise equal to one another so that $A \not=B$ in the notation of (\ref{3d1}) and (\ref{3b5}), the second being that zeros appear in the exponents of functions that do not manifestly admit a reduction to graphs of lower loop order.

\sm

The first issue is readily remedied. The Laplace operator $\Delta$ on the Poincar\'e upper half $\tau$-plane is made to act on {\sl all} modular graph functions of (\ref{3b5}), namely,
\bea
\label{5a3}
\cC \left [ \matrix{A \cr B \cr } \right ] \hskip 1in \sum _{r=1}^R a_r =\sum _{r=1}^R b_r
\eea
It is clear from (\ref{Lap1}) that $\Delta$ on these modular graph functions has the following properties,
\begin{enumerate}
\itemsep=0in
\item $\Delta$ preserves the weight $w$ of the modular graph function;
\item $\Delta$ commutes with complex conjugation;
\item $\Delta $ maps a modular graph function with $\ell$ loops to a linear combination of modular graph functions with $\ell$ loops  and a polynomial of modular graph functions with strictly lower number of loops. 
Therefore, $\Delta$ acts consistently on the space of all modular graph functions of (\ref{5a3}), in a pattern which is analogous to the action of $\Delta$ on the two-loop modular graph functions $C_{a,b,c}$.
\end{enumerate}
Contrarily to $C_{a_1, \cdots, a_n}$ the modular graph functions of (\ref{5a3}) are not necessarily real since we are not requiring $A=B$. In view of the momentum conservation relations (\ref{3d3}) and (\ref{3d4}), they are also not necessarily linearly independent. 

\sm

The second issue, namely the appearance of zeros amongst the exponents in $A$ and $B$,  is more delicate and will be addressed below.

\subsubsection{Reducible and irreducible modular graph functions}

String theory dictates that the appropriate space of modular graph functions must include all functions with non-negative exponents. Only modular graph functions given by convergent sums need to be considered. The presence of vanishing exponents may or may not render the modular graph function reducible. To spell out the reducibility conditions, we shall distinguish between {\sl algebraic reducibility} and {\sl holomorphic subgraph reducibility}.

\sm

{\sl Algebraic reducibility} occurs when a modular graph function, with an arbitrary number of vertices, arbitrary weight, and arbitrary  loop order, exhibits a pair of upper and lower exponents $(a_r,b_r)$  which both vanish for the same index $r$. Using permutation symmetry of the vertices and edges, we place the vanishing pair in the first column, and readily obtain the following reduction formula,
\bea
\cC \left [ \matrix{ 0 ~ a_2 \cdots a_n \cr 0 ~ b_2 \cdots b_n \cr } \Bigg | \matrix{ A \cr B \cr } \right]
= \cC \left [ \matrix{ A \cr B  \cr }  \right] \prod _{i=2}^n \cC \left [ \matrix{ a_i ~ 0 \cr b_i ~ 0  \cr } \right ]
- \cC \left [ \matrix{  a_2 \cdots a_n \cr  b_2 \cdots b_n \cr } \Bigg | \matrix{ A \cr B \cr } \right]
\eea
for exponents $A,B$ associated with an arbitrary number of vertices, loops, and weights.

\sm

{\sl Holomorphic subgraph reducibility} occurs when the modular graph function exhibits a closed  subgraph 
whose lower exponents all vanish  so that the $\tau$-dependence of the subgraph is purely holomorphic (up to an overall factor of a power of $\tau_2$). When all the upper exponents on a closed  subgraph vanish, the complex conjugate graph is holomorphic subgraph reducible. Henceforth we may limit attention to vanishing lower exponents. 

\sm

An {\sl irreducible modular graph function} has non-negative exponents and can be  reduced neither algebraically nor by holomorphic subgraph reduction -- in the sense defined above.

\sm

In dihedral graphs, the holomorphic subgraph contains two vertices, and depends holomorphically on $\tau$ in a single momentum external to the subgraph.  Concretely, dihedral graphs which admit holomorphic subgraph reduction are of the form,
\bea
\cC \left [ \matrix{ a_1 ~ a_2 ~ a_3 \cdots a_n \cr 0 \, ~ \,  0 \, ~ \, b_3 \cdots b_n \cr }  \right]
\eea
with $a_1+a_2 \geq 3$.
For trihedral graphs, the subgraph may involve 2 vertices (as for the case of dihedral graphs), or three vertices with holomorphic dependence on two momenta external to the subgraph, in which case it is of the form,
\bea
\cC \left [ \matrix{ a_1 ~ a_2  \cdots a_{n_1} \cr   0 \, ~ \, b_2 \cdots b_{n_1} \cr } \Bigg |
\matrix{ a_1' ~ a_2'  \cdots a_{n_1}' \cr   0 \, ~ \, b_2' \cdots b_{n_1}' \cr } \Bigg |
\matrix{ a_1'' ~ a_2''  \cdots a_{n_1}'' \cr   0 \, ~ \, b_2'' \cdots b_{n_1}'' \cr }  \right]
\eea
with $a_1+a_1'+a_1'' \geq 3$.
The tetrahedral case may involve two, three or four vertices in its holomorphic subgraph reduction, and so on. The precise formulas for the reductions may be derived by elementary methods, but their combinatorial complexity rapidly increases with the number of vertices. For the cases of dihedral, trihedral, and tetrahedral graphs needed in this paper, they are given explicitly in Appendix A.

\subsubsection{Using momentum conservation relations}

The extended space of all modular graph functions is subject to the momentum conservation relations (\ref{3d3}) and (\ref{3d4}), which imply linear dependences amongst those functions, and may be used  to simplify the set of its independent generators.  For simplicity of exposition, we shall limit the discussion to dihedral graphs with $\ell$ loops, which are of the form,
\bea
\cC \left [ \matrix{ a_1 ~a_2 ~  \cdots ~ a_{\ell+1} \cr b_1 ~ b_2 ~  \cdots ~ b_{\ell+1} \cr} \right ]
\eea
The modular graph function is irreducible in the sense defined above provided $a_i , b_i \geq 0$ with at most one vanishing upper and one vanishing lower exponent, and no vanishing pair of exponents $(a_r, b_r)$ for the same index $r$. The momentum conservation relations of (\ref{3d4}) imply that  all irreducible modular graph functions of weight $w$ and $\ell$ loops may be expressed as linear combinations of modular graph functions of the form, 
\bea 
\mC _{w,\ell} = \left \{  \cC \left [ \matrix{ a_1 & a_2 ~ \cdots ~ a_\ell & 0 \cr 0 & b_2 ~ \cdots ~ b_\ell & b_{\ell+1} \cr} \right ], ~ \hbox{with} ~ a_i, b_i \geq 1, ~ \hbox{and } w=\sum_{i=1}^\ell a_i=\sum_{i=2} ^{\ell+1} b_i  \right \} 
\eea
plus combinations of modular graph functions with strictly fewer than $\ell$ loops. Convergence of the momentum sums  defining the modular graph functions in $\mC_{w,\ell}$ requires $a_1+b_{\ell+1} \geq 3$, which is part of the definition of $\mC_{w,\ell}$. The Laplacian maps $\mC_{w,\ell}$ to itself plus modular graph functions with fewer loops. The  form of this action is explicitly available from (\ref{Lap1}), but will not be exhibited here as it is rather  lengthy. Therefore, the modular graph functions satisfy the hierarchy of Laplace-eigenvalue equations announced in the Introduction.

\subsection{The spectrum of the Laplace operator restricted to $\mC_{w,\ell}$}

Let the modular graph functions $v_i$ for $1 \leq i \leq \dim \mC_{w,\ell}$ span a basis for $\mC_{w,\ell}$ and define the matrix $M$ by the inhomogeneous Laplace eigenvalue equation,
\bea
\Delta v_i  - \sum _{j=1}^{\dim \mC_{w,\ell}}  M_{ij} v_j \equiv 0
\eea
where $\equiv $ stands for equivalence under the addition of a polynomial in modular graph functions of  total weight $w$ and with a number of loops strictly less than $\ell$. The goal of the present section is to give evidence and partial proof for the following proposition.

\bigskip

{\bf \large Proposition} ~
The eigenvalues of the matrix $M$ are of the form $s(s-1)$ for $s \in \NN$.

\bigskip

Understanding the full structure of the spectrum, including the multiplicities of the eigenvalues, requires a more refined description  of the spaces of modular graph functions than we have given so far. We shall build up these spaces as a function of the weight $w$ and the number of loops $\ell$ of the graphs in a manner graded by their {\sl level}, a notion which usefully generalizes loop order, and which will be defined below.

\subsubsection{Level 0}

The simplest non-trivial linear subspace of the space of modular graph functions $\mC_{w,\ell}$ of weight $w$ and loop order $\ell$ is the one-dimensional space $\mC_{k,\ell}^{(0)}$ with $w=k+\ell$, $k \geq 2$ and $\ell \geq 1$, consisting of the following modular graph function, 
\bea
\mC_{k,\ell} ^{(0)} = \{ v_{k,\ell}   \} \hskip 1in
v_{k,\ell}  = \cC \left [ \matrix{ k &  1_{\ell-1} & 0 \cr 0 & 1_{\ell -1} & k \cr} \right ] 
\eea
where $1_n$ is the $n$-dimensional array whose entries are all one. We require $k\geq 2$ for convergence of the momentum sums. The function $v_{k,\ell}$ is real. For $\ell=1$, the middle column is absent, and the modular graph function reduces to the Eisenstein series $(-)^k E_k$.  For arbitrary $\ell \geq 1$, and arbitrary $k \geq 2$,  it is easy to see that $v_{k,\ell}$ satisfies,
\bea
\Big ( \Delta -k(k-1) \Big ) v_{k,\ell}  \equiv 0
\eea
where $\equiv$  again stands for equality modulo the addition of contributions with loop number strictly less than $\ell$. The matrix $M$ is one-dimensional and its eigenvalue $k(k-1)$ is of the form given by the Proposition, with multiplicity one.

\subsubsection{Level 1}

Level 1 consists of the following vector space of functions of weight $w$ and loop order $\ell \geq 2$,
\bea
\mC_{k,\ell} ^{(1)} = \left \{ v_{k,\ell} \left [ {a \atop b} \right ] ~ \hbox{with} ~ a,b \geq 0, ~ a+b >0, ~ k-1 > 2a, 2b \right \} 
\eea
where the functions are defined by,
\bea
v_{k,\ell} \left [ {a \atop b} \right ]   =  \cC \left [ \matrix{ k-a  &  a + 1 & 1_{\ell-2} & 0 \cr 0 & b+1 & 1_{\ell -2} & k-b \cr} \right ] 
\eea
Complex conjugation amounts to interchanging $a$ and $b$, so that the functions with $a=b$ are real, but those with $a \not= b$ are generally complex.  It is straightforward to establish that the Laplacian acts as follows,
\bea
\Delta : \mC_{k,\ell} ^{(1)} & \to & \mC_{k,\ell} ^{(0)} \oplus \mC_{k,\ell} ^{(1)} 
\eea
up to the addition of a polynomial of weight $w$ in modular graph functions with strictly fewer loops than $\ell$.  Actually, the space $\mC_{k,\ell}^{(1)}$ admits a natural grading according to the value of $c=a+b$,  and we shall introduce the spaces defined as follows,
\bea
\mC_{k, \ell} ^{(1;c)} & = & \Big \{ v_{k,\ell} \left [ {a \atop b} \right ] \in \mC_{k,\ell}^{(1)} ~~ \hbox{with}  ~ a+b =c \Big \} 
\eea
which implies the restriction $1 \leq c \leq k-2$. It is now easy to see that the dimension is given by the number of partitions of the integer $c$ into the sum of two non-negative integers, $\dim \mC_{k,\ell}^{(1;c)}  = c+1$. The total space is obtained as follows,
\bea
\mC_{k,\ell} ^{(1)} & = & \bigoplus _{c=1}^{k-2} \, \mC_{k, \ell} ^{(1;c)}
\eea
and the Laplacian acts as follows,
\bea
\Delta : \mC^{(1;c)} _{k,\ell} & \to &  \mC_{k, \ell} ^{(0)} \oplus \bigoplus _{c'=1}^c \mC^{(1;c')} _{k,\ell} 
\eea
Spelling out the above action of $\Delta$, the matrix $M$ is seen to be block upper trigonal, 
\bea
\Delta \left ( \matrix{ \mC^{(1;c)} _{k,\ell}  \cr \mC^{(1;c-1)} _{k,\ell}  \cr \cdots \cr \mC^{(1;2)} _{k,\ell} \cr \mC^{(1;1)} _{k,\ell} \cr } \right )
- \left ( \matrix{ M_{c,c}  & M_{c,c-1} & \cdots & M_{c,2} & M_{c,1} \cr
0 & M_{c-1,c-1} & \cdots & M_{c-1,2} & M_{c-1,1} \cr
\cdots & \cdots & \cdots & \cdots & \cdots \cr 
0 & 0 & \cdots & M_{2,2} & M_{2,1} \cr
0 & 0 & \cdots & 0 & M_{1,1} \cr} \right ) 
\left ( \matrix{ \mC^{(1;c)} _{k,\ell} \cr \mC^{(1;c-1)} _{k,\ell}  \cr \cdots \cr \mC^{(1;2)} _{k,\ell}  \cr \mC^{(1;1)} _{k,\ell} \cr } \right ) \equiv 0
\eea
Therefore, the spectrum of $\Delta$ restricted to the space $\mC_{k,\ell}^{(1)}$ is provided by the eigenvalues of the $(c+1) \times (c+1)$ block-diagonal matrices $M_{c, c}$ for $1 \leq c \leq k-2$ and  is given as follows,
\bea
\label{spec1}
(k- \nu ) (k-\nu -1) \hskip 1in \nu = 0,1,2, \cdots, c
\eea
each eigenvalue occurring with multiplicity one, which accounts for $\dim \mC_{k,\ell}^{(1;c)}  = c+1$.

\subsubsection{Proving the level 1 spectrum}

The Laplace operator on $\mC_{k,\ell}^{(1)}$, up to the addition of modular graph functions of loop order strictly less than $\ell$, is given by,
\bea
\Delta v _{k,\ell} \left [ {a \atop b} \right ]
& \equiv & \Big ( (k-a)(k-b) +(a+1)(b+1)-k-1 \Big ) v_{k,\ell} \left [ {a \atop b} \right ]
\no \\ &&
-(k-a)(b+1) v _{k,\ell} \left [ {a-1 \atop b+1} \right ] -(k-b)(a+1) v _{k,\ell} \left [ {a+1 \atop b-1} \right ]
\no \\ &&
+(k-a)(k-2b-1) v _{k,\ell} \left [ {a-1 \atop b} \right ] + (k-b)(k-2a-1) v _{k,\ell} \left [ {a \atop b-1} \right ]
\no \\ &&
+ (k-a)(k-b) v _{k,\ell} \left [ {a-1 \atop b-1} \right ]
\eea
While the sum of the indices on the left side, and on the first two lines on the right side are all given by $a+b=c$, it is clear that the terms of the third and fourth lines of the right side have a lesser value. Thus, in computing the eigenvalues within the block $\mC_{k,\ell}^{(1;c)}$, the third and fourth lines may be dropped,
as they correspond to off-diagonal upper triangular block entries of the matrix $M$. Therefore, the eigenvalue problem is now reduced to the following system for fixed $k, \ell, c=a+b$, with $a,b \geq 0$ and $1 \leq c \leq k-2$,
\bea
\Delta v _{k,\ell} \left [ {a \atop b} \right ]
& = & \Big ( (k-a)(k-b) +(a+1)(b+1)-k-1 \Big ) v _{k,\ell} \left [ {a \atop b} \right ] 
\no \\ &&
-(k-a)(b+1) v _{k,\ell} \left [ {a-1 \atop b+1} \right ] -(k-b)(a+1) v _{k,\ell} \left [ {a+1 \atop b-1} \right ]
\qquad
\eea
Functions with an upper or a lower exponent equal to $-1$ will arise but they correspond to lower loop level and may effectively be set to zero. The matrix may be diagonalized analytically to give the spectrum of (\ref{spec1}), and we have  checked the result numerically using Maple  for $c$ up to 30.

\subsubsection{Level 2}

Level 2 consists of  the following vector space of modular graph functions of weight $w=k+\ell$ and loop order $\ell \geq 3$, 
\bea
\mC_{k,\ell} ^{(2)} = \left \{ v_{k,\ell} \left [ {a_1 ~ a_2 \atop b_1 \, ~ b_2} \right ] ~ \hbox{with} ~ a_i,b_i \geq 0, ~ a_i+b_i >0,   ~ k-1 > 2a, 2b \right \} 
\eea
where $a=a_1+a_2$, $b=b_1+b_2$,  $i=1,2$,  and  the modular functions are given by,
\bea
v_{k,\ell} \left [ {a_1 ~ a_2 \atop b_1 ~\,  b_2} \right ]  & = & \cC \left [ \matrix{ k-a  &  a_1 + 1 & a_2+1 & 1_{\ell-3} & 0 \cr 0 & b_1+1 & b_2+1 &  1_{\ell -3} & k-b \cr} \right ] 
\eea
The matrix $M$ is independent of $\ell$, and the Laplacian acts as follows,
\bea
\Delta : \mC_{k,\ell} ^{(2)} & \to & \mC_{k,\ell} ^{(0)} \oplus \mC_{k,\ell} ^{(1)} \oplus \mC_{k,\ell}^{(2)}
\eea
up to the addition of modular functions with strictly fewer loops than $\ell$.  The space $\mC_{k,\ell}^{(2)}$ admits a natural grading according to the values of $c_1=a_1+b_1$ and $c_2=a_2+b_2$,  and we shall introduce the spaces  defined as follows,
\bea
\mC_{k, \ell} ^{(2; c_1,c_2)} & = & \Big \{ v_{k,\ell} \left [ {a_1 ~ a_2 \atop b_1 ~\,  b_2 } \right ] \in \mC_{k,\ell}^{(2)} ~~ \hbox{with}  ~ c_1=a_1+b_1, ~ c_2 = a_2+b_2 \Big \} 
\eea
which implies the restriction $1 \leq c_i$ and $c=c_1+c_2 \leq k-2$. The Laplace operator preserves this grading, up to contributions of lower loop order. The dimensions of these spaces is easily computed, and we find, 
\bea
c_2 < c_1 & \hskip 1in & \dim \mC_{k,\ell} ^{(2;c_1,c_2)} = (c_1+1)(c_2+1)
\no \\
c_2 = c_1 & \hskip 1in & \dim \mC_{k,\ell} ^{(2;c_1,c_1)} = \half (c_1+1) (c_1+2)
\eea

\begin{table}[htb]
\label{table1}
\begin{center}
\begin{tabular}[h]{||c||c|c|c|c|c|c|c|c|c|c|c|c|c|c||}
\hline \hline
eigenvalue $\backslash ~ c$      &  $2$  & $3$      & $4$     & $5$     & $6$      & $7$     & $8$     & $9$    & $10$     & $11$     & $12$  &$13$ &$14$ &$15$ \cr  \hline  \hline           
$ k(k-1) $          & 1           & 1          &   2       &    2      &  3         &   3       &    4      &  4       &   5         &    5        &  6     &     6       &7	  &7	\cr \hline
$ (k-1)(k-2) $    & 1           & 2          &   3       &    4      &  5         &   6       &    7      &  8       &   9         &   10      &  11     &   12     &13  	&14	\cr \hline
$ (k-2)(k-3) $    & 1           & 2          &   4       &    5      &   7        &   8       &  10      &  11     &   13       &   14      &  16    &   17     &19	&20	\cr \hline
$ (k-3)(k-4) $    &              & 1          &   3       &    5      &   7        &   9       &  11      &  13     &   15       &   17      &  19    &    21    & 23	& 25 	\cr \hline
$ (k-4) (k-5) $   &              &             &   2       &    4      &  7         &   9       &  12      &  14     &   17       &   19      &  22     &   24     &27	&29\cr \hline
$ (k-5) (k-6) $   &              &             &            &    2      & 5          &   8       &  11      &  14     &   17       &   20      &  23     &    26     &29	&32\cr \hline
$ (k-6) (k-7) $   &              &             &            &            & 3          &   6       &  10      &  13     &   17       &   20      &  24     &     27    &31	&34\cr \hline
$ (k-7) (k-8) $   &              &             &            &            &             &   3       &  7        &  11     &   15       &   19      &  23     &      27 	&31	&35	\cr \hline
$ (k-8) (k-9) $   &              &             &            &            &             &            &  4        &  8       &   13       &   17      &  22      &     26 	&31	&35	\cr \hline
$ (k-9) (k-10) $   &            &             &            &            &             &            &            &  4       &   9         &   14      &  19      &    24 	&29	&34	\cr \hline
$ (k-10) (k-11) $   &          &             &            &            &             &            &            &           &   5         &   10      &  16      &    21 	&27	&32	\cr \hline
$ (k-11) (k-12) $   &          &             &            &            &             &            &            &           &              &    5       &  11       &   17 	&23	&29	\cr \hline
$ (k-12) (k-13) $   &          &             &            &            &             &            &            &           &              &             &  6         &  12	&19	&25	\cr \hline
$ (k-13) (k-14) $   &          &             &            &            &             &            &            &           &              &             &           &  6 		&13	&20	\cr \hline
$ (k-14) (k-15) $   &          &             &            &            &             &            &            &           &              &             &           &    	 	&7  	&14	\cr \hline
$ (k-15) (k-16) $   &          &             &            &            &             &            &            &           &              &             &            &     		&	&7 \cr \hline \hline

\end{tabular}
\caption{Eigenvalues and multiplicities at level 2  for $c\leq 15$.}
\end{center}
\end{table}

\begin{table}[htb]
\label{table2}
\begin{center}
\begin{tabular}[h]{||c||c|c|c|c|c|c||c||}
\hline
eigenvalue    $\backslash ~ c_1$    &  $12$  & $11$      & $10$     & $9$     & $8$      & $7$     &Total  \cr  \hline  \hline           
$ k(k-1) $          & 1           & 1          &   1       &    1      &  1         &   1      &     6        \cr \hline
$ (k-1)(k-2) $    & 2           & 2          &   2       &    2      &  2         &   2        &     12    \cr \hline
$ (k-2)(k-3) $    & 2           & 3          &   3       &    3      &   3       &   3          &     17\cr \hline
$ (k-3)(k-4) $    & 2           & 3          &   4       &    4      &   4        &   4          &      21\cr \hline
$ (k-4) (k-5) $   & 2           &3           &   4       &    5      &  5         &   5         &         24 \cr \hline
$ (k-5) (k-6) $   &  2          & 3          &   4       &    5      & 6          &   6          &         26\cr \hline
$ (k-6) (k-7) $   &  2          & 3          &   4       &   5       & 6          &   7           &        27\cr \hline
$ (k-7) (k-8) $   &  2          & 3          &   4       &   5       & 6          &   7            &     27\cr \hline
$ (k-8) (k-9) $   &  2          & 3          &   4       &    5      & 6          &   6             &        26   \cr \hline
$ (k-9) (k-10) $ & 2           &3           &   4       &    5      &  5         &   5             &       24\cr \hline
$ (k-10) (k-11) $   & 2           & 3          &   4       &    4      &   4        &   4         &      21\cr \hline
$ (k-11) (k-12) $  & 2           & 3          &   3       &    3      &   3       &   3            &     17\cr \hline
$ (k-12) (k-13) $    & 2           & 2          &   2       &    2      &  2         &   2          &       12 \cr \hline
$ (k-13) (k-14) $    & 1           & 1          &   1       &    1      &  1         &   1          &      6 \cr \hline
\end{tabular}
\caption{Eigenvalues and multiplicities at level 2  for $c = 13$, and fixed $c_1$.}
\end{center}
\end{table}

\sm

Maple allows us to compute the spectrum for each value of $c=c_1+c_2$. The results for $c\leq 15$ are listed in Table 1. The combinatorial pattern for the spectrum of the Laplace operator restricted to the level 2 spaces $\mC_{k,\ell} ^{(2,c_1,c_2)}$ may be read off from Tables 2 and 3, and summarized as follows. The values of $c_1, c_2$ are restricted to $1 \leq c_2 \leq  c_1$ and, setting  $c=c_1+c_2$, we must have $c \leq k-2$. The support for the eigenvalues is as follows,
\bea
(k-\nu)(k-\nu-1) \hskip 1in  0 \leq \nu \leq c
\eea
The eigenvalues and their range coincide with those for level 1 with the same value of $c$ but, of course, their multiplicities are different.  When restricted to a subspace $\mC_{k,\ell} ^{(2;c_1,c_2)}$, the multiplicity $n_\nu$ of the eigenvalue $(k-\nu)(k-\nu-1)$ is as follows,
\bea
c_2 < c_1 & \hskip 1in & 
n_\nu = \left \{ \matrix{ 
\nu+1 & \hbox{for} & 0 \leq \nu \leq c_2 \cr
c_2+1 & \hbox{for} & c_2 \leq \nu \leq c_1 \cr
c-\nu+1 & \hbox{for} & c_1 \leq \nu \leq c \cr} \right .
\eea
When $c$ is odd the above spectrum covers all possible values of $c_2\leq c_1$. When $c$ is even, we also have the case $c_2=c_1=c/2$, for which the multiplicities are given by,
\bea
c_2 = c_1 & \hskip 1in & 
n_\nu = \left \{ \matrix{ 
\left [ { \nu \over 2} \right ] +1 & \hbox{for} & 0 \leq \nu \leq c_1 \cr
&& \cr
 \left [ { c - \nu \over 2} \right ] +1 & \hbox{for} & c_1 \leq \nu \leq c \cr} \right .
\eea

\begin{table}[htb]
\label{table3}
\begin{center}
\begin{tabular}[h]{||c||c|c|c|c|c|c|c||c||}
\hline
eigenvalue   $\backslash ~ c_1$     &$13$ &  $12$  & $11$      & $10$     & $9$     & $8$      & $7$     &Total  \cr  \hline  \hline           
$ k(k-1) $         & 1& 1           & 1          &   1       &    1      &  1         &   1      &     7        \cr \hline
$ (k-1)(k-2) $    &2& 2           & 2          &   2       &    2      &  2         &   1        &     13    \cr \hline
$ (k-2)(k-3) $    &2& 3           & 3          &   3       &    3      &   3       &   2         &     19\cr \hline
$ (k-3)(k-4) $    &2& 3           & 4          &   4       &    4      &   4        &   2          &      23\cr \hline
$ (k-4) (k-5) $   &2& 3           &4          &   5       &    5      &  5         &   3        &         27 \cr \hline
$ (k-5) (k-6) $   &2&  3          & 4          &   5       &    6      & 6          &   3          &         29\cr \hline
$ (k-6) (k-7) $   &2&  3          & 4          &   5       &   6      & 7          &   4          &        31\cr \hline
$ (k-7) (k-8) $   &2&  3          & 4          &   5       &   6       & 7          &   4            &     31\cr \hline
$ (k-8) (k-9) $   &2&  3          & 4          &   5       &    6      & 7          &   4             &        31   \cr \hline
$ (k-9) (k-10) $ &2& 3          &4         &   5          &    6      &  6         &   3             &       29\cr \hline
$ (k-10) (k-11) $  &2 & 3           & 4       &   5      &    5      &   5        &   3         &      27\cr \hline
$ (k-11) (k-12) $  &2& 3          & 4          &   4       &    4      &   4       &   2            &     23\cr \hline
$ (k-12) (k-13) $   &2 & 3           &3          &   3       &    3      &  3         &   2          &       19 \cr \hline
$ (k-13) (k-14) $    &2& 2           & 2          &   2       &    2      &  2         &   1          &      13 \cr \hline
$ (k-14) (k-15) $    &1 & 1           & 1          &   1       &    1      &  1         &   1          &      7 \cr \hline
\end{tabular}
\caption{Eigenvalues and multiplicities at level 2  for $c = 14$, and fixed $c_1$.}
\end{center}
\end{table}

\subsubsection{Level 3}

Level 3 consists of  the following family of modular graph functions of weight $w=k+\ell$ and loop order $\ell \geq 4$, 
\bea
\mC_{k,\ell} ^{(3)}  =  \left \{ v_{k,\ell} \left [ {a_1 ~ a_2 ~ a_3 \atop b_1 ~\,  b_2 ~\,  b_3 } \right ] ~ \hbox{with} ~ a_i,b_i \geq 0, ~ a_i+b_i >0, ~ k-1 > 2a,2b  \right \} 
\eea
where $i$ takes the values $i=1,2,3$, and we define $m,n$ by,
\bea
\label{mn}
a & = & a_1 + a_2 + a_3
\no \\
b & = & b_1 + b_2 + b_3
\eea
The modular functions may be cast in terms of $\cC$-functions as follows,
\bea
v_{k,\ell} \left [ {a_1 ~ a_2 ~ a_3 \atop b_1 ~\,  b_2 ~\, b_3} \right ]  
& = & 
\cC \left [ \matrix{ k-a  &  a_1 + 1 & a_2+1 & a_3 + 1 & 1_{\ell-4} & 0 \cr 0 & b_1+1 & b_2+1 & b_3+1 &  1_{\ell -4} & k-b \cr} \right ] 
\eea
The matrix $M$ is independent of $\ell$, and  the Laplacian acts as follows,
\bea
\Delta : \mC_{k,\ell} ^{(3)} & \to & \mC_{k,\ell} ^{(0)} \oplus \mC_{k,\ell} ^{(1)} \oplus \mC_{k,\ell}^{(2)} \oplus \mC_{k,\ell}^{(3)}
\eea
up to the addition of modular functions with strictly fewer loops than $\ell$.  Actually, the space $\mC_{k,\ell}^{(3)}$ admits a natural grading according to the values of $c_i=a_i+b_i$ for $i=1,2,3$,  and we shall introduce the spaces,
\bea
\mC_{k, \ell} ^{(3; c_1, c_2, c_3)} & = & \left \{ v_{k,\ell} \left [ {a_1 ~ a_2 ~ a_3 \atop b_1 ~\,  b_2 ~ \, b_3} \right ] \in \mC_{k,\ell}^{(3)} ~~ \hbox{with}  ~ a_i+b_i =c_i\right  \} 
\eea
where we continue to use the abbreviations $m,n$ of (\ref{mn}). On the sum $c=c_1+c_2+c_3$, the  conditions imply the restriction $c \leq k-2$. For example, the space $\cV^{(3; 1,1,1)}_{k,\ell}$ consists of the following four independent functions, 
\bea
v_{k,\ell} \left [ {1 ~ 1 ~ 1 \atop 0 ~ 0 ~ 0 } \right ] 
\hskip 0.5in 
v_{k,\ell} \left [ {1 ~ 1 ~ 0\atop 0 ~ 0 ~ 1 } \right ] 
\hskip 0.5in 
v_{k,\ell} \left [ {1 ~ 0 ~ 0\atop 0 ~ 1 ~ 1 } \right ] 
\hskip 0.5in 
v_{k,\ell} \left [ {0 ~ 0 ~ 0 \atop 1 ~ 1 ~ 1} \right ] 
\eea
Maple allows us to compute the spectrum for each value of $c_1, c_2, c_3$ and thus $c$. The results for $c\leq 12$ are listed in Table 4 below. Manifestly, the eigenvalues are all of the form $s(s-1)$ for $s \in \NN$, consistent with the Proposition.

\begin{table}[htb]
\label{level3table}
\begin{center}
\begin{tabular}[h]{||c||c|c|c|c|c|c|c|c|c|c|c||}
\hline \hline
eigenvalue  $\backslash ~ c$      &  $2$  & $3$      & $4$     & $5$     & $6$      & $7$     & $8$     & $9$   & $10$ & $11$  & $12$ \cr  \hline  \hline           
$ k(k-1) $          &              & 1          &   1       &    2      &  3         &   4       &   5       &  7      &    8            &10  & 12\cr \hline
$ (k-1)(k-2) $    &              & 1          &   2       &    4      &  6         &   9       &   12     &  16    &     20       &25   &30\cr \hline
$ (k-2)(k-3) $    &              & 1          &   3       &    6      & 10        &   15     &  21      &  28    &    36       &45  & 55\cr \hline
$ (k-3)(k-4) $    &              & 1          &   2       &    6      &  11       &   18     &  26      &  37    &     48      &62   &78\cr \hline
$ (k-4) (k-5) $   &              &             &   1       &    4      &  10       &   18     &  29      &  42    &     58      & 76  &97\cr \hline
$ (k-5) (k-6) $   &              &             &            &    2      & 6          &   15     &  26      &  42    &    60        & 83  &108\cr \hline
$ (k-6) (k-7) $   &              &             &            &            & 3          &   9       &  21      &  37    &     58       & 83  &112\cr \hline
$ (k-7) (k-8) $   &              &             &            &            &             &   4       &  12      &  28    &     48      &  76   &108\cr \hline
$ (k-8) (k-9) $   &              &             &            &            &             &            &  5        &  16    &     36       & 62   &97\cr \hline
$ (k-9) (k-10) $   &            &             &            &            &             &            &            &  7      &     20       &  45  &78\cr \hline
$ (k-10) (k-11) $   &          &             &            &            &             &            &            &          &     8          & 25  &55\cr \hline
$ (k-11) (k-12) $   &          &             &            &            &             &            &            &          &               & 10    &30\cr \hline
$ (k-12) (k-13) $   &          &             &            &            &             &            &            &          &               &         &12\cr \hline \hline
\end{tabular}
\caption{Eigenvalues and multiplicities at level 3  for $c\leq 12$.}
\end{center}
\end{table}

\subsubsection{Arbitrary Level}

Let us now fix an arbitrary level $\lambda \geq 0$. The vector space of modular graph functions of level $\lambda$ consists of   weight $w=k+\ell$ and loop order $\ell \geq \lambda +1$, 
\bea
\mC_{k,\ell} ^{(\lambda)}  =  \left \{ v_{k,\ell} \left [ {a_1 ~ a_2 ~ \cdots ~ a_\lambda \atop b_1 ~\,  b_2 ~\,  \cdots ~ b_\lambda } \right ] ~ \hbox{with} ~ a_i,b_i \geq 0, ~ a_i+b_i >0, ~ k-1 > 2a,2b  \right \} 
\eea
where $i$ takes the values $i=1,\cdots, \lambda$, and we define $m,n$ by,
\bea
\label{mng}
a = \sum _{i=1}^ \lambda  a_i  \hskip 1in b = \sum _{i=1}^\lambda b_i
\eea
The modular functions may be cast in terms of $\cC$-functions as follows,
\bea
v_{k,\ell} \left [ {a_1 ~  \cdots ~ a_\lambda \atop b_1 ~\,   \cdots ~ b_\lambda} \right ]   =  
\cC \left [ \matrix{ k-a  &  a_1 + 1 & \cdots & a_\lambda + 1 & 1_{\ell-\lambda-1} & 0 \cr 0 & b_1+1 & \cdots & b_\lambda +1 &  1_{\ell -\lambda-1} & k-b \cr} \right ] 
\eea
The matrix $M$ is independent of $\ell$, and the Laplacian acts  as follows,
\bea
\Delta : \mC_{k,\ell} ^{(\lambda )} & \to & \bigoplus _{\sigma =0}^\lambda \mC_{k,\ell} ^{(\sigma )} 
\eea
up to the addition of modular functions with strictly fewer loops than $\ell$.  The space $\mC_{k,\ell}^{(\lambda)}$ again admits a natural grading according to the value of $c_i=a_i+b_i$,  and we may introduce the spaces $\mC_{k,\ell} ^{(\lambda; c_1, \cdots, c_\lambda)}$ defined as follows,
\bea
\mC_{k, \ell} ^{(\lambda ; c_1, \cdots,  c_\lambda)} & = & \left \{ v_{k,\ell} \left [ {a_1 ~ \cdots ~ a_\lambda \atop b_1 ~\,  \cdots ~ \, b_\lambda} \right ] \in \mC_{k,\ell}^{(\lambda)} ~~ \hbox{with}  ~ a_i+b_i =c_i \right  \} 
\eea
The pattern of multiplicities in the general case remains to be explored.

\newpage

\section{Discussion and Open Problems}
\setcounter{equation}{0}
\label{sec:6}

In this paper, we have investigated  the structure of the space of modular graph functions and the action of differential operators thereupon. Many of our results were proven here, but for others full proofs remain outstanding open problems. In this final section, we shall provide a brief list of these open questions.

\sm 

\begin{enumerate}
\item 
We have used the holomorphic subgraph reduction and sieve algorithm developed in \cite{D'Hoker:2016a}  to construct and prove all  algebraic  identities between modular graph functions at weight six, and all dihedral and one trihedral modular graph functions at weight 7. We have also related the structure of these identities to the structure of their Laurent polynomial at the cusp. In each case where the Laurent polynomial is available, the corresponding modular graph identity can be predicted uniquely by the cancellation of its complete Laurent polynomial at the cusp. But whether this occurrence is an accident at low weights or true in a more general sense remains open.
\item
In terms of {\sl primitive modular graph functions}, defined by momentum summations in which all non-trivial subsets of momenta entering any vertex sum to a non-zero value, the modular graph identities become linear for all the cases we have examined.  Whether this occurrence holds to higher weight and loop orders is a challenging open problem, and if it holds, a complete proof should be provided.
\item
The extended space of modular graph functions, in which upper and lower exponents are allowed to be pairwise distinct, admits a closed action of the Laplace operator and obeys a hierarchy of inhomogeneous Laplace eigenvalue equations. For the extended family of dihedral modular graph functions the eigenvalues are all of the form $s(s-1)$ for $s \in \NN$, and with interesting patterns of multiplicities governed by the representation theory of the permutation group acting on exponents. However, we have examined only a subset of all dihedral modular functions which, although infinite in dimension, does not span all dihedral modular graph functions. Therefore, a full proof of the validity of the Proposition in section 5.3 remains open, as do general explicit formulas for the multiplicities at each weight and loop orders. For the two-loop functions $C_{a,b,c}$ the proof of the Proposition, and an explicit formula for the multiplicities, was derived by using techniques of generating functions and the representations of the permutation group on the exponents. If the methods of generating functions can be used here, they will require a good deal of innovative adaptation and generalization from the two-loop case, and this provides another open problem.
\item
Without doubt, there must be an intimate and intricate relation between the algebraic  identities between modular graph functions  investigated here and the algebraic identities between multi-zeta-values derived and discussed in \cite{ZagierMZV,Hoffmann,W,BBBL,Zudilin,Blumlein:2009cf,Brown:2013gia}. Making this relation explicit should provide interesting insights into the number theoretic significance of the algebraic  identities between modular graph functions.
\end{enumerate}

\bigskip\bigskip

\noindent{\Large \bf Acknowledgments}

\bigskip

First and foremost, ED would like to thank Michael Green and Pierre Vanhove for the fruitful collaboration on earlier papers out of which the present work has grown. We also acknowledge stimulating discussions with William Duke on related topics. The research of ED is supported in part by  the National Science Foundation under grants PHY-13-13986 and PHY-16-19926. The research of JK is supported in part by the Mani L. Bhaumik Institute for Theoretical Physics, and by a fellowship from the Julian Schwinger Foundation.

\appendix

\section{Holomorphic subgraph reductions}
\setcounter{equation}{0}
\label{holo}

In this appendix, we provide a summary of the holomorphic subgraph reduction formulas needed in this
paper for dihedral, trihedral, and tetrahedral modular graph functions. We shall also spell out the modifications of these formulas which are required for primitive modular graph functions and forms.

\subsection{Dihedral modular graph functions}
\label{sec:A1}

The dihedral case was discussed in detail in \cite{D'Hoker:2016a}, and we shall provide here only a brief summary of relevant results. Holomorphic subgraph reduction can be applied to dihedral modular graphs functions whenever they are of the form,
\bea
\label{holform}
\cC^+ \left[ \matrix {a_+ & a_- & A \cr 0 & 0 & B \cr} \right] = \sum_{p_1,\ldots,p_r \in \Lambda}' \sum_{p_0 \in \Lambda} \delta_{p_0 + p, 0} (\tau_2)^{a_0} \cG_{a_+,a_-}(p_0) \prod_{\rho = 1}^r   { (\tau _2)^{a_\rho}  \pi^{-\half a_\rho - \half b_\rho} \over  (p_\rho) ^{a_\rho} ~ (\bar p _\rho) ^{b_\rho} }
\eea
with the usual exponent row vectors $A = [a_1 \, a_2 \,\ldots\, a_r]$ and $B=[b_1 \, b_2 \,\ldots\, b_r]$. We assume that $a_+, a_- \geq 1$, and $a_++a_- \geq 3$ to ensure the convergence of all subgraphs. 
We have introduced the notation $a_0 = a_+ + a_-$ for the weight of the holomorphic subgraph, $p_0 = p_+ + p_-$ for the momentum through the subgraph, and $p=p_1 + \ldots + p_r$. The function $\cG_{a_+,a_-}(p_0)$ represents the sum over holomorphic momenta $p_+$ and $p_-$, and is given by, 
\bea
\cG_{a_+,a_-}(p_0) ={1\over \pi^{ \half a_0}} \sum_{p_+, p_- \in \Lambda}' \delta_{p_+ + p_-, p_0} {1\over (p_+)^{a_+} (p_-)^{a_-}}
\eea
This sum can in fact be calculated explicitly. For $a_0$ odd,  we have  $\cG_{a_+,a_-}(0)=0$ by parity. For $a_0$ even, we have $\cG_{a_+,a_-}(0) = (-)^{a_+} G_{a_0}$ with the holomorphic Eisenstein function $G_{a_0}$ defined in (\ref{Gk}).  In the case of general $p_0 \neq 0$, it was found in \cite{D'Hoker:2016a} that,  
\bea
\label{Gaa}
\pi^{\half a_0}\cG_{a_+,a_-}(p_0) &=& \sum_{k=3}^{a_+} { a_0-1-k\choose a_+ - k} {\pi^{k/2} G_k \over (p_0)^{a_0 - k}} + \sum_{k=3}^{a_-} { a_0-1-k\choose a_- - k} {\pi^{k/2} G_k \over (p_0)^{a_0 - k}}
\no\\&&
-{1 \over (p_0)^{a_0}} {a_0 \choose a_+} + {a_0 -2 \choose a_+ - 1} \left( {\pi G_2 \over (p_0)^{a_0 - 2}} + { \pi \bar p \over \tau_2 (p_0)^{a_0 -1}}\right)
\eea
Assembling the two contributions, the formula governing the elimination of an arbitrary weight holomorphic subgraph is, 
\bea
\label{apA1}
\cC^+ \left[ \matrix {a_+ & a_- & A \cr 0 & 0 & B \cr} \right] &=& 
\cC^+ \left[ \matrix {a_+ & a_-  \cr 0 & 0 \cr} \right] 
\cC^+ \left[{A \atop B } \right]  - {a_0 \choose a_+}\cC^+ \left[ \matrix {a_0 & A \cr 0 & B \cr} \right]
\\&&
+\sum_{k=2}^{[a_+/2]} {a_0 - 1 - 2k\choose a_+ - 2k} \, \cG_{2k} \, 
\cC^+ \left[ \matrix {a_0 -2k & A \cr 0 & B \cr} \right] + (a_+ \leftrightarrow a_-) 
\no\\&&
+{a_0 -2\choose a_+ - 1} \left\{ \tau_2^2 G_2 \,  \cC^+ \left[ \matrix {a_0 -2 & A \cr 0 & B \cr} \right] 
+ \pi \tau_2 \, \cC^+  \left[ \matrix {a_0 -1 & A \cr -1 & B \cr} \right] \right\}
\no
\eea
The first term on the right-hand side arises from the contribution of $p_0 =0$, while the remaining terms come from $p_0 \neq 0$. In practice, the terms of the form (\ref{holform}) that arise during the calculation of modular graph form identities always appear in certain linear combinations for which the contributions from the last line of (\ref{apA1}) always cancels. A basis for these combinations may be chosen as follows,
\bea
\label{apA2}
&&
 \cC^+ \left [ \matrix{a_+ & a_- & A \cr 0 & 0 & B  \cr} \right ] 
 - \left ( \matrix{ a_0-2  \cr a_+ -1} \right ) 
  \cC^+ \left [ \matrix{a_0-1 & 1 & A \cr 0 & 0 & B  \cr} \right ] 
\no \\ && \hskip 0.1in =
\left ( \left ( \matrix{ a_0-2 \cr a_+ -1} \right ) + (-)^{a_-} \right )
 \cC^+ \left [ \matrix{a_0  ~ 0 \cr 0  ~~ 0  \cr} \right ] \, \cC^+ \left [ \matrix{A  \cr B  \cr} \right ] 
 + \left ( \matrix{ a_0 \cr a_+ } \right ) { (a_+-1) (a_--1) \over a_0-1} \, 
 \cC^+ \left [ \matrix{a_0 ~ A  \cr 0 ~~ B  \cr} \right ] 
 \no \\ && \hskip 0.4in +
 \sum _{k=2}^{ [(a_0-1) /2]} \Lambda _k (a_+, a_-) \, \cG_{2k} \,
 \cC^+ \left [ \matrix{a_0 - 2 k & A  \cr 0 & B  \cr} \right ] 
\eea
where the function $\Lambda _k (a_+, a_-)$ is given by,
\bea
\label{Lambda}
\Lambda _k (a_+, a_-) 
= \left ( \matrix{ a_0 -1 - 2 k \cr a_+ - 2k \cr } \right ) + \left ( \matrix{ a_0 -1 - 2 k \cr a_- - 2k \cr } \right ) 
- \left ( \matrix{ a_0 - 2  \cr a_+ - 1 \cr } \right ) 
\eea
For the lowest non-trivial case $a_0=4$, a special case of the identity was encountered in (\ref{3d5}). For higher values of $a_0 \leq 8$ explicit formulas were listed in equation (5.19-23) of \cite{D'Hoker:2016a}.

\subsubsection{Primitive dihedral modular graph functions}
\label{sec:A11}

For primitive dihedral modular graph functions, the holomorphic subgraph reduction formulas simplify, as the total momentum of the subgraph, namely $p_0$ in (\ref{holform}) is not allowed to vanish. The remaining contributions arising from $p_0\not=0$ are identical to the ones considered earlier, so that the final formula, with $\Lambda _k$ given in (\ref{Lambda}), reads as follows,
\bea
\label{apA3}
&&
\hat  \cC^+ \! \left [ \matrix{a_+ & a_- & A \cr 0 & 0 & B  \cr} \right ] 
 - \left ( \matrix{ a_0-2  \cr a_+ -1} \right ) 
 \hat  \cC^+ \! \left [ \matrix{a_0-1 & 1 & A \cr 0 & 0 & B  \cr} \right ] 
 \\  && \hskip 0.1in  = 
  \left ( \matrix{ a_0 \cr a_+ } \right ) { (a_+-1)(a_--1) \over a_0-1} \, 
  \hat \cC^+ \! \left [ \matrix{a_0 & A  \cr 0 & B  \cr} \right ] 
 + \sum _{k=2}^{ [(a_0-1) /2]} \Lambda _k (a_+, a_-) \, \cG_{2k} \,
\hat \cC^+ \! \left [ \matrix{a_0 - 2 k & A  \cr 0 & B  \cr} \right ] 
 \no
\eea

\subsection{Trihedral modular graph functions}
\label{sec:A2}

Trihedral modular graph forms were defined  in (\ref{7a2}). (Note that  the sign convention used here differs from the one used in \cite{D'Hoker:2016a}.) Holomorphic subgraph reduction can be applied whenever the graph is of either of the following forms,
\bea 
\label{A7}
\cC ^+ \! \left [ \matrix{ a_1 & a_2 &A_1 \cr 0 & 0 & B_1\cr}  \Bigg | \matrix{ A_2 \cr B_2 \cr} 
\Bigg |  \matrix{ A_3  \cr B_3 \cr} \right ]
\hskip 1in
\cC ^+ \! \left [ \matrix{ a_1 &  A_1 \cr 0 & B_1\cr}  \Bigg | \matrix{a_2 & A_2 \cr 0 & B_2 \cr} 
\Bigg |  \matrix{ a_3 & A_3  \cr 0 & B_3 \cr} \right ]
\eea
where $A_i$ and $B_i$ are row vectors of length $R_i$ for $i=1,2,3$. Subgraphs of the first kind  give rise to so-called {\sl two-point holomorphic subgraphs}, while  subgraphs of the second type give rise to  {\sl three-point holomorphic subgraphs}. Two and three point reductions of the trihedral graph $D_{2,2,2}$  are represented in the figure (\ref{fig14}) below, where the holomorphic subgraph is indicated by the dashed edges.
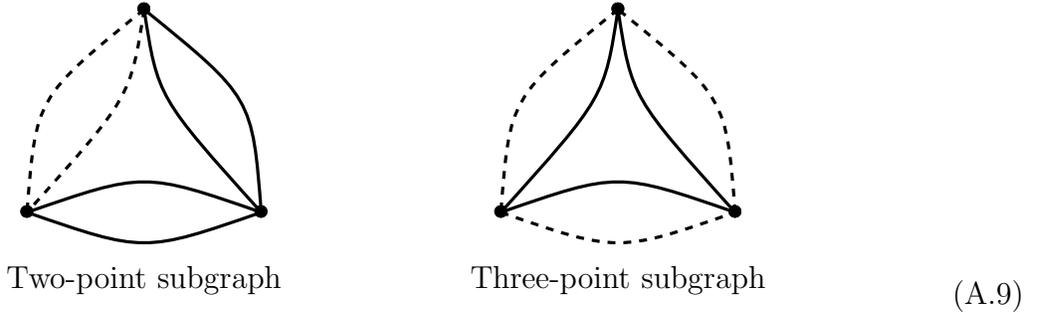
\bea
\label{fig14}
\tikzpicture[scale=0.9]
\scope[xshift=-5cm,yshift=-0.4cm]
\draw[very thick]  (1.73,-1) node{$\bullet$} .. controls (1.6,0.7) .. (0,2) node{$\bullet$};
\draw[very thick]  (1.73,-1) node{$\bullet$} .. controls (0.2,0.7) .. (0,2) node{$\bullet$};
\draw[very thick, dashed]  (0,2)       node{$\bullet$} .. controls (-1.6,0.7) ..(-1.73,-1) node{$\bullet$};
\draw[very thick, dashed]  (0,2)       node{$\bullet$} .. controls (-0.2,0.7) ..(-1.73,-1) node{$\bullet$};
\draw[very thick] (-1.73,-1) node{$\bullet$} .. controls (0,-1.6) .. (1.73,-1) node{$\bullet$};
\draw[very thick] (-1.73,-1) node{$\bullet$} .. controls (0,-0.4) .. (1.73,-1) node{$\bullet$};
\draw[very thick,dashed]  (8.73,-1) node{$\bullet$} .. controls (8.6,0.7) .. (7,2) node{$\bullet$};
\draw[very thick]  (8.73,-1) node{$\bullet$} .. controls (7.2,0.7) .. (7,2) node{$\bullet$};
\draw[very thick, dashed]  (7,2)       node{$\bullet$} .. controls (5.4,0.7) ..(5.27,-1) node{$\bullet$};
\draw[very thick]  (7,2)       node{$\bullet$} .. controls (6.8,0.7) ..(5.27,-1) node{$\bullet$};
\draw[very thick,dashed] (5.27,-1) node{$\bullet$} .. controls (7,-1.6) .. (8.73,-1) node{$\bullet$};
\draw[very thick] (5.27,-1) node{$\bullet$} .. controls (7,-0.4) .. (8.73,-1) node{$\bullet$};
\draw (0,-2) node{Two-point subgraph};
\draw (7,-2) node{Three-point subgraph};
\endscope
\endtikzpicture
\eea

\subsubsection{Two-point holomorphic subgraph reduction}
\label{sec:A21}

The pattern of identities for 2-point holomorphic subgraph reductions essentially coincides with the pattern of identities for dihedral graphs, for which all holomorphic subgraph reductions are of the 2-point type. To obtain these reductions, it suffices to substitute in (\ref{apA1}) and (\ref{apA2}) the following string of exponents, 
\bea
\matrix{ A \cr B \cr}
= \matrix{ A_1 \cr B_1 \cr} \Bigg | \matrix{ A_2 \cr B_2 \cr} \Bigg | \matrix{ A_3 \cr B_3 \cr}
\eea
to obtain the reduction formula for the first expression in (\ref{A7}). For example, the $a_0=4$ formula is given as follows, 
\bea
\label{11b3}
\cC ^+ \! \! \left [  \matrix{ 2 \, 2 \, A_1\cr 0 \, 0\,B_1 \cr} \! \Bigg | \! \matrix{ A_2 \cr B_2 \cr} \! \Bigg | \! \matrix{ A_3 \cr B_3 \cr} \right ] \!
- 2 \, \cC ^+  \! \! \left [ \matrix{ 3 \, 1\, A_1 \cr 0 \, 0\, B_1 \cr} \! \Bigg |  \!  \matrix{ A_2 \cr B_2 \cr} \! \Bigg | \! \matrix{ A_3 \cr B_3 \cr} \right ]
\! = 
2 \, \cC ^+ \! \! \left [ \matrix{ 4\, A_1 \cr 0\, B_1 \cr}  \! \Bigg | \!  \matrix{ A_2 \cr B_2 \cr}  \! \Bigg | \! \matrix{ A_3 \cr B_3 \cr} \right] 
 + 3   \, \cG_4 \, \cC ^+  \! \! \left [  \matrix{ A_1 \cr B_1 \cr}  \! \Bigg | \!  \matrix{ A_2 \cr B_2 \cr}  \! \Bigg | \! \matrix{ A_3 \cr B_3 \cr}  \right ] 
 \no
\eea
while the $a_0=5$ formula takes the from, 
\bea
\cC ^+ \! \! \left [  \matrix{ 3 \, 2 \, A_1\cr 0 \, 0 \, B_1 \cr}  \! \Bigg | \!  \matrix{ A_2 \cr B_2 \cr}  \! \Bigg | \! \matrix{ A_3 \cr B_3 \cr} \right ]
\! - 3 \, \cC ^+ \! \! \left [ \matrix{ 4 \, 1\, A_1 \cr 0 \, 0 \, B_1 \cr}   \! \Bigg | \!   \matrix{ A_2 \cr B_2 \cr}  \! \Bigg | \! \matrix{ A_3 \cr B_3 \cr} \right ] \!  =  
5 \, \cC ^+ \! \! \left [ \matrix{ 5 \, A_1 \cr 0 \, B_1 \cr}  \! \Bigg | \! \matrix{ A_2 \cr B_2 \cr}  \! \Bigg | \! \matrix{ A_3 \cr B_3 \cr} \right] \!
- 3   \, \cG_4 \, \cC ^+ \! \! \left [  \matrix{ 1 \, A_1 \cr 0 \, B_1 \cr}  \! \Bigg | \!  \matrix{ A_2 \cr B_2 \cr}  \! \Bigg | \! \matrix{ A_3 \cr B_3 \cr}  \right ] 
\no
\eea

\subsubsection{Three-point holomorphic subgraph reduction}
\label{sec:A22}

In this appendix, we list some of the three-point holomorphic subgraph reduction formulas for trihedral graphs, given in all generality by the second expression in (\ref{A7}). Their calculation  proceeds in a manner similar to the one in the two-point case, namely by isolating  the three-point subgraph with only holomorphic momentum dependence, and evaluating this subgraph by the methods of holomorphic Eisenstein series in terms of the loop momenta exterior to the subgraph. The final summations over these external momenta then provide the proper structure for the resulting modular graph forms. 

\sm

$\bullet$ The simplest such calculation was carried out explicitly in \cite{D'Hoker:2016a}, showing that,
\bea 
\cL_0 = \cC^+ \left[ \matrix{2 ~  a_2 \cr 0 ~ b_2 \cr} \Bigg | \matrix{2 ~ a_4 \cr 0 ~ b_4 \cr} \Bigg |  \matrix{2  \cr 0 \cr}\right] 
- 2\, \cC^+ \left[ \matrix{3 ~ a_2 \cr 0 ~ b_2 \cr} \Bigg | \matrix{2 ~ a_4 \cr 0 ~ b_4 \cr} \Bigg |  \matrix{1  \cr 0 \cr}\right] 
- 2\, \cC^+ \left[ \matrix{2 ~ a_2 \cr 0 ~ b_2 \cr} \Bigg | \matrix{3 ~ a_4 \cr 0 ~ b_4 \cr} \Bigg |  \matrix{1  \cr 0 \cr}\right]
\eea
reduces to 
\bea 
\cL_0 = 9 \, \cC^+ \left[ \matrix{a_2 + a_4 + 6 & 0 \cr b_2 + b_4 & 0 \cr}\right] 
- 3 \, \cG_4 \cC^+ \left[ \matrix{a_2 + a_4 + 2 & 0 \cr b_2 + b_4 & 0 \cr}\right]  + (-)^{a_4 + b_4} \cL'_0
\eea
where $\cL'_0$ is given by
\bea
\cL' _0&=& 
2 \, \cC^+ \left[ \matrix{a_2 +4 & a_4 & 2 \cr b_2 & b_4 & 0 \cr}\right] 
+ 2 \, \cC^+ \left[ \matrix{a_2  & a_4 + 4 & 2 \cr b_2 & b_4 & 0 \cr}\right]
\no\\&&
-3 \, \cC ^+ \left[ \matrix{a_2 +2 & a_4+2 & 2 \cr b_2 & b_4 & 0 \cr}\right] 
+2\, \cC ^+ \left[ \matrix{a_2 +1 & a_4+1 & 4 \cr b_2 & b_4 & 0 \cr}\right] 
\eea

$\bullet$ 
For the remaining cases of of weight six and seven, there are more edges to the graph and thus more external momenta to take into account. These extra sums complicate the calculation, as illustrated by the simplest cases
\bea
\label{7f1}
\cL_1  & = & 
3 \, \cC ^+ \left [ \matrix{ 4 ~ a_2 \cr 0 ~ b_2 \cr}  \Bigg | \matrix{ 1 ~ a_4 \cr 0 ~ b_4 \cr} \Bigg |  \matrix{ 1 ~ a_6 \cr 0 ~ b_6 \cr} \right ]
+  \cC ^+ \left [ \matrix{ 3 ~ a_2 \cr 0 ~ b_2 \cr}  \Bigg | \matrix{ 2 ~ a_4 \cr 0 ~ b_4 \cr} \Bigg |  \matrix{ 1 ~ a_6 \cr 0 ~ b_6 \cr} \right ]
+  \cC ^+ \left [ \matrix{ 3 ~ a_2 \cr 0 ~ b_2 \cr}  \Bigg | \matrix{ 1 ~ a_4 \cr 0 ~ b_4 \cr} \Bigg |  \matrix{ 2 ~ a_6 \cr 0 ~ b_6 \cr} \right ]
\no \\
\cL_2  & = & 
3 \, \cC ^+ \left [ \matrix{ 2 ~ a_2 \cr 0 ~ b_2 \cr}  \Bigg | \matrix{ 2 ~ a_4 \cr 0 ~ b_4 \cr} \Bigg |  \matrix{ 2 ~ a_6 \cr 0 ~ b_6 \cr} \right ]
+ 2 \, \cC ^+ \left [ \matrix{ 3 ~ a_2 \cr 0 ~ b_2 \cr}  \Bigg | \matrix{ 2 ~ a_4 \cr 0 ~ b_4 \cr} \Bigg |  \matrix{ 1 ~ a_6 \cr 0 ~ b_6 \cr} \right ] +\hbox{5 permutations}
\eea
with the permutations of the indices $2,4,6$ applied only to the second term of the second formula above. These evaluate respectively to  
\bea
\cL_1 = \sum_{n=1}^5 \cL_1^{(n)} \hspace{1.5 in} \cL_2 = \sum_{n=1}^5 \cL_2^{(n)}
\eea
with 
\bea
\label{L1}
\cL_1^{(1)} & = & 5 \, \cG_6 \, \cC ^+ \! \left [  \matrix{ a_2+a_4+a_6 & 0  \cr b_2 + b_4 + b_6 & 0  \cr}  \right ] 
\no \\
\cL_1^{(2)} & = & 
9 \, (-)^{a_4+b_4} \, \cC ^+ \! \left [  \matrix{ 6 & a_4 & a_2+a_6  \cr 0 & b_4 & b_2+b_6  \cr}  \right ] 
- 3 \, (-)^{a_4+b_4} \, \cG_4 \, \cC ^+ \! \left [  \matrix{ 2 & a_4 & a_2+a_6  \cr 0 & b_4 & b_2+b_6  \cr}  \right ] 
\no \\
\cL_1^{(3)} & = & \cL_1^{(2)} (a_4 \leftrightarrow a_6)
\no \\
\cL_1^{(4)} & = & 
-5  \, (-)^{a_2+b_2}                 \, \cC ^+ \! \left [  \matrix{ 6 & a_2 & a_4+a_6  \cr 0 & b_2 & b_4+b_6  \cr}  \right ] 
+ 3 \, (-)^{a_2+b_2} \, \cG_4 \, \cC ^+ \! \left [  \matrix{ 2 & a_2 & a_4+a_6  \cr 0 & b_2 & b_4+b_6  \cr}  \right ] 
\no \\
\cL_1^{(5)} & = & 
\, \cC ^+ \! \left [  \matrix{ 3 & a_2 \cr 0 & b_2 \cr } \Bigg | \matrix{3 & a_4  \cr 0 & b_4 \cr} \Bigg | \matrix{ a_6 \cr b_6 \cr}  \right ] 
- 3 \, \cG_4 \, \cC^+  \! \left [  \matrix{ 1 & a_2 \cr 0 & b_2 \cr } \Bigg | \matrix{1 & a_4  \cr 0 & b_4 \cr} \Bigg | \matrix{ a_6 \cr b_6 \cr}  \right ] 
\no \\ &&
+ 3 \, \cC ^+ \! \left [  \matrix{ 4 & a_2 \cr 0 & b_2 \cr } \Bigg | \matrix{2 & a_4  \cr 0 & b_4 \cr} \Bigg | \matrix{ a_6 \cr b_6 \cr}  \right ] 
- 9 \, (-)^{a_6+b_6}  \, \cC ^+ \! \left [  \matrix{ 6 & a_6 & a_2+a_4  \cr 0 & b_6 & b_2+b_4  \cr}  \right ] 
\no \\ &&
+ 5 \, \cC ^+ \! \left [  \matrix{ 5 & a_2 \cr 0 & b_2 \cr } \Bigg | \matrix{1 & a_4  \cr 0 & b_4 \cr} \Bigg | \matrix{ a_6 \cr b_6 \cr}  \right ] 
+ 3 \, (-)^{a_6+b_6} \cG_4 \, \cC ^+ \! \left [  \matrix{ 2 & a_6 & a_2+a_4  \cr 0 & b_6 & b_2+b_4  \cr}  \right ]  
+ (4 \leftrightarrow 6) \qquad
\eea
and
\bea
\label{L2}
\cL_2^{(1)} & = & 15 \, \cG_6 \, \cC ^+ \! \left [  \matrix{ a_2+a_4+a_6 & 0  \cr b_2 + b_4 + b_6 & 0  \cr}  \right ] 
\no \\
\cL_2^{(2)} & = & 
- \, (-)^{a_4+b_4} \, \cC ^+ \! \left [  \matrix{ 6 & a_4 & a_2+a_6  \cr 0 & b_4 & b_2+b_6  \cr}  \right ] 
+ 3 \, (-)^{a_4+b_4} \, \cG_4 \, \cC ^+ \! \left [  \matrix{ 2 & a_4 & a_2+a_6  \cr 0 & b_4 & b_2+b_6  \cr}  \right ] 
\no \\
\cL_2^{(3)} & = & \cL_2^{(2)}(a_4 \leftrightarrow a_6) 
\no \\
\cL_2^{(4)} & = &  \cL_2^{(2)}(a_2 \leftrightarrow a_4) 
\no\\
\cL_2^{(5)} & = & 
2 \, \cC ^+ \! \left [  \matrix{ 3 & a_2 \cr 0 & b_2 \cr } \Bigg | \matrix{3 & a_4  \cr 0 & b_4 \cr} \Bigg | \matrix{ a_6 \cr b_6 \cr}  \right ] 
+ 2 \, \cC ^+ \! \left [  \matrix{ 4 & a_2 \cr 0 & b_2 \cr } \Bigg | \matrix{2 & a_4  \cr 0 & b_4 \cr} \Bigg | \matrix{ a_6 \cr b_6 \cr}  \right ] 
-  \, \cC ^+ \! \left [  \matrix{ 2 & a_2 \cr 0 & b_2 \cr } \Bigg | \matrix{4 & a_4  \cr 0 & b_4 \cr} \Bigg | \matrix{ a_6 \cr b_6 \cr}  \right ] 
\no \\ &&
+  (-)^{a_6+b_6}  \, \cC ^+ \! \left [  \matrix{ 6 & a_6 & a_2+a_4  \cr 0 & b_6 & b_2+b_4  \cr}  \right ] 
- 4 \, \cC ^+ \! \left [  \matrix{ 1 & a_2 \cr 0 & b_2 \cr } \Bigg | \matrix{5 & a_4  \cr 0 & b_4 \cr} \Bigg | \matrix{ a_6 \cr b_6 \cr}  \right ] 
+ (4 \leftrightarrow 6)
\qquad
\eea

\subsubsection{Primitive trihedral modular graph functions}
\label{sec:A23}

The modifications required  for primitive trihedral modular graph functions are analogous to the ones required in the dihedral case. For example, instead of the relations of section \ref{sec:A21} for two-point reduction, we have,
\bea
\label{11b9}
\hat \cC^+ \left [  \matrix{ 2 \, 2 \, A_1\cr 0 \, 0\,B_1 \cr} \Bigg |  \matrix{ A_2 \cr B_2 \cr} \Bigg | \matrix{ A_3 \cr B_3 \cr} \right ] 
- 2 \, \hat \cC^+   \left [ \matrix{ 3 \, 1\, A_1 \cr 0 \, 0\, B_1 \cr}  \Bigg |    \matrix{ A_2 \cr B_2 \cr} \Bigg | \matrix{ A_3 \cr B_3 \cr} \right ] 
& = & 
2 \, \hat \cC^+ \left [ \matrix{ 4\, A_1 \cr 0\, B_1 \cr} \Bigg |  \matrix{ A_2 \cr B_2 \cr} \Bigg | \matrix{ A_3 \cr B_3 \cr} \right] 
\no \\
\hat \cC^+ \left [  \matrix{ 3 \, 2 ~A_1\cr 0 \, 0~ B_1 \cr} \Bigg |  \matrix{ A_2 \cr B_2 \cr} \Bigg | \matrix{ A_3 \cr B_3 \cr} \right ]
\! - 3 \, \hat \cC^+ \left [ \matrix{ 4 \, 1~A_1 \cr 0 \, 0~B_1 \cr}  \Bigg |    \matrix{ A_2 \cr B_2 \cr} \Bigg | \matrix{ A_3 \cr B_3 \cr} \right ] 
& = & 
5 \, \hat \cC^+  \! \left [ \matrix{ 5~A_1 \cr 0~B_1 \cr} \Bigg |  \matrix{ A_2 \cr B_2 \cr} \Bigg | \matrix{ A_3 \cr B_3 \cr} \right] 
\eea
For the three-point reductions discussed in \ref{sec:A22}, we replace all $\cC$ by $\hat \cC$: for $\cL_0$ no other modifications are required;  for $\cL_1$ we drop the $\cL_1^{(1)}$ term;   for $\cL_2$ we drop the $\cL_2^{(1)}$ term.

\subsection{Tetrahedral modular graph function}
\label{sec:A3}

The holomorphic subgraph reduction for the unique weight six tetrahedral graph $\cD_T$ of subsection \ref{sec:35} involves holomorphic subgraphs with 3 points and 4 points, but none with two points, as no 2-point closed subgraphs exist. Although the calculations of 3-point and 4-point reductions follow in the spirit of the previous ones,  the holomorphic subgraph reduction procedure for  tetrahedral diagram has not yet been explored  in earlier papers. Therefore, we shall present its derivation below in some detail. In the figure (\ref{fig15}) we represent the two cases separately, and have indicated in dashed line the edges corresponding to the holomorphic subgraph.
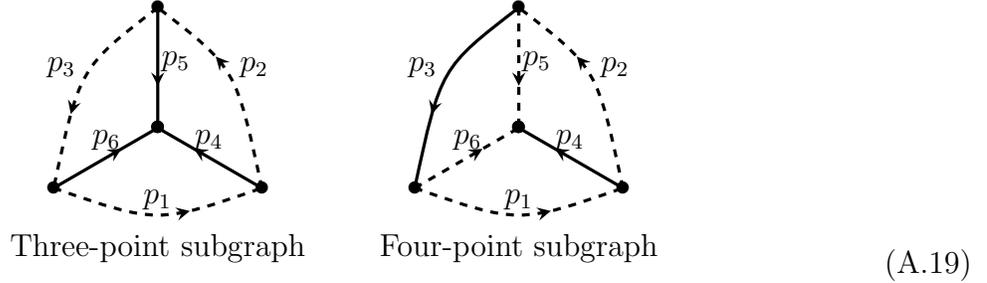
\bea
\label{fig15}
\tikzpicture[scale=0.8]
\scope[xshift=-5cm,yshift=-0.4cm]
\draw[directed, very thick, dashed]  (1.73,-1) ..controls (1.3,1) .. (0,2);
\draw[directed, very thick, dashed] (0,2) ..controls (-1.3,1) ..(-1.73,-1);
\draw[directed, very thick,dashed] (-1.73,-1)  ..controls (0,-1.6) .. (1.73,-1);
\draw[ directed, very thick] (0,2) node{$\bullet$} --  (0,0) node{$\bullet$};
\draw[ directed, very thick] (-1.73,-1) node{$\bullet$} --  (0,0) node{$\bullet$};
\draw[ directed, very thick] (1.73,-1) node{$\bullet$} --  (0,0) node{$\bullet$};
\draw (1.6,1) node{$p_2$};
\draw (-1.6,1) node{$p_3$};
\draw (0,-1.2) node{$p_1$};
\draw (0.3,1.1) node{$p_5$};
\draw (-0.85,-0.2) node{$p_6$};
\draw (0.85,-0.2) node{$p_4$};
\draw[directed, very thick, dashed]  (7.73,-1) ..controls (7.3,1) .. (6,2);
\draw[directed, very thick] (6,2) ..controls (4.7,1) ..(4.27,-1);
\draw[directed, very thick,dashed] (4.27,-1)  ..controls (6,-1.6) .. (7.73,-1);
\draw[ directed, very thick,dashed] (6,2) node{$\bullet$} --  (6,0) node{$\bullet$};
\draw[ directed, very thick,dashed] (4.27,-1) node{$\bullet$} --  (6,0) node{$\bullet$};
\draw[ directed, very thick] (7.73,-1) node{$\bullet$} --  (6,0) node{$\bullet$};
\draw (7.6,1) node{$p_2$};
\draw (4.4,1) node{$p_3$};
\draw (6,-1.2) node{$p_1$};
\draw (6.3,1.1) node{$p_5$};
\draw (5.15,-0.2) node{$p_6$};
\draw (6.85,-0.2) node{$p_4$};
\draw (0,-2) node{Three-point subgraph};
\draw (6,-2) node{Four-point subgraph};
\endscope
\endtikzpicture
\eea

\subsubsection{Three-point reduction identity}

From the structure of $\nabla ^3 \cD_T$ given in (\ref{D3DT}), and the use of the symmetries of the graphs given in (\ref{symtet}), all 3-point reductions that are required may be put in the following form, 
\bea
\cL_3 & = & 
+\cD ^+ \left [ \matrix{ 4 ~ 1 ~ 1 ~ a_4 ~ a_5 ~ a_6 \cr 0 ~  0 ~  0 ~ b_4 ~  b_5 ~ b_6 \cr } \right ]
+ 
\hbox{2 permutations of (411)}
\no \\ &&
+ 
\cD ^+ \left [ \matrix{ 3 ~ 2 ~ 1 ~ a_4 ~ a_5 ~ a_6 \cr 0 ~ 0 ~  0 ~ b_4 ~  b_5 ~b_6 \cr } \right ]
+ 
\hbox{5 permutations of (321)}
\no \\ &&
+ 
\cD ^+ \left [ \matrix{ 2 ~ 2 ~ 2 ~ a_4 ~ a_5 ~ a_6 \cr 0 ~  0 ~  0 ~ b_4 ~  b_5 ~ b_6 \cr } \right ]
\eea
The holomorphic 3-point subgraph, which has been made to lie always on the edges labelled $1,2,3$, may be isolated
in the total sum over momenta as follows,
\bea
\cL_3 = \sum _{p_4, p_5, p_6 \in \Lambda} ' { \tau_2 ^6 \over \pi^3}  \, X (p_4,p_5,p_6) \prod _{i=4,5,6} { \tau _2  \over \pi \, p_i \, \bar p_i}
\eea
The value of the 3-point holomorphic subgraph is given by,
\bea
X(p_4,p_5,p_6) & = & \sum ' _{p_1,p_2,p_3} \Bigg  ( 
{1 \over p_1^4 p_2 p_3} 
+ { 1 \over p_1 p_2^4 p_3} 
+ { 1 \over p_1 p_2 p_3^4 } 
+ { 1 \over p_1^3 p_2^2 p_3 } 
+ { 1 \over p_1^3 p_2 p_3^2 } 
\no \\ && \hskip 0.5in
+ { 1 \over p_1^2 p_2^3 p_3 } 
+ { 1 \over p_1^2 p_2 p_3^3 } 
+ { 1 \over p_1 p_2^3 p_3^2 } 
+ { 1 \over p_1 p_2^2 p_3^3 } 
+ { 1 \over p_1^2 p_2^2 p_3^2 } 
\Bigg )
\eea
Parametrizing the independent loop momentum by $p_1$, we have,
\bea
p_2 & = & p_1-p_4 \hskip 1in p_1\not=p_4
\no \\
p_3 & = & p_1+p_6 \hskip 1in p_1\not=-p_6
\eea
Expressing $p_2, p_3$ in terms of $p_1$ and the loop momenta $p_4, p_5, p_6$ external to the holomorphic subgraph, and then decomposing into partial fractions, we find, 
\bea
X(p_4,p_5,p_6) = \sum _{p_1 \not= 0, p_4, -p_6} 
\left ( 
-{ 1 \over p_4 p_6 p_1^4} - { 1 \over p_4 p_5 (p_1-p_4)^4}  - { 1 \over p_5 p_6 (p_1+p_6)^4} \right )
\eea
We use the summation formulas,
\bea
\sum _{p_1 \not= 0, p_4, - p_6} { 1 \over p_1^4} & = & \pi ^2 G_4 -{ 1 \over p_4^4} - { 1 \over p_6^4} 
\no \\
\sum _{p_1 \not= 0, p_4, - p_6} { 1 \over (p_1-p_4)^4} & = & \pi ^2 G_4 -{ 1 \over p_4^4} - { 1 \over p_5^4} 
\no \\
\sum _{p_1 \not= 0, p_4, - p_6} { 1 \over (p_1+p_6)^4} & = & \pi ^2 G_4 -{ 1 \over p_5^4} - { 1 \over p_6^4} 
\eea
The terms in $G_4$ cancel  in view of the relation $p_4+p_5+p_6=0$, and we obtain, 
\bea
X(p_4,p_5,p_6) =
{ 1 \over p_4^5 p_5}+ { 1 \over p_4^5 p_6}+ { 1 \over p_5^5 p_4}+ { 1 \over p_5^5 p_6}+ { 1 \over p_6^5 p_4}+ { 1 \over p_6^5 p_5}
\eea
Putting it all together, we find, 
\bea
\cL_3 =  \cC^+ \left [ \matrix{ a_4+5 & a_5+1 & a_6 \cr b_4 & b_5 & b_6 \cr } \right ]
+ \hbox{ 5 permutations of 4, 5, 6}
\eea
For the specific graph $\cD_T$ needed here, we set $a_4=a_5=a_6=b_4=b_5=b_6=1$.

\subsubsection{Four-point reduction identity}

We evaluate  only the combination encountered for the weight six tetrahedral  graph $\cD_T$,
\bea
\cL_4 & = & 
\cD^+ \left [ \matrix{ 2 ~ 2 ~ 1 ~ 1 ~ 2 ~ 2 \cr 0 ~ 0 ~ 1 ~ 1 ~ 0 ~ 0 \cr } \right ]
+ 4 \, \cD^+ \left [ \matrix{ 3 ~ 2 ~ 1 ~ 1 ~ 1 ~ 2 \cr 0 ~ 0 ~ 1 ~ 1 ~ 0 ~ 0 \cr } \right ]
- 8 \, \cD^+ \left [ \matrix{ 3 ~ 2 ~ 1 ~ 1 ~ 2 ~ 1 \cr 0 ~ 0 ~ 1 ~ 1 ~ 0 ~ 0 \cr } \right ]
\no \\ &&
- 2 \, \cD^+ \left [ \matrix{ 3 ~ 1 ~ 1 ~ 1 ~ 3 ~ 1 \cr 0 ~ 0 ~ 1 ~ 1 ~ 0 ~ 0 \cr } \right ]
-4 \, \cD^+ \left [ \matrix{ 3 ~ 3 ~ 1 ~ 1 ~ 1 ~ 1 \cr 0 ~ 0 ~ 1 ~ 1 ~ 0 ~ 0 \cr } \right ]
\no \\ &&
- 4 \, \cD^+ \left [ \matrix{ 4 ~ 1 ~ 1 ~ 1 ~ 2 ~ 1 \cr 0 ~ 0 ~ 1 ~ 1 ~ 0 ~ 0 \cr } \right ]
- 8 \, \cD^+ \left [ \matrix{ 4 ~ 2 ~ 1 ~ 1 ~ 1 ~ 1 \cr 0 ~ 0 ~ 1 ~ 1 ~ 0 ~ 0 \cr } \right ]
- 4 \, \cD^+ \left [ \matrix{ 5 ~ 1 ~ 1 ~ 1 ~ 1 ~ 1 \cr 0 ~ 0 ~ 1 ~ 1 ~ 0 ~ 0 \cr } \right ]
\eea
We  parametrize the momenta $p_1, p_2,p_5,p_6$ of the 4-point holomorphic subgraph in terms of $p_1$ and the loop momenta $p_3, p_4$ external to the holomorphic subgraph,
\bea
p_2 & = & p_1 - p_4 \hskip 1.3in p_1 \not= p_4
\no \\
p_6 & = & -p_1 + p_3 \hskip 1.2in p_1 \not= p_3
\no \\
p_5 & = & p_1 - p_3- p_4 \hskip 1in p_1 \not= p_3+p_4
\eea
One divides up the contributions into those with $p_3 \pm p_4 \not=0 $ and those with $p_3\pm p_4=0$,
\bea
\cL_4=\cL_4'+\cL_4''
\eea
For $p_3 \pm p_4 \not= 0$, the sum over $p_1$ reduces to,
\bea
X(p_3,p_4) & = & \sum _{p_1 \not= p_3, p_4, p_3+p_4}
\left ( 
-{ 1 \over p_3 p_4 (p_3+p_4) p_1^5} -{ 1 \over p_3 p_4 (p_3-p_4) (p_1-p_3)^5} \right .
\no \\ && \hskip 0.5in \left . 
+ { 1 \over p_3 p_4 (p_3+p_4) (p_1-p_3-p_4)^5} +{ 1 \over p_3 p_4 (p_3-p_4) (p_1-p_4)^5} 
\right )
\qquad
\eea
The sums are evaluated as follows,
\bea
\sum _{p_1 \not= p_3, p_4, p_3+p_4} { 1 \over p_1^5} & = &
- { 1 \over p_3^5} - { 1 \over p_4^5} - { 1 \over (p_3+p_4)^5} 
\no \\
\sum _{p_1 \not= p_3, p_4, p_3+p_4} { 1 \over (p_1-p_3)^5} & = &
+ { 1 \over p_3^5} - { 1 \over p_4^5} + { 1 \over (p_3-p_4)^5} 
\no \\
\sum _{p_1 \not= p_3, p_4, p_3+p_4} { 1 \over (p_1-p_4)^5} & = &
- { 1 \over p_3^5} + { 1 \over p_4^5} - { 1 \over (p_3-p_4)^5} 
\no \\
\sum _{p_1 \not= p_3, p_4, p_3+p_4} { 1 \over (p_1-p_3-p_4)^5} & = &
+ { 1 \over p_3^5} + { 1 \over p_4^5} + { 1 \over (p_3+p_4)^5} 
\eea
so that $X$ evaluates to,
\bea
X(p_3,p_4) & = & 
{ 2 \over p_3^6 p_4 (p_3+p_4)} 
+{ 2 \over p_3 p_4^6 (p_3+p_4)} 
+{ 2 \over p_3 p_4 (p_3+p_4)^6} 
\no \\ &&
-{ 2 \over p_3^6 p_4 (p_3-p_4)} 
-{ 2 \over p_3 p_4^6 (p_4-p_3)} 
-{ 2 \over p_3 p_4 (p_3-p_4)^6} 
\eea
Translating the sums over $p_3,p_4$ into standard notations, we get after some rearrangements,
\bea
\cL_4' = 
 8 \, \cC^+ \left [ \matrix{ 7 ~ 2 ~ 1 \cr 1 ~ 0 ~ 1 \cr } \right ] 
 - { 195 \over 16 } \, \cC^+ \left [ \matrix{ 10 &  0 \cr 2 & 0  \cr } \right ]
\eea
The contributions from $p_3+p_4=0$ and $p_3-p_4=0$ are equal to one another, and we find, 
\bea
\cL_4'' = - 10 \, \cG_6 \, \cC^+ \left [ \matrix{ 4 &  0 \cr 2 & 0  \cr } \right ]
+ {353 \over 16 } \, \cC^+ \left [ \matrix{ 10 &  0 \cr 2 & 0  \cr } \right ]
\eea
In total, we get,
\bea
\cL_4=  8 \, \cC^+ \left [ \matrix{ 7 ~ 2 ~ 1 \cr 1 ~ 0 ~ 1 \cr } \right ] 
- 10 \, \cG_6 \, \cC^+ \left [ \matrix{ 4 &  0 \cr 2 & 0  \cr } \right ]
+ 18 \, \cC^+ \left [ \matrix{ 10 &  0 \cr 2 & 0  \cr } \right ]
\eea
More general 4-point reductions may be evaluated with the same methods.


\end{document}